\documentclass[11pt,a4paper]{article}
\pdfoutput=1
\usepackage{jheppub_kim}

\usepackage{multirow, graphicx,amssymb,url,mathrsfs,amsmath}
\usepackage{wrapfig,boxedminipage,setspace,subfigure,epsfig}
\usepackage{amsxtra,amstext,latexsym,dsfont,amsfonts}
\usepackage{color,eucal}

\newcommand{\dd}{\mathrm{d}}

\renewcommand{\Re}{{ \rm Re}}
\renewcommand{\Im}{{ \rm Im}}

\newcommand{\nJ}{\mbox{${J}$}}  
\newcommand{\nT}{\mbox{${T}$}} 
\newcommand{\nS}{\mbox{${S}$}}  
\newcommand{\nQ}{\mbox{${Q}$}}

\newcommand{\tT}{\mbox{$\tilde{T}$}}
\newcommand{\trho}{\mbox{$\tilde{\rho}$}}

\title{Ward Identity and Homes' Law in a Holographic Superconductor with Momentum Relaxation}

\author[a]{Keun-Young Kim,}
\author[a,b]{Kyung Kiu Kim,}
\author[c]{and Miok Park}

\emailAdd{fortoe@gmail.com}
\emailAdd{kimkyungkiu@gmail.com}
\emailAdd{miokpark76@gmail.com}

\affiliation[a]{ School of Physics and Chemistry, Gwangju Institute of Science and Technology,
Gwangju 61005, Korea
}
\affiliation[b]{Department of Physics, College of Science, Yonsei University, Seoul 120-749, Korea
}
\affiliation[c]{School of Physics, Korea Institute for Advanced Study, Seoul 130-722, Korea
}

\abstract{
We study three properties of a holographic superconductor related to conductivities, where momentum relaxation plays an important role.  First, we find that there are constraints between electric, thermoelectric and thermal conductivities. The constraints are analytically derived by the Ward identities regarding diffeomorphism from field theory perspective. We confirm them by numerically computing all two-point functions from holographic perspective. Second, we investigate Homes' law and Uemura's law for various high-temperature and conventional superconductors. They are empirical and  (material independent) universal relations between the superfluid density at zero temperature, the transition temperature, and the electric DC conductivity right above the transition temperature. In our model, it turns out that the Homes' law does not hold but the Uemura's law holds at small momentum relaxation related to coherent metal regime. Third, we explicitly show that the DC electric conductivity is finite for a neutral scalar instability while it is infinite for a complex scalar instability. This shows that the neutral scalar instability has nothing to do with superconductivity as expected.
 }


\subheader{
$  $
          }

\begin{document}

\maketitle

\section{Introduction}

Holographic methods have provided novel and effective tools to study strongly correlated systems~\cite{CasalderreySolana:2011us,Hartnoll:2009sz,Herzog:2009xv, Iqbal:2011ae} and  they have been applied to many condensed matter problems. In particular, holographic understanding 
of high $T_c$ superconductor is one of the important issues.  After the first holographic superconductor model
proposed by Hartnoll, Herzog, and Horowitz (HHH)\footnote{The HHH model is a class of Einstein-Maxwell-complex scalar action with negative cosmological constant.} \cite{Hartnoll:2008vx, Hartnoll:2008kx}, there have been extensive development and extension of the model. For reviews and references, we refer to \cite{Hartnoll:2009sz, Herzog:2009xv, Horowitz:2010gk, Cai:2015cya}.

The HHH model is a  translationally invariant system with finite charge density. Therefore, it cannot relax momentum and exhibits an infinite electric DC conductivity even in normal phase not only in superconducting phase. To construct more realistic superconductor models, a few methods incorporating momentum relaxation were proposed. 
One way of including momentum relaxation is to break translational invariance explicitly by imposing inhomogeneous (spatially modulated) boundary conditions on a bulk field~\cite{Horowitz:2012ky, Horowitz:2012gs,Ling:2013nxa,Chesler:2013qla,Donos:2014yya}. 
Massive gravity models~\cite{Vegh:2013sk,Davison:2013jba,Blake:2013bqa,Blake:2013owa, Amoretti:2014zha,Amoretti:2014mma,Amoretti:2015gna} give some gravitons mass terms, which breaks bulk diffeomorphism and translation invariance in the boundary field theory. Holographic Q-lattice models~\cite{Donos:2013eha,Donos:2014uba,Ling:2014bda,Ling:2015exa,Ling:2016ewj} take advantage of a global symmetry of the bulk theory. For example, a global phase of a complex scalar plays a role of breaking translational invariance. Models with massless scalar fields linear in spatial coordinate~\cite{Andrade:2013gsa,Gouteraux:2014hca,Taylor:2014tka, Kim:2014bza,Bardoux:2012aw,Iizuka:2012wt, Cheng:2014qia, Fang:2015dia,Seo:2015pug, Andrade:2015iyf, Andrade:2016tbr} utilize the shift symmetry. Some models  with a Bianchi VII$_0$ symmetry are dual to helical lattices~\cite{Donos:2012js, Donos:2014oha,Donos:2014gya}. Based on these models, holographic superconductor incorporating  momentum relaxation have been developed~\cite{Horowitz:2013jaa, Zeng:2014uoa, Ling:2014laa, Andrade:2014xca, Kim:2015dna, Erdmenger:2015qqa, Baggioli:2015zoa,Baggioli:2015dwa,Koga:2014hwa, Bai:2014poa}.

In this paper, we study the HHH holographic superconductor model with massless scalar fields  linear in spatial coordinate~\cite{Kim:2015dna}, where  the strength of momentum relaxation is identified with the proportionality constant to spatial coordinate. 
The property of the normal phase of this model such as thermodynamics and transport coefficients were studied in \cite{Bardoux:2012aw,Andrade:2013gsa,Donos:2014cya,Taylor:2014tka,Kim:2014bza,Kim:2015sma,Kim:2015wba}. The superconducting phase was analysed in \cite{Andrade:2014xca,Kim:2015dna}.
In particular, optical  electric, thermoelectric and thermal conductivities of the model have been extensively studied in \cite{Kim:2014bza,Kim:2015dna,Kim:2015sma,Kim:2015wba}. Building on them, we further investigate interesting properties related to conductivities and momentum relaxation. 
There are three issues that we want to address in this paper: (1) conductivities with a neutral scalar hair instability, (2) Ward identities: constraints between conductivities, (3) Homes' law and Uemura's law. 
We explain each issue in the following. 

(1) In a holographic superconductor model of a Einstein-Maxwell-scalar action~\cite{Hartnoll:2008kx, Horowitz:2010gk}, a superconducting state is characterized by the formation of a complex scalar hair below some critical temperature. In essence, 
the complex scalar is turned on by coupling between the maxwell field and complex scalar through the covariant derivative. 
Interestingly, it was also observed~\cite{Hartnoll:2008kx, Horowitz:2010gk}  that a different mechanism for the instability forming neutral scalar hair\footnote{A neutral scalar may arise from  a top-down setting~\cite{Banks:2015aca,Banks:2016fab}.} is possible. This instability was not associated with superconductivity because it does not break a $U(1)$ symmetry, but at most breaks a $\mathbb Z_2$ symmetry  $\Phi \rightarrow -\Phi$. 
Therefore, in this system with a neutral scalar hair, it is natural to expect that DC electric conductivity will be finite contrary to the case with a complex scalar hair (superconductor).  However, to our knowledge, it has not been checked yet. In the early models without momentum relaxation, this question is not well posed since electric DC conductivity is always infinite due to translation invariance and finite density.
In this paper,  in a model with momentum relaxation, we show that DC electric conductivity is indeed finite with a neutral scalar hair.

(2) It was shown~\cite{Herzog:2009xv, Hartnoll:2007ip}, in normal phase without momentum relaxation, there are two constraints relating three transport coefficients: electric conductivity($\sigma$), thermoelectric conductivity($\alpha$) and thermal conductivity($\bar{\kappa}$). The constraints can be derived by the Ward identity regarding diffeomorphism. Thanks to these two constraints, $\alpha$ and $\bar{\kappa}$ can be obtained algebraically once $\sigma$ is computed numerically. This is why only $\sigma$ is presented in the literature~\cite{Hartnoll:2009sz}. In our model, there is an extra field, a massless scalar for momentum relaxation, and it turns out  there are three Ward identities of six two-point functions: $\sigma$, $\alpha$ and $\bar{\kappa}$ and  three more two-point functions related to the operator dual to a scalar field. Therefore, the information of $\sigma$ alone cannot determine $\alpha$ and $\bar{\kappa}$. If we know three two-point functions then the Ward identities   enable us to compute the other three two-point functions. In this paper, following the method in~\cite{Herzog:2009xv}, we first derive the Ward identities for two-point functions analytically from field theory perspective. Next, we confirm them numerically from holographic perspective. This confirmation of the Ward identities also demonstrates 
the faithfulness of our numerical method.

(3) Homes' law and Uemura's law are empirical and material independent universal laws for high-temperature and some conventional superconductors~\cite{Homes:2005aa,Homes:2004wv}. 
The law states that,  for various superconductors, there is a {\it universal} material independent  relation between the superfluid density ($\trho_{s}$) at near zero temperature and the transition temperature ($\tT_c$) multiplied by the electric DC conductivity ($\sigma_{DC}$) in the normal state right above the transition temperature $\tT_c$. 
\begin{equation}
\trho_{s}(\tT = 0) = C \sigma_{DC}(\tT_{c}) \, \tT_{c} \,,
\end{equation}
where $\trho_s$, $\tT_c$ and $\sigma_{DC}$ are scaled to be dimensionless, and $C$ is a dimensionless universal constant:  $C \approx 4.4$ or $8.1$. They are computed in~\cite{Erdmenger:2015qqa} from the experimental data in \cite{Homes:2005aa,Homes:2004wv}.  
 For in-plane high $T_c$ superconductors and clean BCS superconductors  $C \approx 4.4$. For c-axis high $T_c$ superconductors and BCS superconductors in the dirty limit $C \approx 8.1$.  Notice that momentum relaxation is essential here because without momentum relaxation $\sigma_{DC}$ is infinite. There is another similar universal relation, Uemura's law, which holds only for underdoped cuprates~\cite{Homes:2005aa,Homes:2004wv}:

\begin{equation}
\trho_{s}(\tT = 0) = B \, \tT_{c} \,.
\end{equation}
where $B$ is another universal constant. 
 In the context of holography Homes' law  was studied in~\cite{Erdmenger:2012ik, Erdmenger:2015qqa}. It was 
 motivated~\cite{Erdmenger:2012ik} by holographic bound of the ratio of  shear viscosity to entropy density ($\eta/s$) in strongly correlated plasma~\cite{CasalderreySolana:2011us} and its understanding in terms of quantum criticality \cite{Sachdev:2011cs} or Planckian dissipation~\cite{Zaanen:2004aa},where the time scale of dissipation is shortest possible. Since Homes' law also may arise in systems of Planckian dissipation~\cite{Zaanen:2004aa} there is a good chance to find universal physics in condensed matter system as well as in quark-gluon plasma. 
In \cite{Erdmenger:2015qqa}  Homes' law was observed in a holographic superconductor model in a helical lattice for some restricted parameter regime of momentum relaxation,  while Uemura's law did not hold in that model.  However, physic behind Homes' law in this model has not been clearly understood yet. For further understanding on Homes' law, in this paper, we have checked Homes' law and Uemuras' law in our holographic superconductor model. We find that Homes' law does not hold but Uemura's law holds at small momentum relaxation region, related to coherent metal regime.

This paper is organised as follows. In section \ref{sec2}, we introduce our holographic superconductor model incorporating momentum relaxation by massless real scalar fields. 
The equilibrium state solutions and the method to compute AC conductivities are briefly reviewed.
In section \ref{sec3}, the conductivities with a neutral scalar  instability are computed and compared with the ones with a complex hair instability. In section \ref{sec4}, we first derived Ward identities giving constraints between conductivities analytically from field theory perspective. These identities are confirmed numerically by holographic method. In section \ref{sec5},  after analysing conductivities at small frequency, we discuss the Home's law and Uemura's law in our model.
In section \ref{sec6} we conclude.

\section{AC conductivities: holographic model and method}\label{sec2}

\subsection{Equilibrium state}

In this section we  briefly review the holographic superconductor model we study, referring to  \cite{Andrade:2013gsa, Kim:2014bza,Kim:2015sma, Kim:2015wba, Bianchi:2001kw} for more complete and detailed analysis. 
We consider the action\footnote{The complete action includes also the Gibbons Hawking term and some boundary terms for holographic renormalization, which are explained in \cite{Andrade:2013gsa, Kim:2014bza, Kim:2015sma, Kim:2015wba, Bianchi:2001kw} in more detail.}
\begin{equation}\label{model A}
S = \int \dd^4 x \sqrt{-g} \left[  R - 2\Lambda   - \frac{1}{4} F^2  -|  D\Phi |^2  - m^2 |\Phi|^2 -\frac{1}{2} \sum_{I=1}^2 (\partial \psi_I)^2      \right]
\,, 
\end{equation}
where $x^M=\{t,x,y,r\}$ and $r$ is the holographic direction.  $R$ is the Ricci scalar and $\Lambda = -3/L^2$ is the cosmological constant with the AdS radius $L=1$. 
We have included the field strength $F = \dd A$ for a $U(1)$ gauge field $A$,  the complex scalar field $\Phi$ with mass $m$, two massless scalar fields, $\psi_I(I=1,2)$.  
The covariant derivative is defined by  $D_M\Phi \equiv \nabla_M \Phi - i q A_M \Phi $ with the charge $q$ of the complex scalar field.
The action \eqref{model A} yields equations of motion
\begin{align}\label{EOMs}
& R_{MN} - \frac{1}{2}g_{MN} \left(  R + 6 -\frac{1}{4}F^2   -  |D\Phi|^2 - m^2  |\Phi| ^2  - \frac{1}{2}\sum_{I=1}^{2} (\partial\psi_I)^2 \right) \nonumber \\ & \qquad = \frac{1}{2} F_{MQ}{F_N}^Q + \frac{1}{2}\left(  D_M \Phi D_N \Phi^* + D_N \Phi D_M \Phi^* \right)  + \frac{1}{2}  \sum_{I=1}^{2}  \partial_M \psi_I \partial_N \psi_I  \,, \\
&\nabla_M  F^{MN} = - i q ( \Phi^* D^N \Phi - \Phi D^N \Phi^*   )\,, \\
&\left(D^2 - m^2 \right) \Phi =0 \,, \qquad \nabla^2 \psi_I =0 \,, 
\end{align}
for which we make the following ansatz:
\begin{align}\label{g ansatz}
&A =A_t(r)\dd t  + \frac{1}{2}B \left( x \dd y- y \dd x\right) \,, \qquad \Phi =\Phi(r) \,, \qquad \psi_I =  \left(\beta  x, \beta y \right) \,,  \\
&\dd s^2 = -U(r) e^{-\chi(r)} \dd t^2  +\frac{\dd r^2}{U(r)}  +  r^2  (\dd x^2+ \dd y^2) \,.
\end{align}
In the gauge field, $A_t(r)$ encodes a finite chemical potential or charge density and $B$ plays a role of an external magnetic field.  $\Phi(r)$ is dual to a superconducting phase order parameter, condensate. Near boundary $(r\to \infty)$, $\Phi \sim \frac{J^\Phi}{r^{3-\Delta}} +\frac{\left<\mathcal O^\Phi\right>}{r^\Delta} +\ldots$ with two undetermined coefficients $J^\Phi$ and $\left<\mathcal O^\Phi\right> $, which are identified with the source and condensate respectively. The dimension $\Delta$ of the condensate is related to the bulk mass of the complex scalar by  $m^2 = \Delta (\Delta- 3)$.  In this paper, we take $m^2 = -2$ and $\Delta =2$ to perform numerical analysis. $\psi_I$ is introduced to give momentum relaxation effect where $\beta$ is the parameter for the strength of momentum relaxation. 
For $\beta=0$, the model becomes the original holographic superconductor proposed by  Hartnoll, Herzog, and Horowitz (HHH) \cite{Hartnoll:2008vx, Hartnoll:2008kx}.

\begin{figure}[]
 \centering
 \subfigure[$\Phi(r)/\mu^{2}$]
{\includegraphics[width=4.8cm]{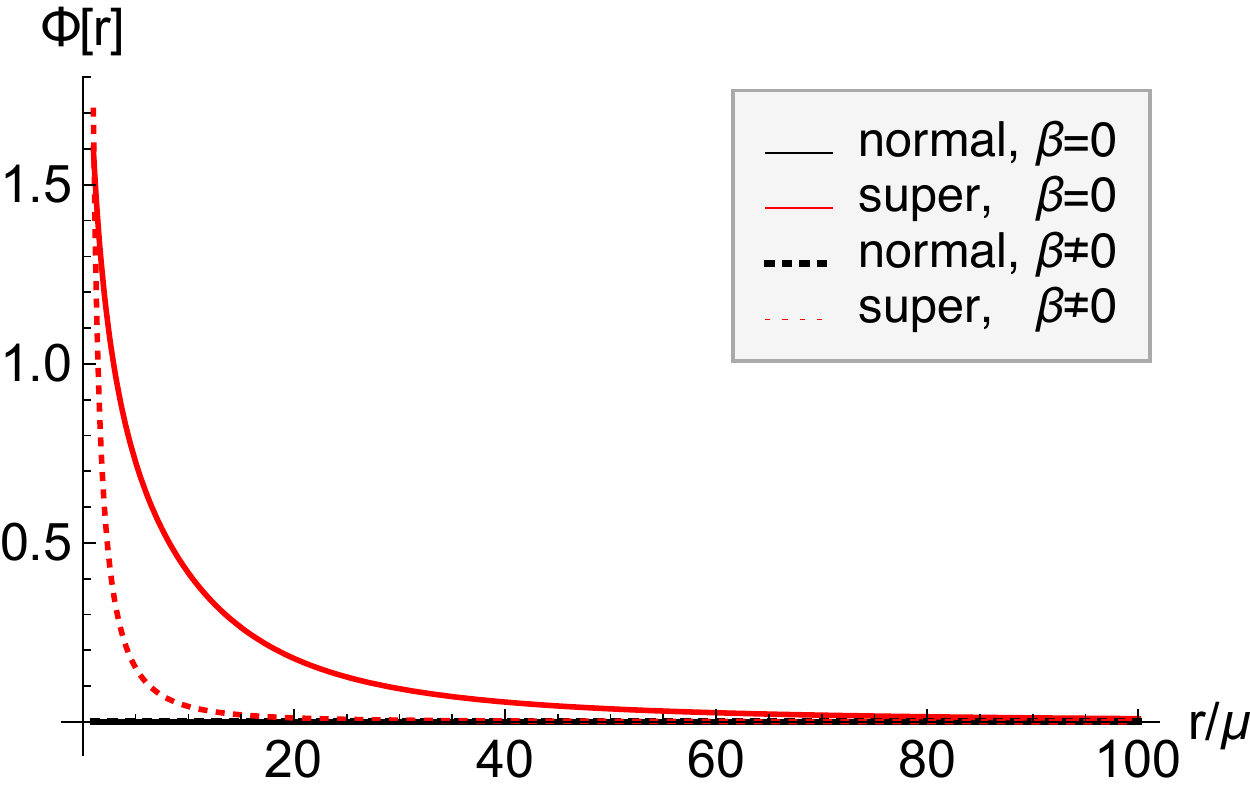} \label{}} \; \; \; \;
\subfigure[$\chi(r)$] 
{\includegraphics[width=4.8cm]{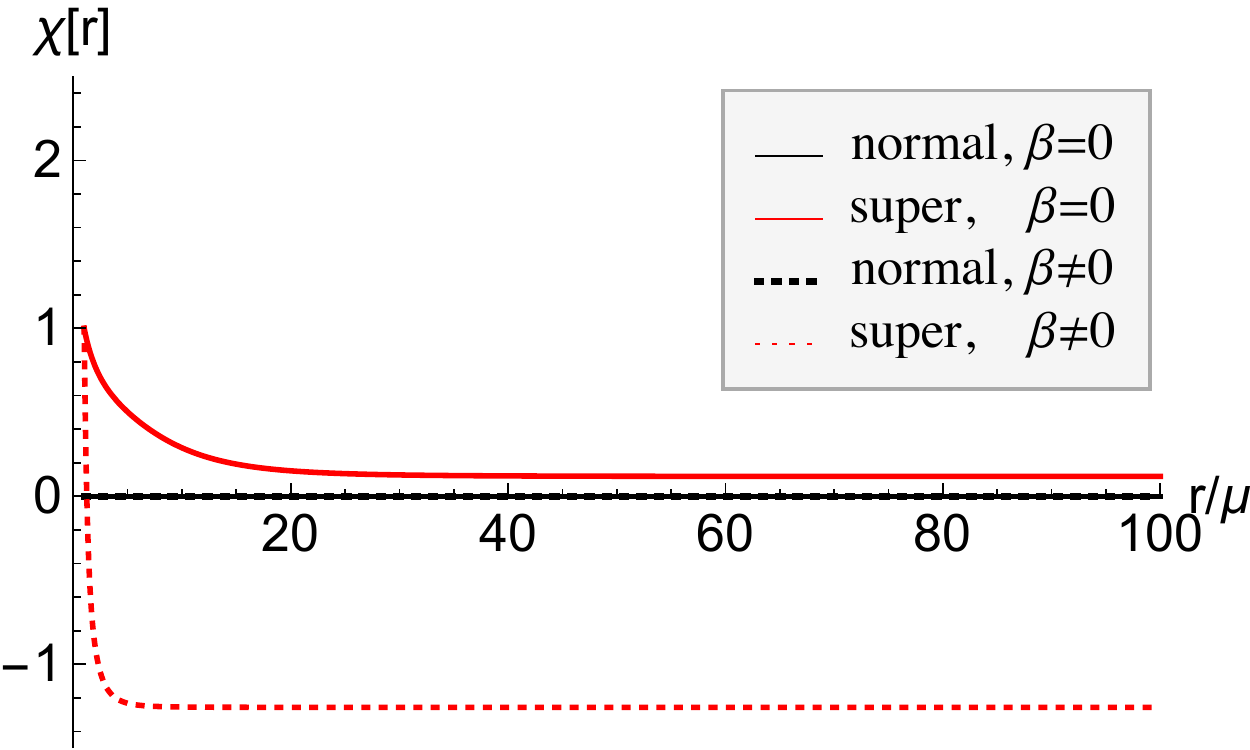} \label{}} 
\subfigure[$A_t(r)/\mu$]
{\includegraphics[width=4.8cm]{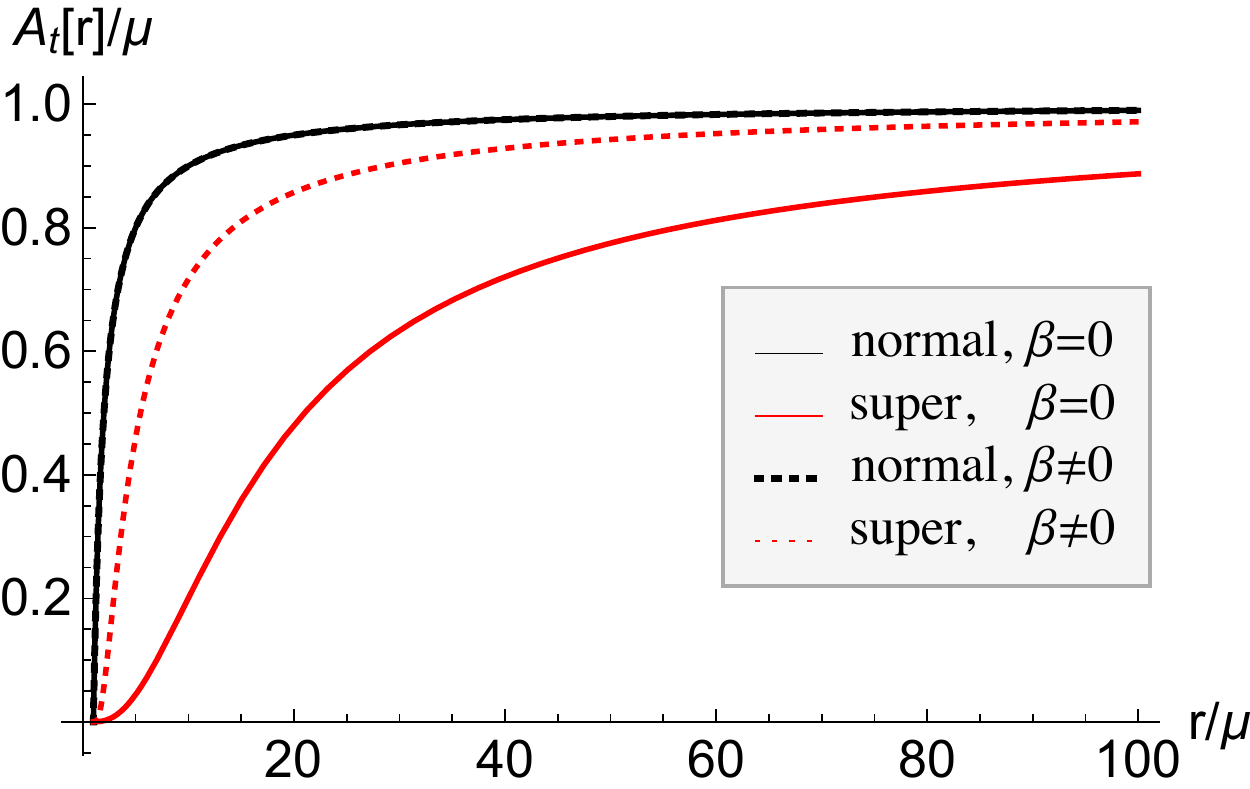} \label{}} \; \; \; \;
\subfigure[$U(r)/\mu^2$]
{\includegraphics[width=4.8cm]{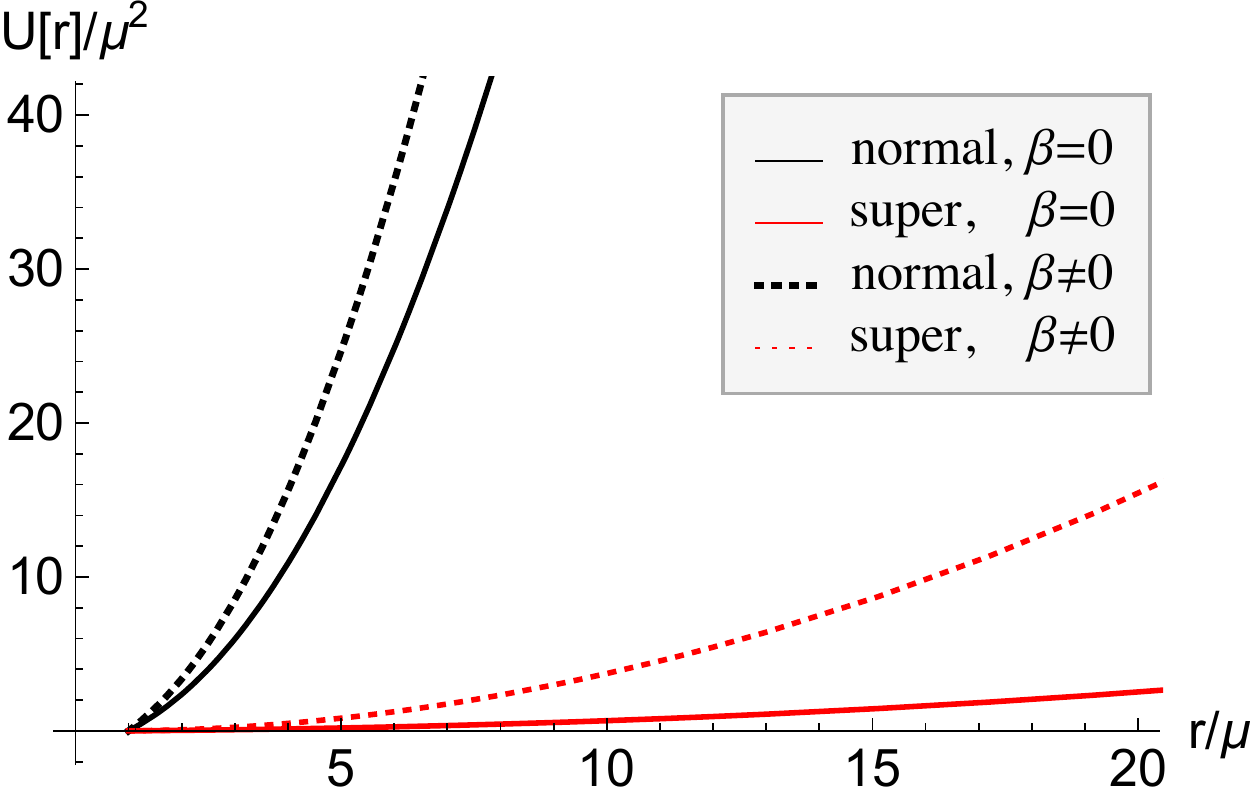} \label{}} 
 \caption{Numerical solutions of bulk background functions, which are numerically integrated from the black hole horizon ($r_{h} = 1$).  
 The solid curves are for the case without momentum relaxation ($\beta =0$) while the dotted curves are for momentum relaxation ($\beta/\mu = 0.5$). The black curves are for normal phase ($\left<\mathcal O^\Phi\right>  = 0$) while the red curves are for superconducting phase ($\left<\mathcal O^\Phi\right>  \neq 0$).   In (a), (b) and (c) the black solid and dotted curves are coincide, but in (d) they are different.  The black curves agree to the analytic formula in \eqref{bgform}, where $\beta$ enters only into $U(r)$. } 
            \label{fig:numbg}
\end{figure}

First, if $\Phi(r) =0$ (no condensate), the solution corresponds to a normal state and its analytic formula is given by 
 \begin{equation} \label{bgform}
 \begin{split}
&U(r) = r^2 - \frac{\beta^2}{2}  - \frac{m_0}{r} +\frac{  {n}^2 + B^2}{4 r^2}  \,, \qquad \chi(r) = 0, \\
&A_t = {n}\left( \frac{1}{r_h}- \frac{1}{r} \right)   \,,
\end{split}
\end{equation}
where $r_h$ is the location of the black brane horizon defined by $U(r_h)=0$, $m_0 \equiv r_h^3 - \frac{\beta^2 r_h}{2} + \frac{n^2 +B^2}{4 r_h}$, and $n$ is interpreted as charge density.  It is the dyonic black brane~\cite{Hartnoll:2007ai} modified by $\beta$ due to $\psi_I$~\cite{Kim:2015wba}. The thermodynamics and transport coefficients(electric, thermoelectric, and thermal conductivity) of this system was analysed in detail in \cite{Kim:2015wba}. In the case without magnetic field, see \cite{Kim:2014bza}.
Next, if  $\Phi(r) \ne 0$, the solution corresponds to a superconducting state with finite condensate and its analytic formula is not available\footnote{A nonzero $\Phi(r)$ induces a nonzero $\chi(r)$, which changes the definition of `time' at the boundary so field theory quantities should be defined accordingly.}.  For $B=0$,  the solutions are numerically obtained in \cite{Hartnoll:2008kx} for $\beta=0$ and in \cite{Kim:2015dna} for $\beta \ne 0$.  For example we display numerical 
solutions for some cases in Figure \ref{fig:numbg}, where we set $r_{h} =1$ and plot dimensionless quantities scaled by $\mu$: $U(r)/\mu^{2}$, $A_{t}/\mu$, and $\chi$.
 For $B\ne0$, due to the generation of vortex our ansatz \eqref{g ansatz} should be modified. In this paper we will not consider this case and refer to \cite{Hartnoll:2009sz,Albash:2009iq,Maeda:2009vf}.

\subsection{AC conductivities}

The purpose of this subsection is to briefly describe the essential points of a method to compute the AC thermo-electric conductivities. For more details and clarification  regarding our model at $B=0$, see \cite{Kim:2015wba,Kim:2015sma} for normal phase and \cite{Kim:2015dna} for superconducting phase. At $B \ne 0$ see \cite{Kim:2014bza} for normal phase. 

In order to study transport phenomena holographically, we introduce small bulk fluctuations around the background obtained in the previous subsection. 
For example, to compute electric, thermoelectric, and thermal conductivities it is enough to consider
\begin{equation} \label{flucs}
\begin{split}
 \delta A_i(t,r) &= \int^{\infty}_{-\infty} \frac{\dd \omega}{2\pi}  e^{-i\omega t}  a_{i}(\omega,r) \,,   \\
 \delta g_{ti}(t,r) &=  \int^{\infty}_{-\infty} \frac{\dd \omega}{2\pi} e^{-i\omega t} r^2 h_{ti}(\omega,r)\,,   \\ 
\delta \psi_i(t,r) &= \int^{\infty}_{-\infty} \frac{\dd \omega}{2\pi} e^{-i\omega t}  \xi_i (\omega,r) \,, 
\end{split}
\end{equation}
where $i=x,y$ for $B \ne 0$ and $i=x$ is enough for $B = 0$ thanks to a rotational symmetry in $x-y$ space. For the sake of illustration of our method, we consider the case for $B=0$~\cite{Kim:2015dna} and refer to \cite{Kim:2015wba} for $B \ne 0 $. 
In momentum space, the linearized equations of motion around the background are\footnote{For $B \ne 0$ case, the bulk fluctuations to $y$ direction should be turned on so the number of equations of motion are doubled too.}
\begin{equation} \label{flucEQ}
\begin{split}
& a_x''+ \left(\frac{U'}{U}-\frac{\chi'}{2}\right)a_x' + \left( \frac{\omega ^2}{U^2} e^{\chi}- \frac{2 q^2 \Phi^2}{U} \right) a_x +\frac{r^2 e^{\chi } A_t' }{U}h_{tx}'=0 \;,\\
& h_{tx}' +\frac{ A_t'}{r^2} a_x +\frac{i \beta  U e^{-\chi }}{r^2 \omega } \xi'=0 \; ,  \\
& \xi ''+\left(\frac{U'}{U}-\frac{\chi'}{2}+\frac{2}{r}\right) \xi'  -\frac{i \beta  \omega  e^{\chi } }{U^2} h_{tx} +\frac{\omega ^2  e^{\chi }}{U^2} \xi  =0 \;. 
\end{split}
\end{equation}
Near boundary ($r \rightarrow \infty$) the asymptotic solutions are
\begin{equation} \label{nearb}
\begin{split}
h_{tx} &=   h^{(0)}_{tx} + \frac{1}{r^2} h^{(2)}_{tx} + \frac{1}{r^3}h_{tx}^{(3)}+\cdots \,,  \\
a_x&=a_x^{(0)} + \frac{1}{r}a_x^{(1)}+ \cdots \,, \qquad  \quad \\
 \xi &= \xi^{(0)} + \frac{1}{r^2} \xi^{(2)}+ \frac{1}{r^3}\xi^{(3)} + \cdots \,,
\end{split}
\end{equation}

The on-shell quadratic action in momentum space reads
\begin{equation} \label{sb}
S_{\mathrm{ren}}^{(2)}  
= \frac{1}{2} \int \frac{\dd \omega}{2\pi}  \left[ J_{-\omega}^a \mathbb{A}_{a b}(\omega) J_\omega^b
+  J_{-\omega}^a \mathbb{B}_{a b}(\omega) {R_\omega^b} \right]  ,
\end{equation}
where 
\begin{align}
J^a =
\begin{pmatrix}
    a_x^{(0)}  \\
    h_{tx}^{(0)} \\
   \xi^{(0)} \\
\end{pmatrix}\,, \quad
R^a =
\begin{pmatrix}
    a_x^{(1)}  \\
    h_{tx}^{(3)} \\
   \xi^{(3)} \\
\end{pmatrix}\,, \quad
\mathbb{A} = \begin{pmatrix}
    0 & \ -n \ & 0  \\
    0 & \ 2U^{(1)} \ & 0  \\
   0 & 0 & 0  \\
\end{pmatrix}\,, \quad
   \mathbb{B}= \begin{pmatrix}
 1 & 0 & 0 \\
 0 & -3 & 0 \\
 0 & 0 & 3 \\
\end{pmatrix} \,.
\end{align}
Here $U^{(1)}$ is the coefficient of $1/r$ when $U(r)$ is expanded near boundary and $n$ is charge density.  The index $\omega$ in $J^a$ and $R^a$ are suppressed.  

The remaining task for reading off the retarded Green's function is to express $R^b$ in terms of $J^a$. It can be done by the following procedure.  First let us denote small fluctuations in momentum space by $\Phi^a$ collectively. i.e.
 \begin{equation}
\Phi^a =  \left( \delta a_i \,,  \delta h_{t i} \,,  \delta \xi_i  \right) \,.
 \end{equation}
Near black brane horizon ($r=1$), solutions may be expanded as
\begin{equation} \label{incoming}
\Phi^a(r) = (r-1)^{-\frac{i\omega}{4\pi T} + n^a} \left( \varphi^{a} + \tilde{\varphi}^{a} (r-1) + \cdots \right) \,,
\end{equation}
which corresponds to incoming boundary conditions for the retarded Green's function \cite{Son:2002sd} and $n^a$ is some integer depending on specific fields, $\Phi^a$.  The leading terms $\varphi^a$ are only free parameters and the higher order coefficients such as  $\tilde{\varphi}^a$ are determined by the equations of motion.  A general choice of $\varphi^a$ can be written as a linear combination of independent basis $\varphi^{a}_{i}$, ($i=1,2,\cdots, N$), i.e. $\varphi^a = \sum_{i=1}^N \varphi^a_i c_i $. For example, $\varphi^a_i$ can be chosen as
\begin{equation} \label{init}
\begin{pmatrix}
    \varphi^{a}_{1} \ & \varphi^{a}_{2}\ &  \ldots \ &  \varphi^{a}_{N}
\end{pmatrix}
   =
\begin{pmatrix}
    1 & 1&   \ldots & 1 \\
    1 & -1&  \ldots & 1 \\
    \vdots & \vdots & \ddots & \vdots \\
    1 & 1 & \ldots & -1
\end{pmatrix} \,.
\end{equation}
 Every  $\varphi_i^a$ yields a  solution ${\Phi}_i^a(r)$, which is expanded near boundary as 
\begin{equation}
\Phi_i^a(r) \ \  \rightarrow  \ \  \mathbb{S}_{i}^{a}  + \cdots +  \frac{\mathbb{O}_{i}^{a}}{r^{\delta_a}}  + \cdots \,,
\end{equation}
where $\delta_a \ge 1$  and the leading terms $\mathbb{S}_i^a$ are the {\it{sources}} of $i$-th  solutions and  $\mathbb{O}_{i}^{a}$ are the corresponding {\it{operator }}expectation values.  $\mathbb{S}$ and $\mathbb{O}$ can be regarded as regular matrices of order $N$, where $a$ is for row index and $i$ is for column index.
A general solution may be constructed from a basis solution set $\{\Phi_{i}^{a}\}$:
\begin{align} \label{GS} 
\Phi^a(r) = \Phi_{i}^{a}(r) c^i & \  \  \rightarrow  \  \  \mathbb{S}_{i}^{a} c^i  + \cdots +  \frac{\mathbb{O}_{i}^{a}c^i }{r^{\delta_a}}  + \cdots \\
 &\  \  \equiv \  \  \  \  J^a   \  + \cdots + \frac{R^a}{r^{\delta_a}} \ \  + \cdots \,,
\end{align}
with arbitrary constants $c^i$'s. For a given $J^a$, we always can find $c^i$\footnote{There is one subtlety in our procedure. The matrix $\mathbb{S}$ of solutions with incoming boundary condition are not invertible and we need to add some constant solutions, which is related to 
a residual gauge fixing $\delta g_{rx} =0$~\cite{Kim:2015sma}.}  
\begin{equation}
 c^i  = (\mathbb{S}^{-1})^{i}_{a} J^a \,,
\end{equation}
so the corresponding response $R^a$ may be expressed in terms of the sources $J^b$
\begin{equation} \label{response1}
R^a =  \mathbb{O}_{i}^{a} c^i =  \mathbb{O}_{i}^{a} (\mathbb{S}^{-1})^{i}_{b} J^b \,.
\end{equation} 

With \eqref{response1}, the action \eqref{sb}  becomes
\begin{equation} \label{Gab}
\begin{split}
S_{\mathrm{ren}}^{(2)} 
 &= \frac{1}{2}  \int_{\omega \ge 0} \frac{\dd \omega}{ 2\pi}  \left[ J_{-\omega}^a \left[\mathbb{A}_{a b}(\omega) + \mathbb{B}_{ac}\mathbb{O}_{i}^{c} (\mathbb{S}^{-1})^{i}_{b}(\omega)\right]J_\omega^b  \right] \\
& \equiv \frac{1}{2}  \int_{\omega \ge 0} \frac{\dd \omega}{ 2\pi}  \left[ J_{-\omega}^a  G_{ab} J_\omega^b  \right] \,,
\end{split}
\end{equation}
where the range of $\omega$ is chosen to be positive following the prescription in  \cite{Son:2002sd} and the retarded Green's functions are explicitly denoted as
\begin{equation}
G_{ab} \equiv 
\left(
\begin{array}{ccc}
 G_{\nJ\nJ} &  G_{\nJ\nT}  &  G_{\nJ\nS}   \\
  G_{\nT\nJ}  &   G_{\nT\nT} &   G_{\nT\nS}  \\
 G_{\nS\nJ}   &  G_{\nS\nT}   &    G_{\nS\nS}
\end{array}
\right) \,.
\end{equation}
%
Finally, the thermo-electric conductivities are related to the retarded Green's functions as 
\begin{equation}
\begin{split}
&\left( 
\begin{array}{cc}
\sigma &\alpha T \\
\bar \alpha T & \bar \kappa T
\end{array}
 \right) \\ 
 &= 
-\frac{i}{\omega} \left(
\begin{array}{cc}
   G_{ {\nJ\nJ}}  & - \mu G_{ {\nJ\nJ} }  + G_{ {\nJ\nT} }   \\
- \mu G_{ {\nJ\nJ}}  + G_{ {\nT\nJ}}   &  \ \ \    G_{ {\nT\nT}}  -G_{ {\nT\nT}}(\omega=0) - \mu \left(G_{ {\nJ\nT}}  + G_{ {\nT\nJ}} -\mu    G_{ {\nJ\nJ}}   \right) 
\end{array}
 \right) \,.
\end{split}
\end{equation}

\section{Conductivities with a neutral scalar hair instability}\label{sec3}

\begin{figure}[]
 \centering
      {\includegraphics[width=4.5cm]{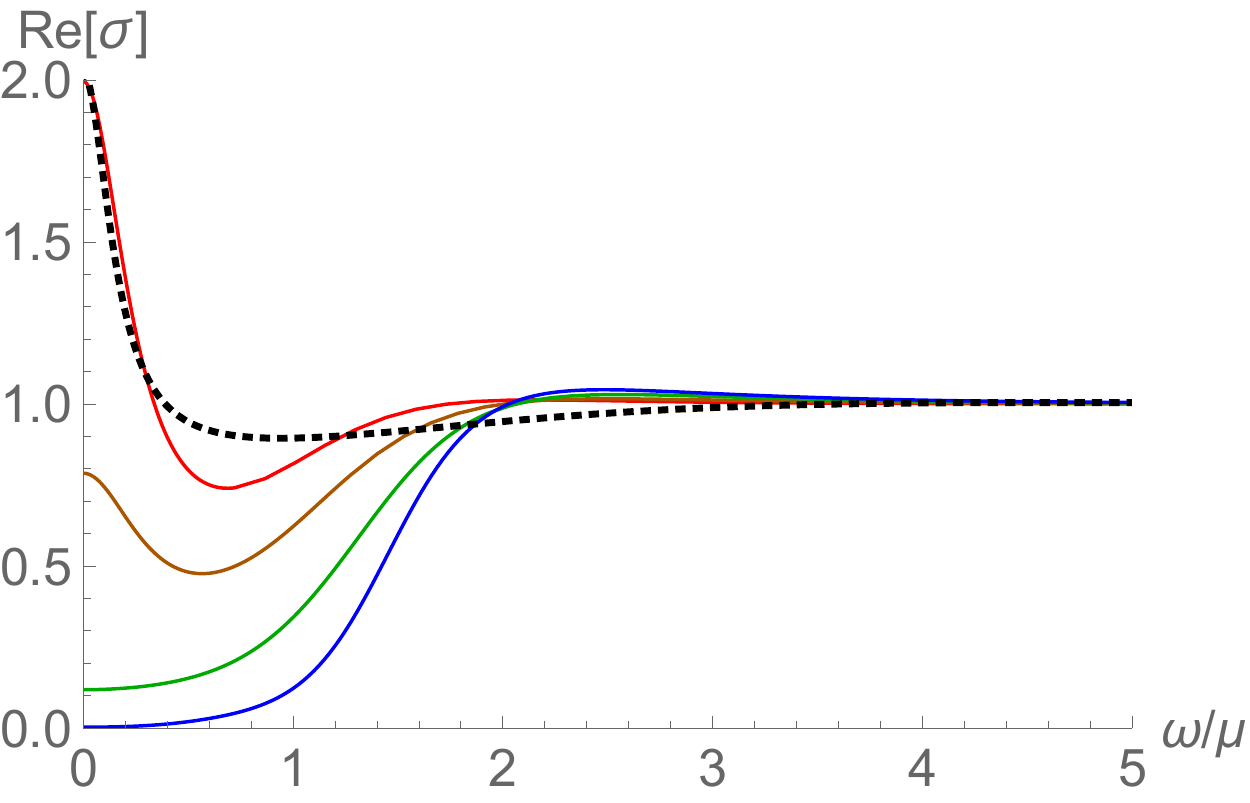} \label{}}\hspace{3mm}
   {\includegraphics[width=4.5cm]{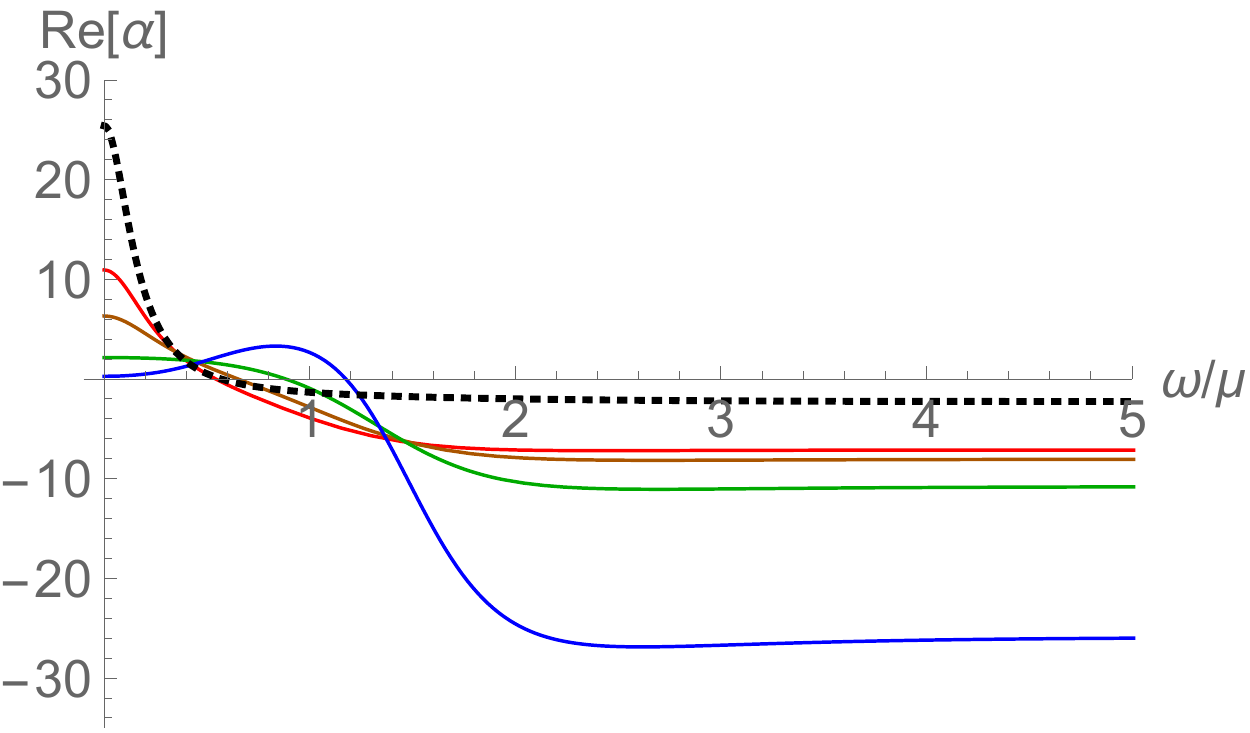} \label{}}\hspace{3mm}
   {\includegraphics[width=4.5cm]{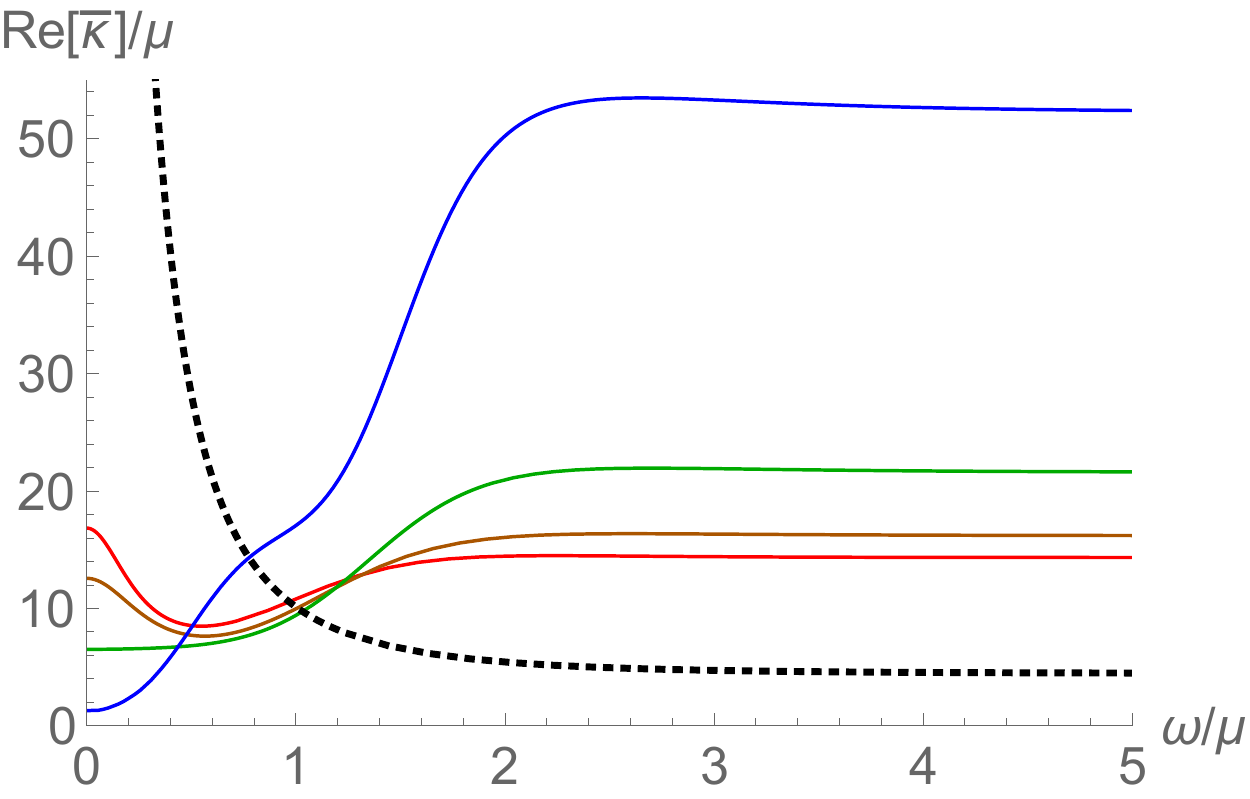} \label{}}
      \subfigure[Electric conductivity]
 {\includegraphics[width=4.5cm]{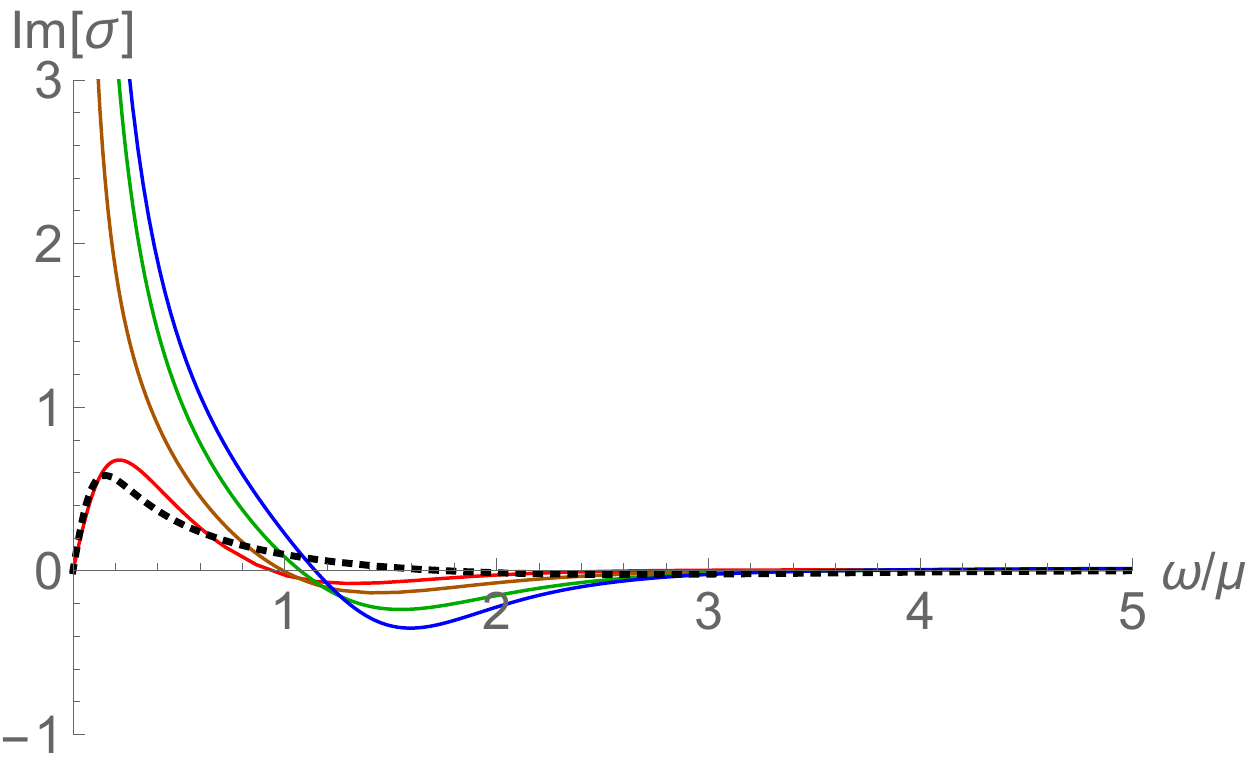} \label{}}\hspace{3mm}
    \subfigure[Thermoelectric conductivity]
   {\includegraphics[width=4.5cm]{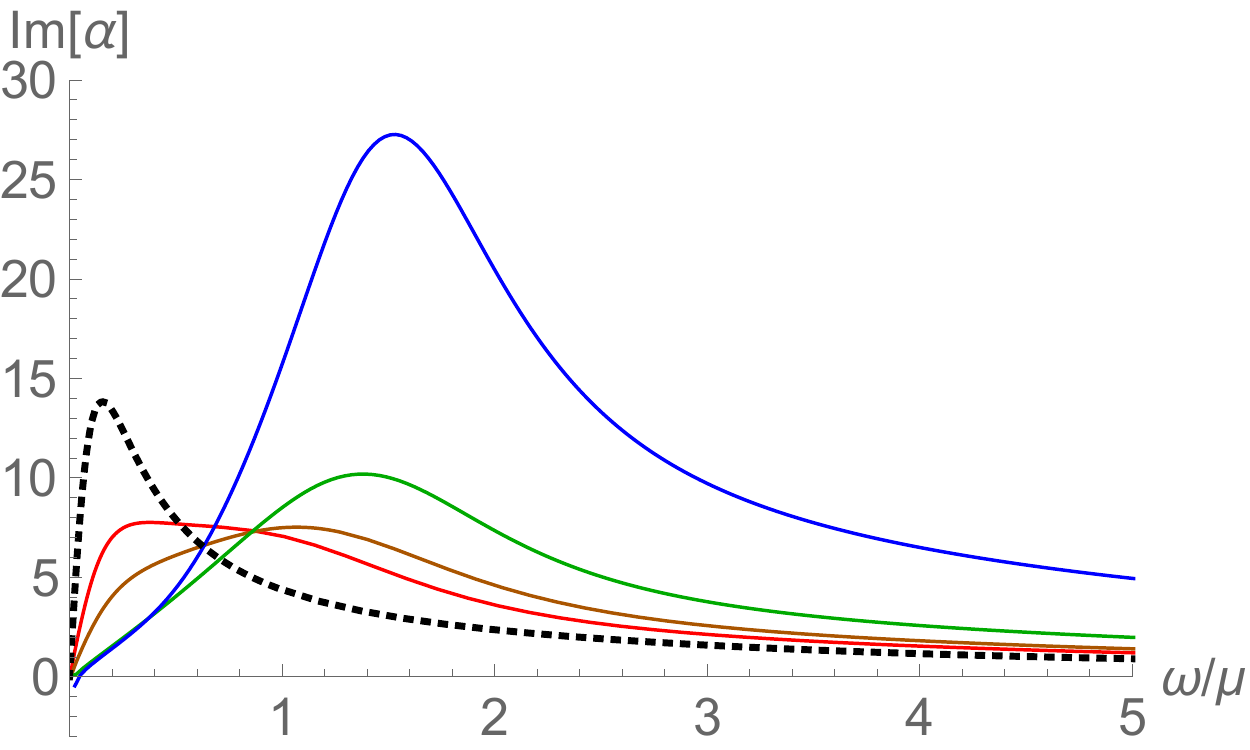} \label{}}\hspace{3mm}
     \subfigure[Thermal conductivity]
   {\includegraphics[width=4.5cm]{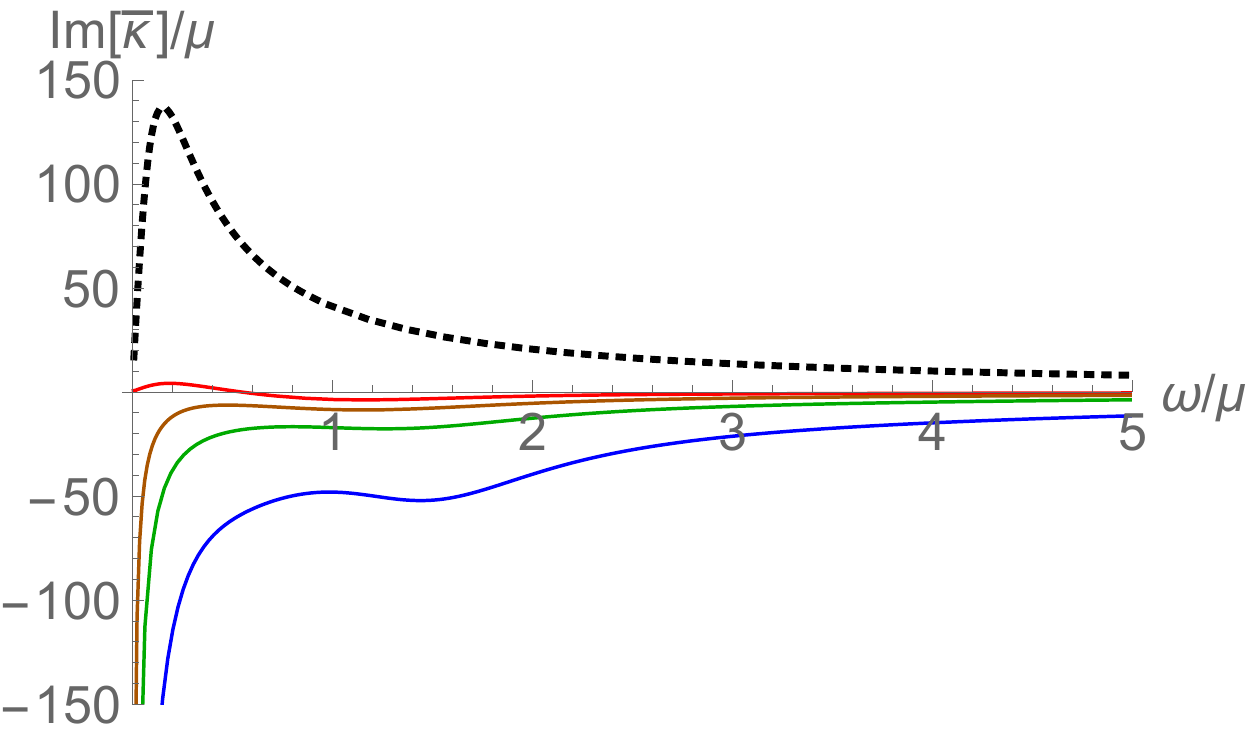} \label{}}
    \caption{AC electric conductivity($\sigma(\omega)$), thermoelectric conductivity($\alpha(\omega)$), and thermal conductivity($\bar{\kappa}(\omega)$) for $\beta/\mu =  1$ and $q=3$ at different temperatures: $T/T_c = 3.2,1,0.89, 0.66, 0.27$ (dotted, red, orange,green, blue).  Top is the real part and bottom is the imaginary part of conductivities.} 
            \label{oldresult}
\end{figure}

By the numerical method reviewed in the previous subsection, the electric, thermoelectric and thermal conductivities 
of the model \eqref{model A} have been computed in various cases \cite{Kim:2014bza,Kim:2015wba,Kim:2015dna}. 
As an example, in Figure \ref{oldresult}, we show the results for $B=0, q=3$ \cite{Kim:2015dna}, which is reproduced here for easy comparison with new results in this paper.

%

Figure \ref{oldresult} shows AC electric conductivity ($\sigma(\omega)$), thermoelectric conductivity ($\alpha(\omega)$), and thermal conductivity ($\bar{\kappa}(\omega)$) for $\beta/\mu =  1$ and $q=3$ at different temperatures. The colors of curves represent the temperature ratio,  $T/T_c $, where $T_c $ is the critical temperature of metal/superconductor phase transition. $T/T_c = 3.2,1,0.89, 0.66, 0.27$ for dotted, red, orange,green, and blue curves respectively. In particular, the dotted curve is the case above $T_c$ and the red curve corresponds to the critical temperature. The first row is the real part and the second row is  the imaginary part of conductivities.   

One feature we want to focus on in Figure \ref{oldresult} is $1/\omega$ pole in Im[$\sigma$] below the critical temperature. There is no $1/\omega$ pole above the critical temperature.  By the Kramers-Kronig relation, the $1/\omega$ pole in Im[$\sigma$]  implies the existence of the delta function at $\omega=0$ in Re[$\sigma$].  It means that in superconducting phase the DC conductivity is infinite while in normal phase the DC conductivity is finite due to momentum relaxation.

Unlike the studies in \cite{Kim:2015dna}, here we set $q=0$. 
Between finite $q$ and zero $q$, there is a qualitative difference in the instability of a Reissner-Nordstrom AdS black hole~\cite{Hartnoll:2008kx}.
The origin of the superconductor (or superfluidity) instability responsible for the complex scalar hair $\Phi$ may be 
understood as the coupling of the charged scalar to the charge of the black hole through the covariant derivative 
$D_M\Phi \equiv \nabla_M \Phi - i q A_M \Phi $. 
In other words,  the effective mass of $\Phi$  defined by $m^2_\mathrm{eff} \equiv m^2 - q^2 |g^{tt}| A_t^2$ 
can be compared with the Breitenlohner-Freedman (BF) bound.  The BF bound for AdS$_{d+1}$ is  $- \frac{d^2}{4} \equiv m_\mathrm{BF}^2$. 
The effective mass $m^2_\mathrm{eff}$ may be sufficiently negative near the horizon to destabilize the scalar field since $|g^{tt}|$ becomes bigger at low temperature\footnote{As the temperature of a charged black hole is decreased, $g_{tt}$ develops a double zero at the horizon.}.
Based on this argument one may expect that when $q = 0$ the instability would turn off. 
However, it turns out that a Reissner-Nordstrom AdS black hole may still be unstable to forming {\it neutral} scalar hair, if $m^2$ is a little bit bigger than the BF bound for AdS$_4$.   It can be understood by the near horizon geometry of an extremal Reissner-Nordstrom AdS black hole. It is AdS$_2 \times$ R$^2$ so scalars above the BF bound for AdS$_4$ may be below the bound for AdS$_2$. These two instability conditions can be summrized by one ineqaulity~\cite{Kim:2015dna}
\begin{equation} \label{BF01}
m_\mathrm{eff}^2 = \left[m^2-  \frac{2q^2}{1+\frac{\beta^2}{\mu^2}}\right]\left[\frac{1}{6} \left( 1+ \frac{\frac{\beta^2}{\mu^2}}{1+\frac{\beta^2}{\mu^2}}  \right)\right] < -\frac{1}{4} = m_\mathrm{BF}^2 \,,
\end{equation}
which reproduces the result for $\beta=0$ in~\cite{Hartnoll:2009sz}
\begin{equation}
m_\mathrm{eff}^2 = \left(m^2- 2q^2\right)\left(\frac{1}{6}\right) < -\frac{1}{4} = m_\mathrm{BF}^2 \,.
\end{equation}
Here, we see $m_{\mathrm{eff}}^2$ can be below the BF bound when $q=0$.

However, it was discussed in~\cite{Hartnoll:2008kx, Horowitz:2010gk}  that  the instability to forming neutral scalar hair for $q=0$ is not associated with superconductivity because it does not break a $U(1)$ symmetry, but at most breaks a $\mathbb Z_2$ symmetry  $\Phi \rightarrow -\Phi$. 
Therefore, it would be interesting to see if the DC conductivity is infinite or not in the background with a neutral scalar hair.\footnote{We thank Sang-Jin Sin for suggesting this.} Without momentum relaxation ($\beta=0$) this question is not well posed since the DC conductivity is always infinite with or without a neutral scalar hair due to translation invariance and finite density. Now we have a model with momentum relaxation ($\beta \ne 0$), we can address this issue properly. 

%

 \begin{figure}[] 
 \centering
     {\includegraphics[width=4.5cm]{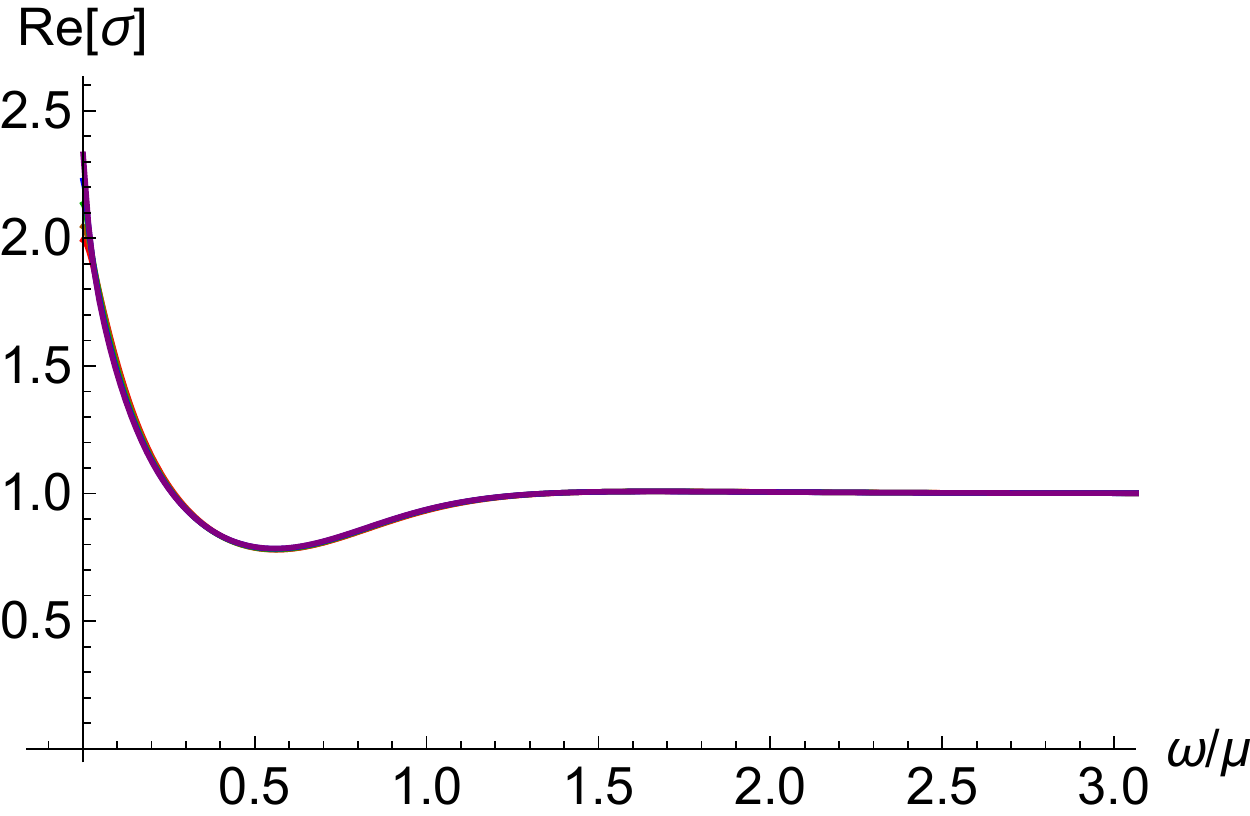} \label{}}\hspace{3mm}
   {\includegraphics[width=4.5cm]{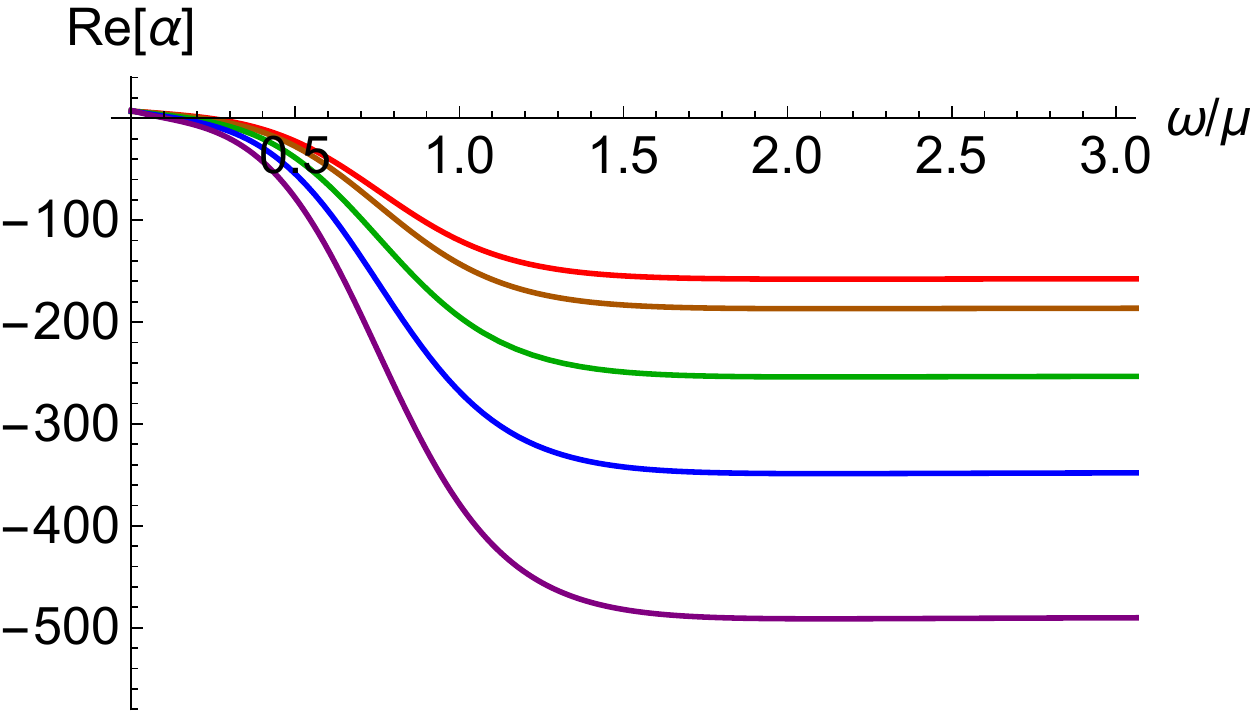} \label{}}\hspace{3mm}
   {\includegraphics[width=4.5cm]{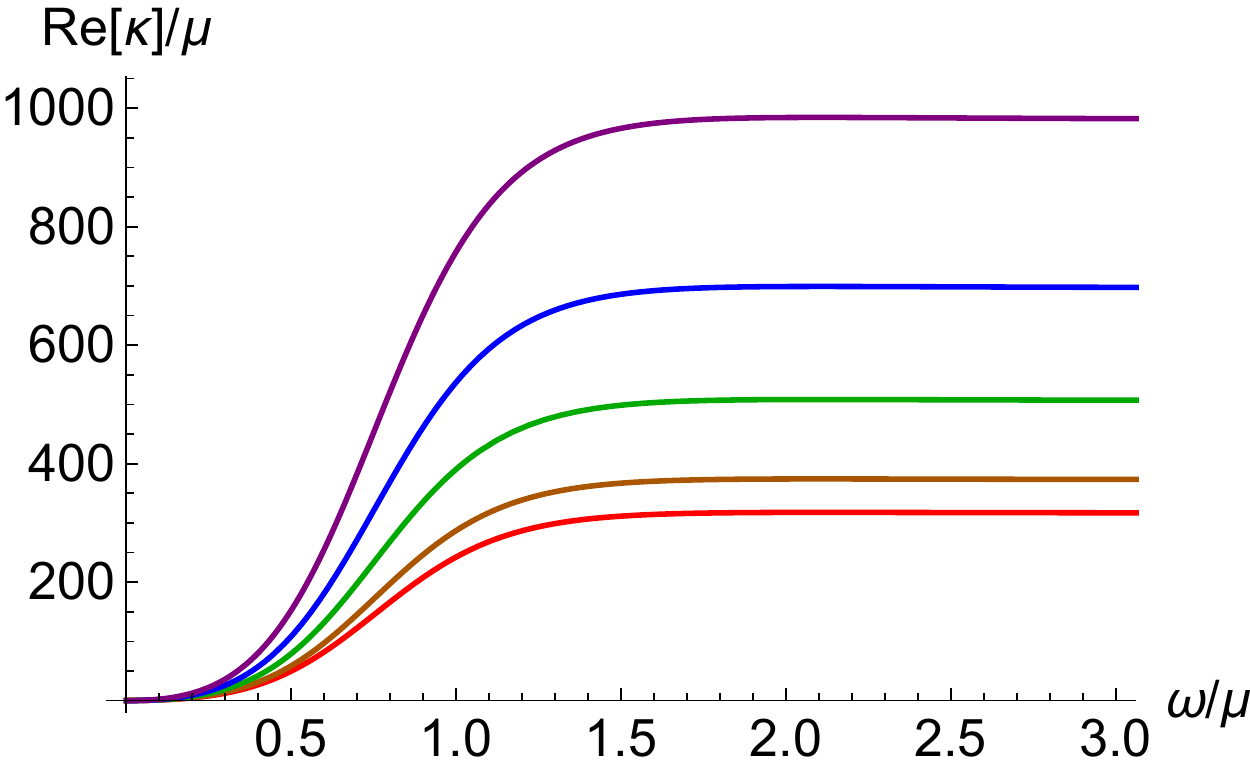} \label{}}
      \subfigure[Electric conductivity]
 {\includegraphics[width=4.5cm]{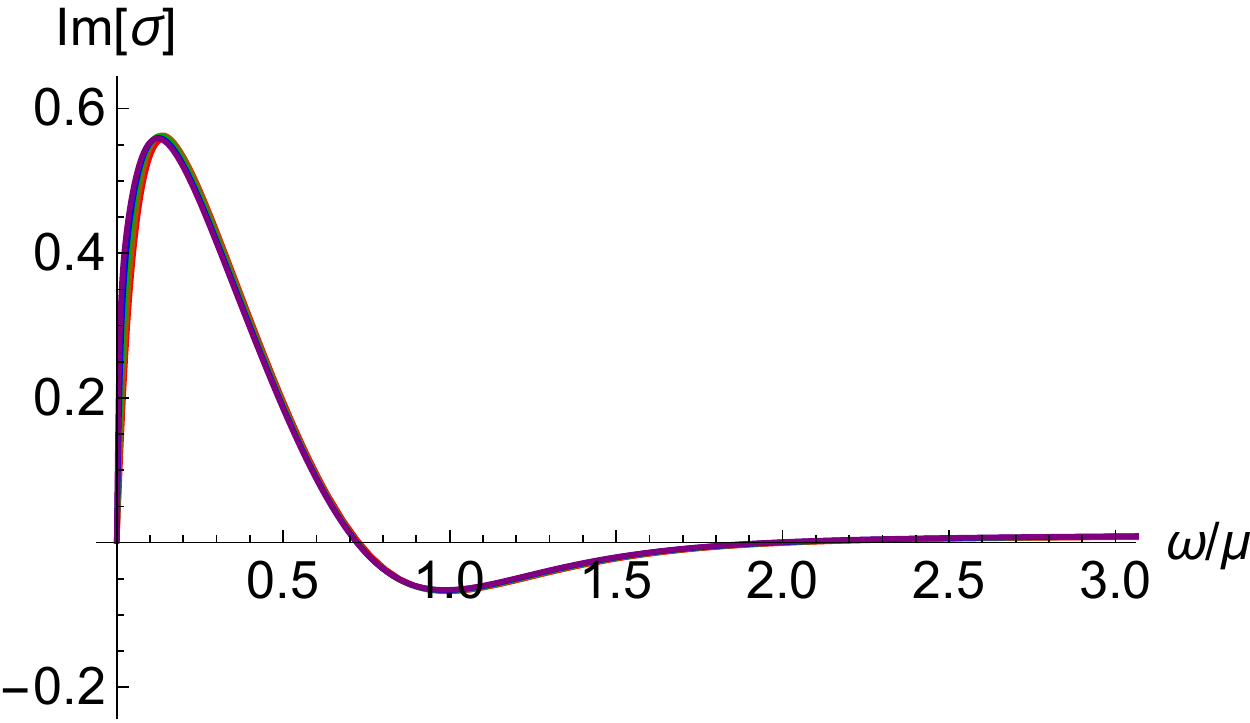} \label{}}\hspace{3mm}
    \subfigure[Thermoelectric conductivity]
   {\includegraphics[width=4.5cm]{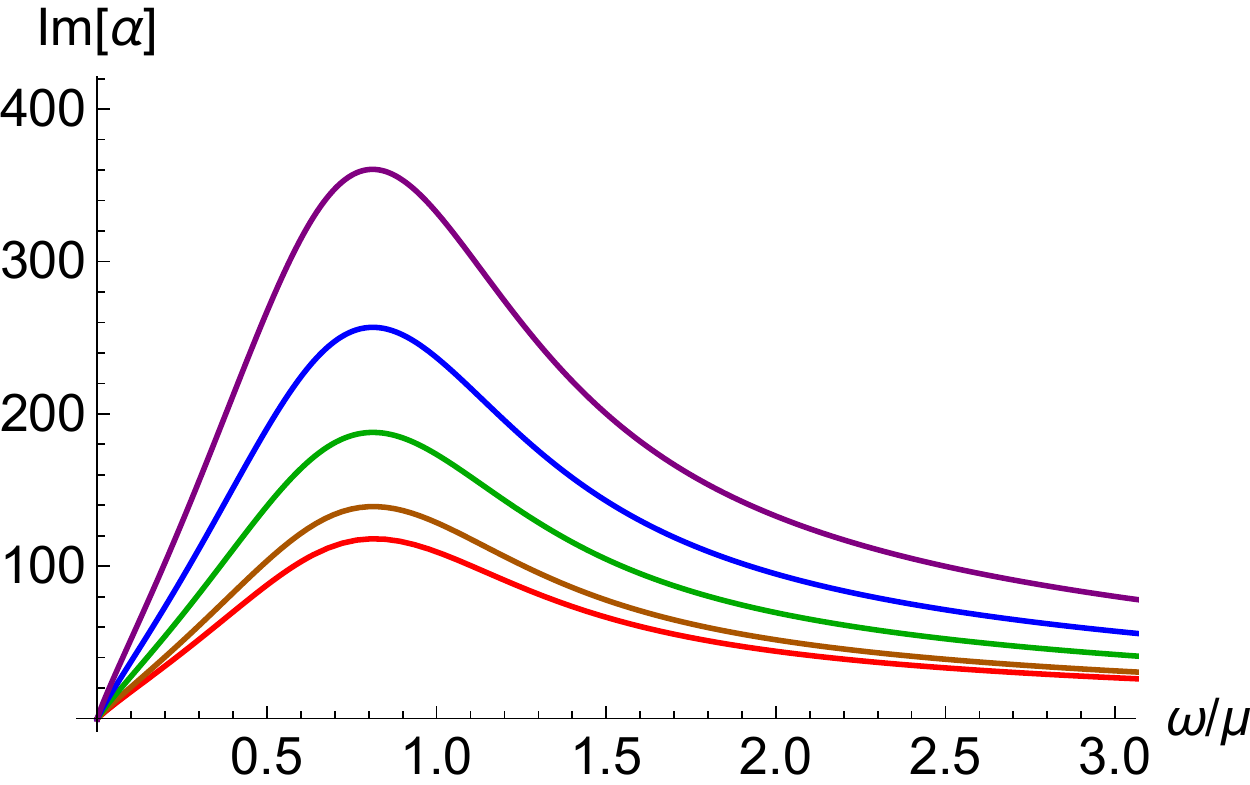} \label{}}\hspace{3mm}
     \subfigure[Thermal conductivity]
   {\includegraphics[width=4.5cm]{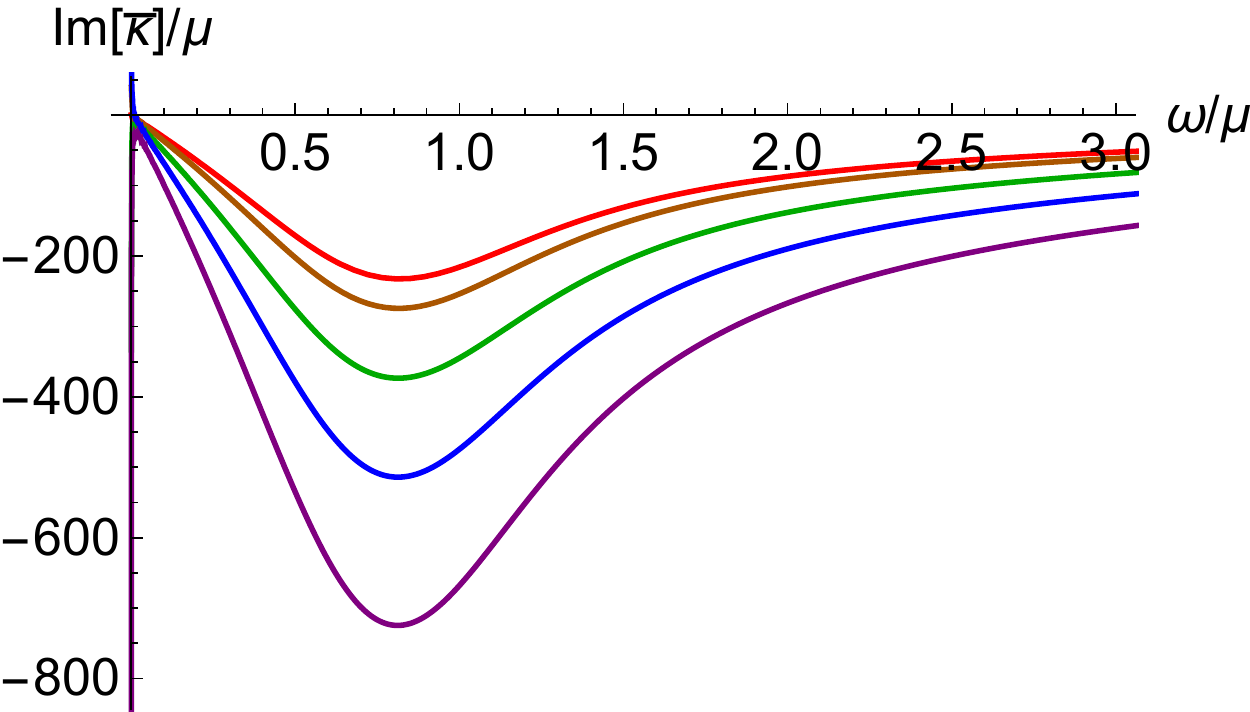} \label{}}
  \caption{AC electric conductivity($\sigma(\omega)$), thermoelectric conductivity($\alpha(\omega)$), and thermal conductivity($\bar{\kappa}(\omega)$) for $\beta/\mu =  1$ and $q=0$ at different temperatures: $T/T_c = 1, 0.84, 0.62, 0.45, 0.32$ (red, orange,green, blue). Top is the real part and bottom is the imaginary part of conductivities. } 
            \label{fig:q0}
\end{figure}
 \begin{figure}[]
 \centering
   {\includegraphics[width=4.5cm]{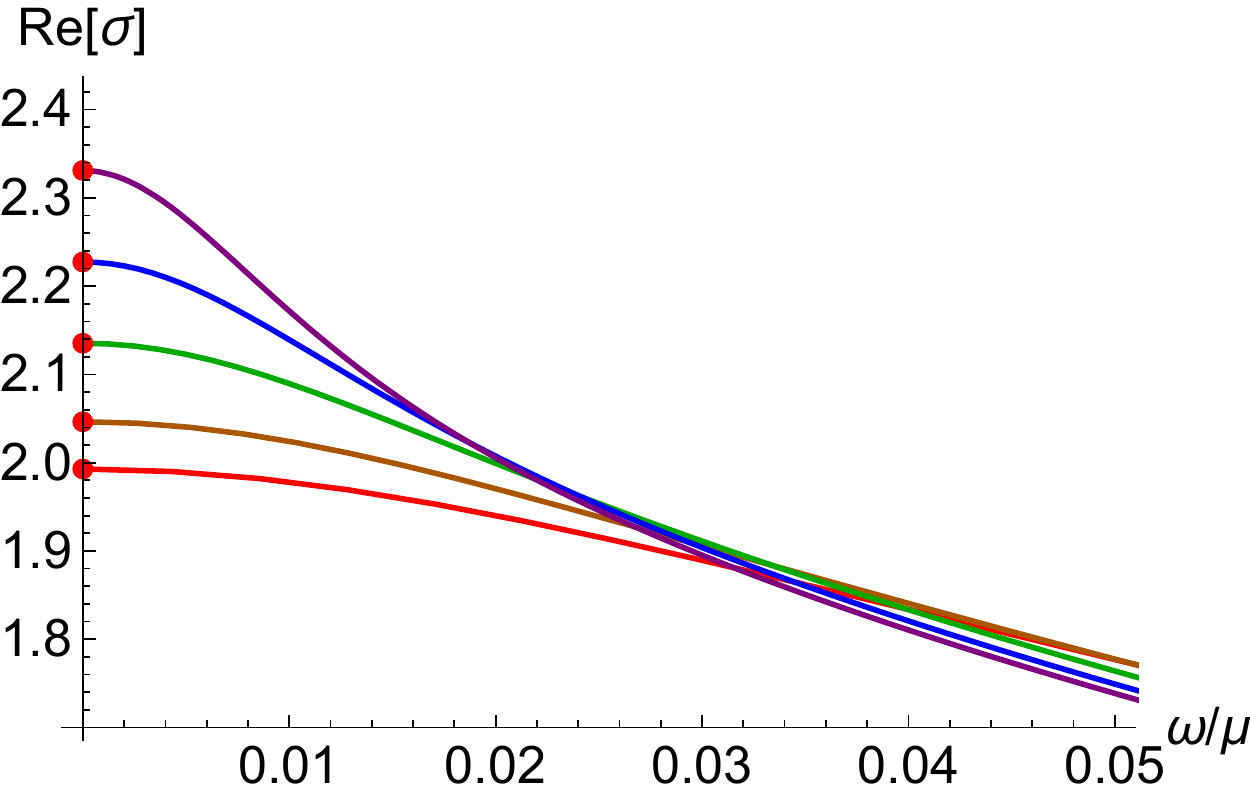} \label{}}\hspace{3mm}
   {\includegraphics[width=4.5cm]{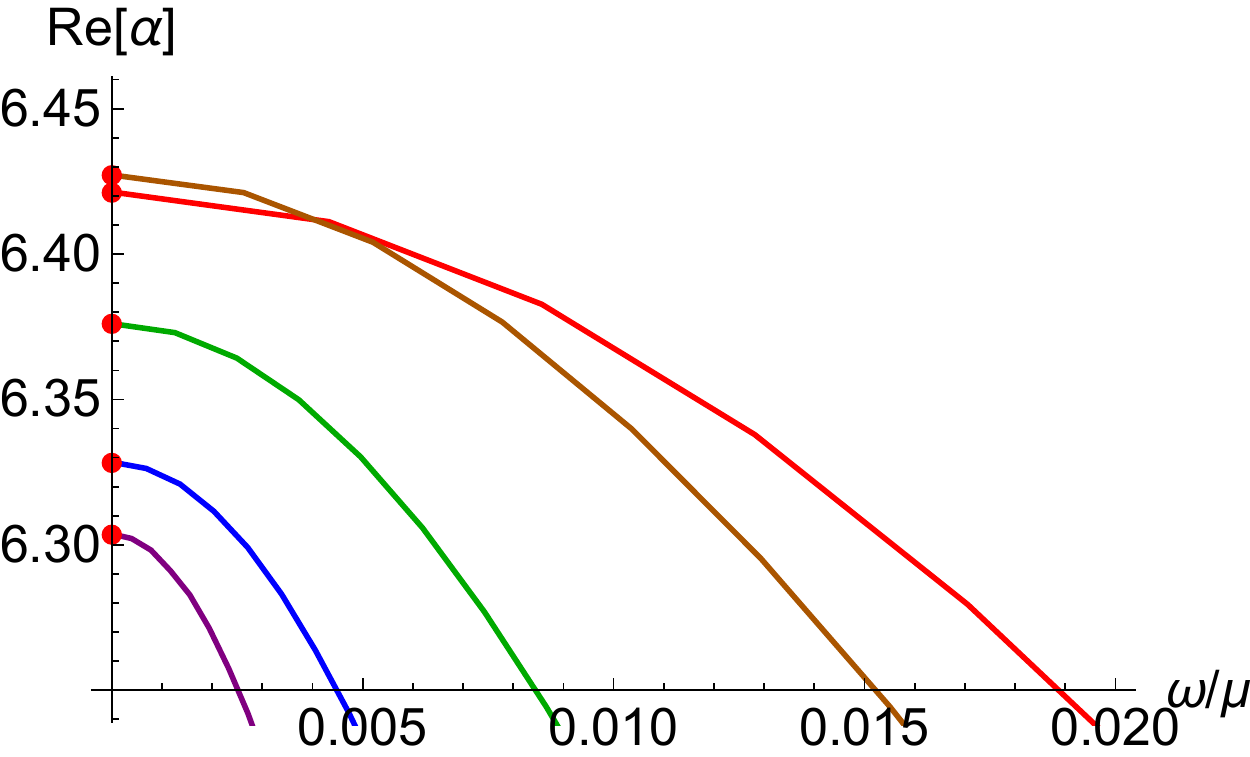} \label{}}\hspace{3mm}
   {\includegraphics[width=4.5cm]{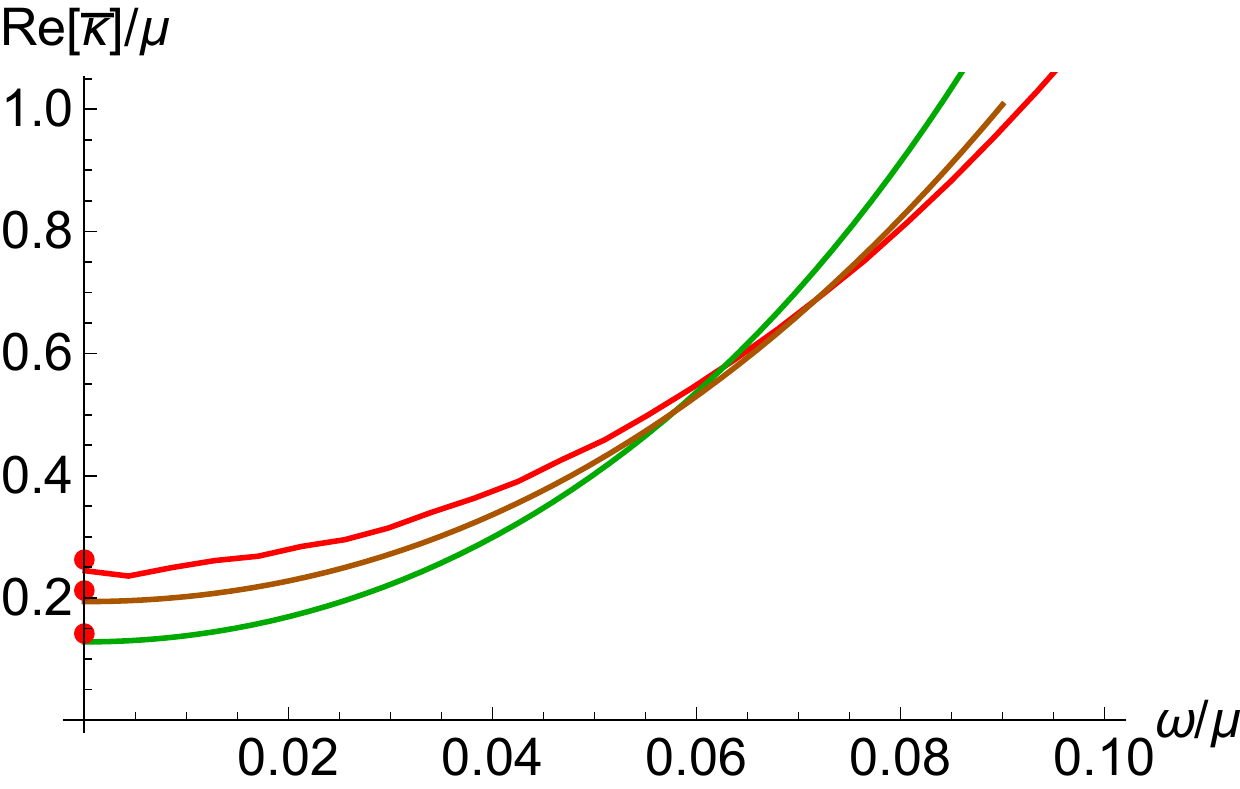} \label{}}\hspace{3mm}
     \caption{ Near $\omega=0$ behaviours of the real part of conductivities in Figure \ref{fig:q0}  }  
            \label{fig:q0zoom}
\end{figure} 


To have an instability at $q=0$ we choose the same parameters as Figure \ref{oldresult}: $m^2 = -2$ and $\beta/\mu=1$.
For $q=0$, $m_\mathrm{eff}^2 = -1/2$, which is below the BF bound \eqref{BF01}.  Figure \ref{fig:q0} shows our numerical results of conductivites, where all temperatures are below $T_c$: $T/T_c = 1, 0.84, 0.62, 0.45, 0.32$ for red, orange,green, and blue curves respectively. 
A main difference of Figure \ref{fig:q0} from Figure \ref{oldresult} is the disappearance of $1/\omega$ pole in Im[$\sigma$] below $T_c$. It confirms that the neutral scalar hair has nothing to do with superconductivity as expected. 

\begin{figure}[] 
 \centering
     {\includegraphics[width=4.5cm]{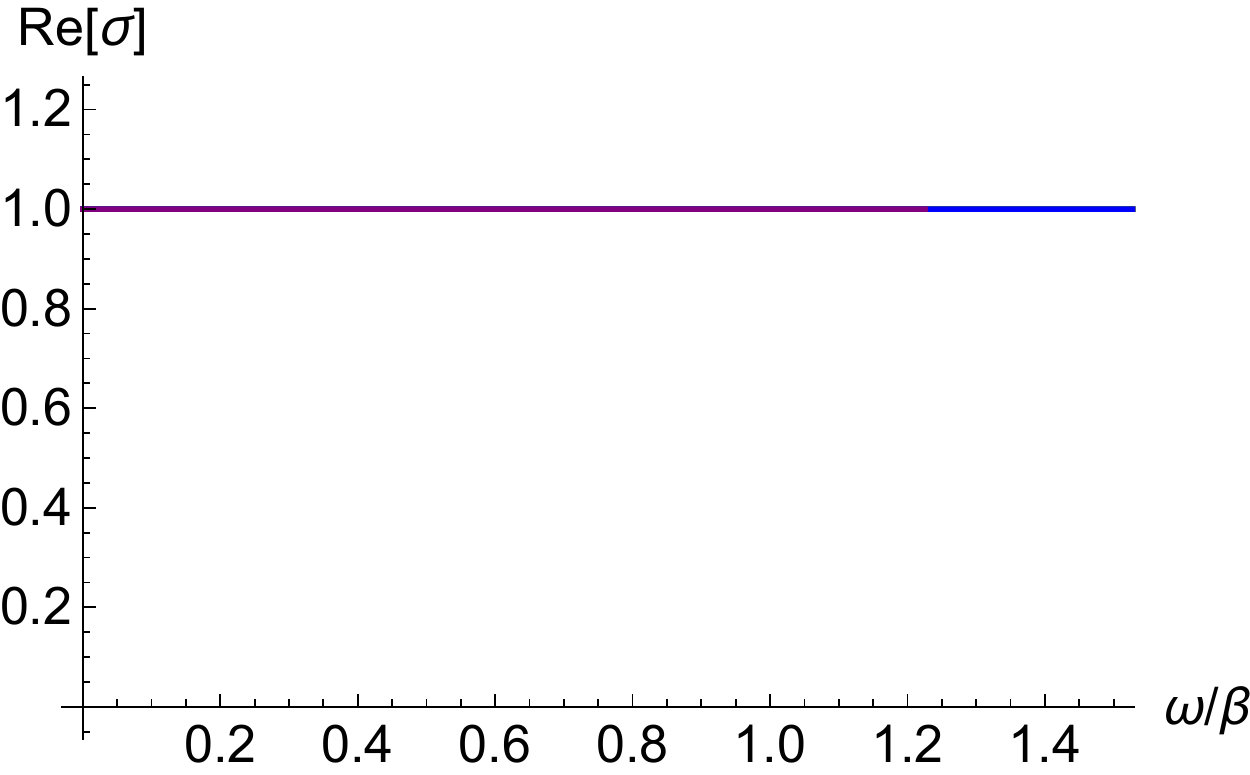} \label{}}\hspace{3mm}
   {\includegraphics[width=4.5cm]{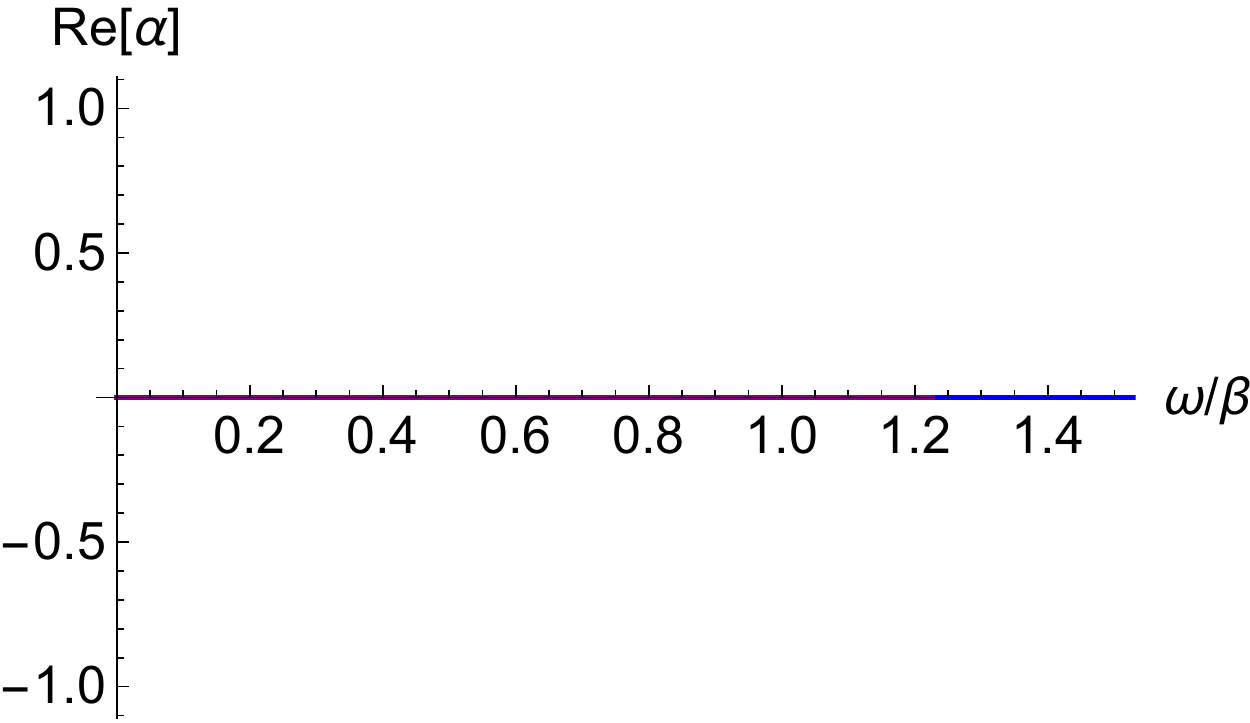} \label{}}\hspace{3mm}
   {\includegraphics[width=4.5cm]{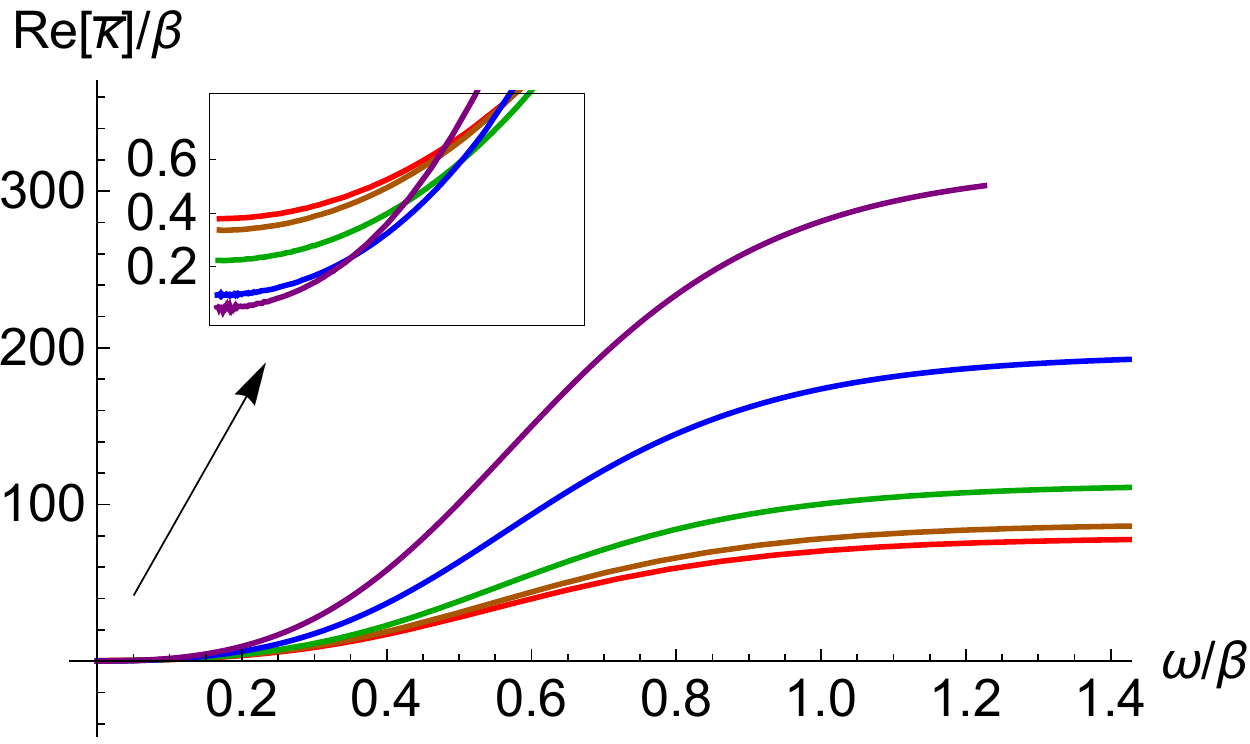} \label{}}
      \subfigure[Electric conductivity]
 {\includegraphics[width=4.5cm]{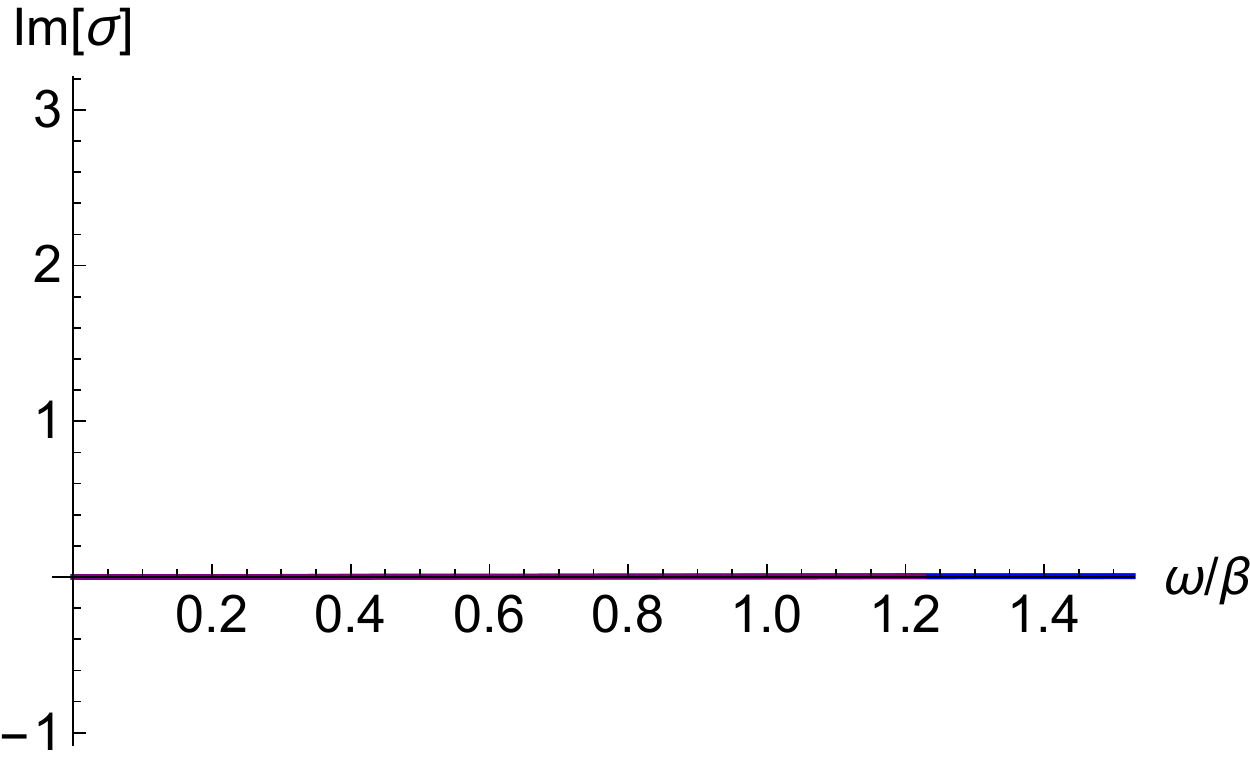} \label{}}\hspace{3mm}
    \subfigure[Thermoelectric conductivity]
   {\includegraphics[width=4.5cm]{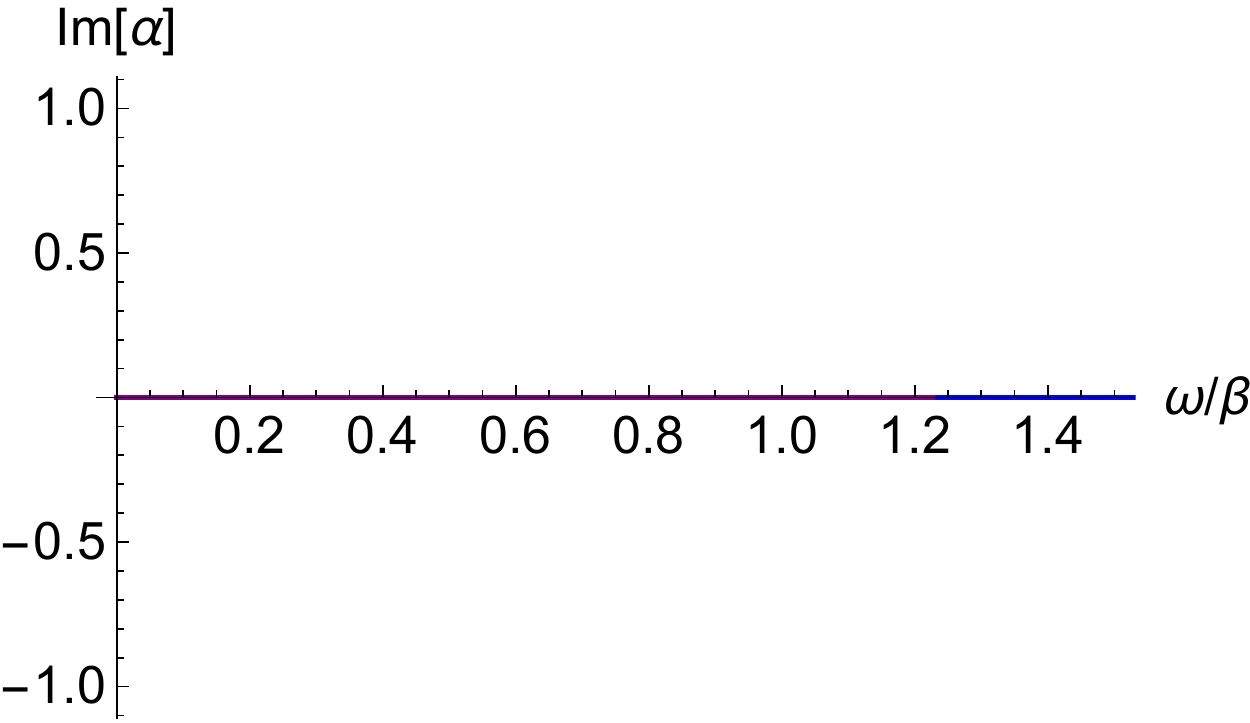} \label{}}\hspace{3mm}
     \subfigure[Thermal conductivity ]
   {\includegraphics[width=4.5cm]{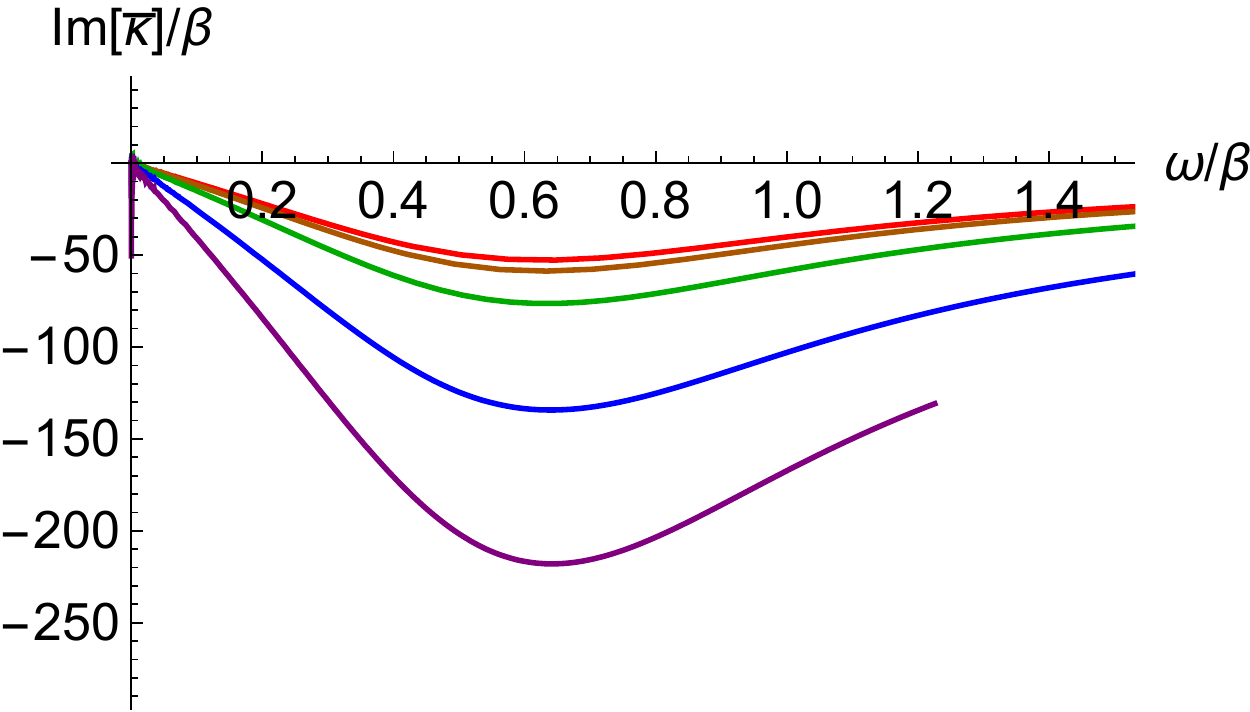} \label{}}
  \caption{AC electric conductivity($\sigma(\omega)$), thermoelectric conductivity($\alpha(\omega)$), and thermal conductivity($\bar{\kappa}(\omega)$) for $\mu =  0$ and $q=0$ at different temperatures: $T/T_c = 1, 0.9, 0.7, 0.4, 0.25$ (red, orange,green, blue, purple). Top is the real part and bottom is the imaginary part of conductivities. } 
            \label{fig:q0mu0}
\end{figure}

In Figure \ref{fig:q0} it is not easy to see the conductivities in small $\omega$ regime, so we zoom in there in Figure \ref{fig:q0zoom}.  Contrary to the conductivity of normal component in superconducting phase, the DC electric conductivity is not so sensitive to temperature and increases as temperature decreases, which is the property of metal. The thermoelectric and thermal conductivities decrease as temperature increases except a small increase of thermoelectric conductivity near the critical temperature. 
As a cross check, we have also computed these DC conductivities analytically by using the black hole horizon data according to the method developed in \cite{Donos:2014cya}. Since there is no singular behavior in the conductivities as $\omega \to 0$ we may regard the real scalar field here as the dilaton in \cite{Donos:2014cya} and the conductivities read
\begin{equation}
\sigma = 1+ \frac{4\pi {\mathcal Q}^2}{\beta^2 s}\,, \qquad \alpha= 4\pi \frac{  {\mathcal{Q}}}{\beta^2}\,, \qquad \bar \kappa = 4 \pi  \frac{s T}{\beta^2} \,,
\end{equation}
where $s$, $\mathcal Q$ and $T$ are the entropy density, charge density and temperature in the dual field theory. They are given by $s= 4\pi r_h^2$, $\mathcal Q = \lim_{r\to \infty}r^2 e^{\chi/2} A_t'(r)$ and $T= \frac{U'(r_h)}{4\pi}e^{(\chi(\infty)-\chi(r_h))/2}$. The analytic values are designated by the red dots in Figure \ref{fig:q0zoom} and they agree to the numerical values very well. For a special case with $\mu=0$, in Figure \ref{fig:q0mu0},  we see that $\sigma(\omega) = 1$, different from superconducting case ($q\ne0$ shown in \cite{Kim:2015dna}), but $\alpha(\omega)=0$, same as superconducting case.  
             
\newpage

\subsection{Superfluid density with a complex scalar hair}

We have found that for $q=0$ there is no $1/\omega$ pole in Im[$\sigma$], of which strength corresponds to superfluid density. To understand it better, we derive an expression for superfluid density for $q \ne 0$. 
Let us start with the Maxwell equation,
\begin{equation}\label{MaxEq}
\nabla_M F^{MN} = - i q \left(  \Phi^* D^N \Phi - \Phi D^N \Phi^*  \right)~~.
\end{equation}
Once we assume that all fields depend on $r$ and $t$ and the fluctuations are allowed  only for the $x$-direction, the   $x$-component of the Maxwell equation reads
\begin{equation} \label{3535}
\partial_r \sqrt{-g} F^{xr} = \partial_t \sqrt{-g} F^{tx} + i q \sqrt{-g} \left(  \Phi^* D^x \Phi - \Phi D^x \Phi^*   \right)~~.
\end{equation}  
The integration of \eqref{3535} from horizon to boundary gives the boundary current
\begin{equation}\label{currentJx}
\begin{split}
J^x &\equiv  \lim_{r \to \infty} \sqrt{-g} F^{xr} \\
&= \lim_{r\to r_h} \sqrt{-g} F^{xr}  + \int_{r_h}^{\infty} \dd r  \partial_t \sqrt{-g} F^{tx} +i q \int_{r_h}^{\infty} \dd r \sqrt{-g} \left(  \Phi^* D^x \Phi - \Phi D^x \Phi^*   \right) \,.
\end{split}
\end{equation}
By hydrodynamic expansion for small $\omega$, it turns out that  the first term and the second term goes to zero as
\begin{equation} \label{omegac}
\lim_{r\to r_h} \sqrt{-g} F^{xr} \sim \mathcal{O} (\omega)  e^{-i\omega t}\,, \qquad \int_{r_h}^{\infty} \dd r  \partial_t \sqrt{-g} F^{tx}  \sim \mathcal{O} (\omega^2)  e^{-i\omega t}\,,
\end{equation}
while the last term goes to constant. 
Here we used the expansions of the fileds near horizon
\begin{equation}
\begin{split}
&\delta A_x = \left( \frac{U(r)}{r^2} \right)^{-i \frac{\omega}{4\pi T}} \hat a_x ( \omega,r) e^{-i \omega t} ~~,~~\\
&\delta g_{tx} = r^2\left( \frac{U(r)}{r^2} \right)^{-i \frac{\omega}{4\pi T}} \hat h_x ( \omega,r) e^{-i \omega t} + i\omega r^2 \zeta e^{-i\omega t}~~,\\
&\delta \psi_x = \left( \frac{U(r)}{r^2} \right)^{-i \frac{\omega}{4\pi T}} \hat \chi_x ( \omega,r) e^{-i \omega t} + \beta \zeta e^{-i\omega t}~~,
\end{split}
\end{equation}
where $\zeta$ is a constant residual gauge parameter fixing $\delta g_{rx} =0$ \cite{Kim:2015sma}, and  $\hat{a}_x$, $\hat{h}_x$ and $\hat{\chi}_x$ can be expanded as
\begin{equation}
\begin{split}
&\hat a_x ( \omega,r) \sim   \mathcal{A}_0 +  \mathcal{A}_1 (r-r_h) +\cdots \,, \\
&\hat h_x ( \omega,r) \sim     \mathcal{H}_1 (r-r_h) +\cdots  \,, \\
&\hat \chi_x (\omega,r) \sim   \tilde\chi_0 +\tilde\chi_1 (r-r_h) +\cdots  \,.
\end{split}
\end{equation}
%
%

With the following source-vanishing-boundary conditions
\begin{equation}
\lim_{r\to \infty} \frac{1}{r^2}  \delta g_{tx} =\lim_{r\to \infty} (  {\hat h}_{x}- i\omega \zeta) e^{-i \omega t}=0\,, \quad \lim_{r\to \infty}\delta \psi_{x}=\lim_{r\to \infty}  \left( {\hat\chi}_{x}+ \beta \zeta \right)e^{-i\omega t}=0\,,
\end{equation}
except $\delta A_x$, the current (\ref{currentJx}) can be interpreted as
\begin{equation}
J^x =- i \omega \sigma_{xx}(\omega) \delta A_x|_{r=\infty} ~~.
\end{equation}
Because only the last term of  (\ref{currentJx}) contribute to $J^x$ for $\omega =0$, as discussed in \eqref{omegac},  the superfluid density $K_s$(the strength of the $1/\omega$ pole of Im[$\sigma$]) is given by
\begin{equation}
\begin{split}
K_s &\equiv \lim_{\omega \rightarrow  0} \omega \mathrm{Im}[\sigma] =  \lim_{\omega \to 0} \frac{J^x}{\delta A_x(r=\infty)}  \\ &=  \lim_{\omega \to 0}  \frac{i q}{\delta A_x(r=\infty)} \int_{r_h}^{\infty} \dd r \sqrt{-g} \left(  \Phi^* D^x \Phi - \Phi D^x \Phi^*   \right)   \\
&=  \lim_{\omega \to 0}  \frac{2 q^2}{\delta A_x(r=\infty)} \int_{r_h}^{\infty} \dd r  e^{-\chi/2} |\Phi|^2  \left(    \delta A_x + \frac{A_t}{U(r)} e^{\chi} \delta g_{tx}  \right)\,.  
\end{split}
\end{equation}
This shows how the hairy configuration $\Phi$ contributes to  $K_s$. If $q=0$, $K_s$ vanishes, which confirms our numerical analysis.

\section{Ward identities: constraints between conductivities}\label{sec4}

In this section, we first analytically derive the Ward identities regarding diffeomorphism from field theory perspective. It gives constraints between conductivities($\sigma, \alpha, \bar{\kappa}$) and two-point functions related to the operator dual to the real scalar field. Next, these identities are confirmed by computing all two-point functions numerically from holographic perspective. 

\subsection{Analytic derivation: field theory}

To derive the Ward identities, we closely follow the procedure in \cite{Herzog:2009xv}\footnote{See \cite{Lindgren:2015lia} for a holographic derivation.} and extend the results therein to the case with real and complex scalar fields, which are $\bar{\psi}_I$ and $\bar{\Phi}$ in \eqref{start1}.   Our final results are \eqref{Ward001}-\eqref{Ward002}  and \eqref{Ward1}-\eqref{Ward3}.

Let us start with a generating functional for Euclidean time ordered correlation functions: 
\begin{equation} \label{start1}
e^{W[\bar h_{\alpha\beta},\bar{A}_\mu, \bar\psi_I,\bar{\Phi},\bar{\Phi}^*]} = \int  DX e^{-S[X,\bar h_{\alpha\beta},\bar{A}_\mu,\bar\psi_I,\bar{\Phi},\bar{\Phi}^*]} \,,
\end{equation}
where $\bar h_{\alpha\beta}$, $\bar{A}_\mu$,  $\bar\psi_I$, $ \bar{\Phi}$, and  
$\bar{\Phi}^* $ are the non-dynamical external sources of the stress-energy tensor $T^{\alpha\beta}$, $U(1)$ current  $J^\mu$, real scalar operators $\mathcal{O}^J$, and complex operators ${\mathcal O_\Phi}^\dagger, {\mathcal O_\Phi}$ respectively. We  define the one-point functions by functional derivatives of $W$:
\begin{equation}
\left< J^\mu(x) \right> = \frac{\delta W}{\delta \bar{A}_\mu(x)}~,~~ \left< T^{\mu\nu}(x) \right> = 2\frac{\delta W}{\delta \bar{h}_{\mu\nu}(x)}~,~~\left< \mathcal{O}^I (x) \right> = \frac{\delta W}{\delta \mathcal{\bar\psi}_I(x)}~,~~\left<  \mathcal O_\Phi  (x)\right>= \frac{\delta W}{\delta \bar\Phi(x)^*}~,
\end{equation}
where these expectation values are not tensors but tensor densities under diffeomorphism. One more functional derivatives acting on one-pint functions give us Euclidean time ordered two-point functions:
\begin{align}
&G_E^{\mu\nu,\alpha\beta} (x,y) \equiv \left<  \mathcal P_t( T^{\mu\nu}(x) T^{\alpha\beta}(y)  ) \right> = 4\frac{\delta^2 W}{ \delta \bar{h}_{\mu\nu}(x) \delta \bar h_{\alpha\beta} (y)    }\,,\\
&
G_E^{\mu\nu,\alpha} (x,y) \equiv \left<  \mathcal P_t( T^{\mu\nu}(x) J^{\alpha}(y)  ) \right> = 2\frac{\delta^2 W}{ \delta \bar{h}_{\mu\nu}(x) \delta \bar{A}_{\alpha} (y)    }\,, \\
&
G_E^{\mu\nu,I} (x,y) \equiv \left<  \mathcal P_t( T^{\mu\nu}(x)
\mathcal{O}^{I}(y)  ) \right> = 2\frac{\delta^2 W}{ \delta \bar{h}_{\mu\nu}(x) \delta \bar\psi_{I} (y)    }\,,
\\&
G_E^{\mu\nu,\Phi} (x,y) \equiv \left<  \mathcal P_t( T^{\mu\nu}(x)
\mathcal{O}^{\Phi}(y)  ) \right> = 2\frac{\delta^2 W}{ \delta \bar{h}_{\mu\nu}(x) \delta \bar{\Phi}^* (y)    }\,,
\\&
G_E^{\mu,\alpha} (x,y) \equiv \left<  \mathcal P_t( J^{\mu}(x) J^{\alpha}(y)  ) \right> = \frac{\delta^2 W}{ \delta\bar{A}_{\mu}(x) \delta \bar{A}_{\alpha} (y)    }\,,
\\
&G_E^{\mu,I} (x,y) \equiv \left<  \mathcal P_t( J^{\mu}(x) \mathcal{O}^{I}(y)  ) \right> = \frac{\delta^2 W}{ \delta\bar{A}_{\mu}(x) \delta \bar\psi_I (y)    }\,, \\
&G_E^{\mu,\Phi} (x,y) \equiv \left<  \mathcal P_t( J^{\mu}(x) \mathcal{O}^{\Phi}(y)  ) \right> = \frac{\delta^2 W}{ \delta \bar{A}_{\mu}(x) \delta \bar{\Phi}^* (y)    }\,,
\\
&G_E^{J,I} (x,y) \equiv \left<  \mathcal P_t( \mathcal{O}^{J}(x) \mathcal{O}^{I}(y)  ) \right> = \frac{\delta^2 W}{ \delta \bar{\psi}_J(x) \delta \bar\psi_I (y)    }\,,
\\&
G_E^{J,\Phi} (x,y) \equiv \left<  \mathcal P_t( \mathcal{O}^{J}(x) \mathcal{O}^{\Phi}(y)  ) \right> = \frac{\delta^2 W}{ \delta \bar\psi_J(x) \delta \bar{\Phi}^* (y)    }\,,
\\&
G_E^{\Phi,\Phi^*} (x,y) \equiv \left<  \mathcal P_t( \mathcal{O}^{\Phi}(x) {\mathcal{O}^{\Phi}}^\dagger(y)  ) \right> = \frac{\delta^2 W}{ \delta \bar{\Phi}^*(x) \delta \bar\Phi (y)    }\,.
\end{align}

We consider the generating functional ${W[\bar h_{\alpha\beta},\bar{A}_\mu, \bar\psi_I,\bar{\Phi},\bar{\Phi}^*]}$ invariant under diffeomorphism, $x^\mu \rightarrow x^\mu + \zeta^\mu$,
and the variation of the fields can be expressed in terms of a Lie derivative with respect to the vector field $\zeta_\mu$
\begin{align}
&\delta \bar{h}_{\mu\nu} = (\mathcal{L}_\zeta \bar h)_{\mu\nu} = \nabla_\mu \zeta_\nu  +  \nabla_\nu \zeta_\mu~~,\\
&\delta \bar{A}_{\mu} = (\mathcal{L}_\zeta \bar{A})_{\mu} = \zeta^\lambda \nabla_\lambda \bar{A}_\mu +(\nabla_\mu \zeta^\nu) \bar{A}_\nu  ~~,\\
&\delta \bar\psi_{I} = (\mathcal{L}_\zeta \bar\psi_I) = \zeta^\lambda \nabla_\lambda \psi_I  ~~,\\
&\delta \bar{\Phi} =  (\mathcal{L}_\zeta \bar\Phi) = \zeta^\lambda \nabla_\lambda \bar\Phi  ~~.
\end{align}
For diffeomorphism invariance, the variation of $W$ should vanish:
\begin{equation}
\begin{split}
&\delta W =\int d^3 x \left(  \frac{\delta W}{\delta \bar{h}_{\mu\nu}(x)  }  (\mathcal{L}_\zeta \bar h)_{\mu\nu}    +  \frac{\delta W}{\delta \bar{A}_{\mu}(x)  }  (\mathcal{L}_\zeta A)_{\mu}  +    \frac{\delta W}{\delta \bar\psi_{I}(x)  }    (\mathcal{L}_\zeta \bar\psi_I)   \right.  \\
& \qquad \qquad \qquad   \left.   +    \frac{\delta W}{\delta \bar{\Phi}(x)  }  (\mathcal{L}_\zeta \bar\Phi)+    \frac{\delta W}{\delta \bar{\Phi^*}(x)  }    (\mathcal{L}_\zeta \bar\Phi^*)\right) =0  \,,
\end{split}
\end{equation}
which, after integration by parts, yields the Ward identity for one-point  functions regarding diffeomorphism. 
\begin{equation}\label{Ward Identity 01}
 D_\mu \big< T^{\mu\nu} \big>  +   {\bar F_\lambda}^{~\nu}\big< J^\lambda \big>+ \big<  \mathcal{O}^I \big> \bar{h}^{\nu\lambda} \partial_\lambda \bar\psi_I +  \big<  \mathcal{O}^\Phi\big> \bar{h}^{\nu\lambda} \partial_\lambda \bar\Phi^* + \big<  {\mathcal{O}^\Phi}^\dagger\big> \bar{h}^{\nu\lambda} \partial_\lambda \bar\Phi  =0 \,,
\end{equation}
where  $D_\mu \left<T^{\mu\nu} (x)\right> \equiv  \partial_\mu \left<  T^{\mu\nu}(x) \right> + \Gamma^\nu_{\alpha\beta} \left<  T^{\alpha\beta} (x)\right> $.

\vspace{0.2cm}

By taking a derivative of (\ref{Ward Identity 01})  with respect to either $ \bar h_{\alpha\beta}(y)$, $\bar A_\alpha (y)$, $\bar \psi^J(y)$ or $\bar\Phi^*(y)$, we obtain the Ward identities for the two-point functions:
\begin{equation} \label{w21}
\begin{split}
&D_\mu \left<  \mathcal P_t ( J^\alpha (y)  T^{\mu\nu} (x) )   \right> + {\bar F_\mu}^{~\nu}  \left<  \mathcal P_t (  J^\alpha (y)  J^\mu(x) ) \right> -\bar{h}^{\nu\beta} \left<  J^\alpha (x) \right>\frac{\partial }{\partial x^\beta}\delta(x-y) \\& + \bar{h}^{\nu\alpha} \left<  J^\mu (x) \right>\frac{\partial }{\partial x^\mu}\delta(x-y) +  \bar{h}^{\mu\nu}  \partial_\mu \bar\psi  \left<  \mathcal P_t ( J^\alpha(y)\mathcal{O}^I (x)  )  \right>\\
&+2\Re\left\{\bar{h}^{\mu\nu}  \partial_\mu \bar{\Phi}^*  \left<  \mathcal P_t ( J^\alpha(y)\mathcal{O}^\Phi (x)  )  \right>\right\} =0\,,
\end{split}
\end{equation}
\begin{equation}
\begin{split}
&D_\mu \left<  \mathcal P_t ( T^{\alpha\beta}(y) T^{\mu\nu} (x) )  \right> +  \delta(x-y) \left(  \bar{h}^{\nu\alpha} D_\mu\left<  T^{\mu\beta}(x) \right> + g^{\nu\beta} D_\mu\left<  T^{\mu\alpha}(x) \right>\right)\\
&+\left(  g^{\nu\alpha} \left< T^{\mu\beta}(x) \right>   +  \bar{h}^{\nu\beta} \left< T^{\mu\alpha}(x) \right>  -  \bar{h}^{\mu\nu} \left< T^{\alpha\beta}(x) \right> \right) \frac{\partial}{\partial x^\mu} \delta(x-y) \\
&-\left(    \bar{h}^{\nu\alpha} \Gamma^\beta_{\mu\lambda}  +  \bar{h}^{\nu\beta} \Gamma^\alpha_{\mu\lambda}  \right)\left< T^{\mu\lambda} (x) \right> \delta(x-y)
\\&+\bar{h}^{\nu\mu} \left(  \bar F_{\mu\lambda} \left<  \mathcal P_t (T^{\alpha\beta}(y) J^\lambda(x)    )  \right> -\partial_\mu \bar \psi_I  \left<   \mathcal P_t (  T^{\alpha\beta}(y) \mathcal{O}^I(x)    )  \right>     \right) \\
&-2\Re \left\{ {\bar h}^{\nu\mu}\partial_\mu \bar{\Phi}^* \left<   \mathcal P_t (  T^{\alpha\beta}(y) \mathcal{O}^\Phi(x)    )  \right>\right\} =0\,,
\end{split}
\end{equation}
\begin{equation}
\begin{split}
&D_\mu \left<  \mathcal P_t (\mathcal{O}^J(y)T^{\mu\nu}(x) )  \right> - \bar{h}^{\nu\mu} \bar F_{\mu \lambda} \left<  \mathcal P_t (\mathcal{O}^J (y) J^\lambda(x)  ) \right>
+\bar{h}^{\nu\lambda}\left<  \mathcal P_t (  \mathcal{O}^J(y) \mathcal{O}^I(x) ) \right>\partial_\lambda \psi_I \\
&+\bar{h}^{\nu\lambda}\left<  \mathcal{O}^J(x) \right>  \frac{\partial}{\partial x^\lambda} \delta(x-y) +2\Re\left\{\bar{h}^{\nu\lambda}\left<  \mathcal P_t (  \mathcal{O}^J(y) \mathcal{O}^\Phi(x) ) \right>\partial_\lambda \bar{\Phi}^* \right\}=0\,,
\end{split}
\end{equation}
\begin{equation} \label{w22}
\begin{split}
&D_\mu \left<  \mathcal P_t (\mathcal{O}^\Phi(y)T^{\mu\nu}(x) )  \right> - \bar{h}^{\nu\mu} \bar F_{\mu \lambda} \left<  \mathcal P_t (\mathcal{O}^\Phi (y) J^\lambda(x)  ) \right>
+\bar{h}^{\nu\lambda}\left<  \mathcal P_t (  \mathcal{O}^\Phi(y) \mathcal{O}^I(x) ) \right>\partial_\lambda \bar \psi_I\\
&+\bar{h}^{\nu\lambda}\left<  \mathcal{O}^\Phi(x) \right>  \frac{\partial}{\partial x^\lambda} \delta(x-y) + \bar{h}^{\nu\lambda}\left<  \mathcal P_t (  \mathcal{O}^\Phi(y) \mathcal{O}^\Phi(x) ) \right>\partial_\lambda \bar{\Phi}^*  \\
&+ \bar{h}^{\nu\lambda}\left<  \mathcal P_t (  \mathcal{O}^\Phi(y) {\mathcal{O}^\Phi}^\dagger(x) ) \right>\partial_\lambda \bar{\Phi}=0\,,
\end{split}
\end{equation}
where the covariant derivatives act only on the operators of $x$. 

From here we consider a flat space, $\bar{h}_{\mu\nu}=\delta_{\mu\nu}$, and assume external fields such as $\bar F_{\mu\nu}$, $\partial_\mu \bar \psi_I$ and $ \bar \Phi$, are constant in space-time.    We further assume translation invariance is not spontaneously broken in the equilibrium state, so all the one point functions, $\big< T^{\mu\nu}\big>$, $\big< J^\mu \big>$, $\big< \mathcal{O}^I \big>$ and $\big< \mathcal{O}^\Phi \big>$, should be constant in space-time. In momentum space, the Ward identities \eqref{w21}-\eqref{w22} read
 \begin{align}
0=&-k_\mu \tilde G_E^{\alpha, \mu\nu}(k)-i {{\bar F}_\mu}^{~\nu}\tilde G_E^{\alpha,\mu} (k) +k^\nu \left< J^\alpha \right>- k_\mu \delta^{\alpha\nu} \left< J^\mu \right>- i \delta^{\mu\nu} \partial_\mu \bar \psi_I \tilde G_E^{\alpha,I}(k)
\,,  \label{w2p1}\\
0=&~k_\mu \left(  \tilde G_E^{\alpha\beta,\mu\nu}(k) + \delta^{\nu\alpha} \left< T^{\mu\beta} \right>   + \delta^{\nu\beta} \left< T^{\mu\alpha} \right> -  \delta^{\mu\nu} \left< T^{\alpha\beta} \right>  \right)\nonumber\\
& + i\left(   {\bar F_\lambda}^{~\nu}  \tilde{G}_E^{\alpha\beta,\lambda}(k)  + \delta^{\nu\mu} \partial_\mu  \bar \psi_I \tilde G_E^{\alpha\beta,I}(k)  
\right) 
\nonumber \\&
-i \delta^{\beta\nu}  \left(    {\bar F_\mu}^{~\alpha}  \left< J^\mu \right>+ \delta^{\alpha \lambda} \left<  \mathcal{O}^I \right>  \partial_\lambda \bar \psi_I 
\right) 
-i \delta^{\alpha\nu}  \left(    {\bar F_\mu}^{~\beta}  \left< J^\mu \right>+ \delta^{\beta \lambda} \left<  \mathcal{O}^I \right>  \partial_\lambda \bar \psi_I 
\right) \,, \\
0=&
-k_\mu \tilde G_E^{J,\mu\nu}(k) - i {\bar F_\mu}^{~\nu}  \tilde G_E^{J,\mu} - i \tilde G_E^{J,I}(k)  \delta^{\nu\lambda} \partial_\lambda \bar \psi_I   
-\left<  \mathcal{O}^J \right>k^\nu \,,
\\0=&
-k_\mu \tilde G_E^{\Phi,\mu\nu}(k) - i {\bar F_\mu}^{~\nu}  \tilde G_E^{\Phi,\mu} - i \tilde G_E^{\Phi,I}(k)  \delta^{\nu\lambda} \partial_\lambda \bar \psi_I   
-\left<  \mathcal{O}^\Phi \right>k^\nu \,. \label{w2p2}
\end{align}

Since we want to study transport coefficients, we analytically continue to Minkowski space, so that the Euclidian Green's functions can be continued to the Retarded Green's functions. Thus, the Ward identities \eqref{w2p1}-\eqref{w2p2} become
\begin{align}
0=
&-k_\mu \tilde G_R^{\alpha, \mu\nu}(k)-i {\bar F_\mu}^{~\nu}\tilde G_R^{\alpha,\mu} (k) +k^\nu \left< J^\alpha \right>- k_\mu \eta^{\alpha\nu} \left< J^\mu \right> + i \eta^{\mu\nu} \partial_\mu \bar \psi_I \tilde G_R^{\alpha,I}(k)
\label{WF1}
\\
0=&k_\mu \left(  \tilde G_R^{\alpha\beta,\mu\nu}(k) + \eta^{\nu\alpha} \left< T^{\mu\beta} \right>   + \eta^{\nu\beta} \left< T^{\mu\alpha} \right> -  \eta^{\mu\nu} \left< T^{\alpha\beta} \right>  \right) \nonumber\\
&\nonumber+ i\left(   {\bar F_\lambda}^{~\nu}  \tilde{G}_R^{\alpha\beta,\lambda}(k)  - \eta^{\nu\mu} \partial_\mu  \bar \psi_I \tilde G_R^{\alpha\beta,I}(k)  
\right)\nonumber\\
&-i \eta^{\beta\nu}  \left(    {\bar F_\mu}^{~\alpha}  \left< J^\mu \right> - \eta^{\alpha \lambda} \left<  \mathcal{O}^I \right>  \partial_\lambda \bar\psi_I 
\right)
-i \eta^{\alpha\nu}  \left(    {\bar F_\mu}^{~\beta}  \left< J^\mu \right>-\eta^{\beta \lambda} \left<  \mathcal{O}^I \right>  \partial_\lambda \bar \psi_I
\right)~, \\
0=&
k_\mu \tilde G_R^{J,\mu\nu}(k) + i {\bar F_\mu}^{~\nu}  \tilde G_R^{J,\mu} - i \tilde G_R^{J,I}(k)  \eta^{\nu\lambda} \partial_\lambda \bar \psi_I   
+\left<  \mathcal{O}^J \right>k^\nu \,, \label{WF11}
\\
0=&
-k_\mu \tilde G_R^{\Phi,\mu\nu}(k) -i {\bar F_\mu}^{~\nu}  \tilde G_R^{\Phi,\mu} + i \tilde G_R^{\Phi,I}(k)  \eta^{\nu\lambda} \partial_\lambda \bar \psi_I   
-\left<  \mathcal{O}^\Phi \right>k^\nu \,. \label{WF2}
\end{align}

In particular, we consider 2+1 dimensional system in an equilibrium state with the constant expectation values for the energy-momentum and current
\begin{equation}
\left<  T^{\mu\nu} \right> =  \left( \begin{array}{ccc}
  \epsilon & 0 & 0 \\
0 &  p  & 0\\
0 & 0 &  p\end{array} \right)\,, \qquad \left< J^\mu \right> =  \left( n , 0 , 0  \right) \,,
\end{equation}
with finite or zero condensate  $\left<  \mathcal{O}^\Phi \right> $  and $\left<\mathcal O^I\right>=0$.  
To this system we apply a constant external magnetic field with a background scalar $\bar{\psi}_I$:
\begin{equation}
\bar F =  B \dd x \wedge \dd y \,, \qquad \bar \psi_I =(\beta x,\beta y)\,.
\end{equation} 
We take $k^\mu=(\omega , 0, 0)$ to focus on the spatially homogeneous AC conductivity induced by the small external electric field and  temperature gradient along $i(=x,y)$ direction.

Under these conditions the Ward identities \eqref{WF1}-\eqref{WF11} becomes 
\begin{align}\label{Ward 00}
&\omega \tilde G_R^{j, 0 k} -i B\epsilon^{ik} \tilde G_R^{j,i}  + \omega \delta^{j k} n  + i   \beta \delta_{I}^k \tilde G_R^{j,I} =0\,,\\
&\omega \left(  \tilde G_R^{0j,0k}     + \delta^{kj} \epsilon   \right)- i\left(   B\epsilon^{ik}  \tilde{G}_R^{0 j,i}  - \beta \delta^k_I \tilde G_R^{0 j,I}   \right) =0 \,,\\&
-\omega \tilde G_R^{J,0 j}  +  i B \epsilon^{ij} \tilde G_R^{J,i} - i \tilde G_R^{J,I}      \beta \delta^j_I=0 \,, \label{Ward 01}
\end{align}
where $i,j,k$ run over $x$ and $y$.  It is convenient to introduce the complexified combinations defined as 
 \begin{equation}\label{check1}
 \begin{split}
&\left< {\nJ}{\nT} \right> _\pm \equiv \pm \tilde G^{x,0x}_R - i \tilde G^{x,0y}_R  \,,  \quad \ \ \ 
\left< {\nJ}{\nJ} \right> _\pm \equiv \pm \tilde G^{x,x}_R - i \tilde G^{x,y}_R \,,   \\
& \left< {\nT}{\nT} \right> _\pm \equiv \pm \tilde G^{0x,0x}_R - i \tilde G^{0x,0y}_R\,, \quad  \  
\left< {\nS}{\nS} \right> _\pm \equiv \pm \tilde G^{1,1}_R - i \tilde G^{1,2}_R  \,,   \\
& \left< {\nS}{\nJ} \right> _\pm \equiv \pm \tilde G^{1,x}_R - i \tilde G^{1,y}_R \,,  \qquad \ \ \ 
  \left< {\nS}{\nT} \right> _\pm \equiv \pm \tilde G^{I=1,0x}_R - i \tilde G^{I=1,0y}_R  \,.
 \end{split}
 \end{equation} 
With this notation, \eqref{Ward 00}-\eqref{Ward 01}  can be rewritten as  
\begin{align}
&\pm \omega \left<\nJ\nT \right>_{\pm }-B\left<\nJ \nJ \right>_{\pm }+\omega \, n  \pm i \beta\left<\nJ \nS \right>_{\pm }= 0  \,, \label{tt1} \\
&\pm \omega \left<\nT\nT \right>_{\pm }-B\left<\nT\nJ \right>_{\pm }+\omega \, \epsilon   \pm i \beta \left<\nT \nS \right>_{\pm }= 0 \,, \\
&\mp \omega \left<\nS\nT \right>_{\pm }-B\left<\nS\nJ \right>_{\pm }  \mp i \beta \left<\nS \nS \right>_{\pm }= 0 \,, \label{tt2}
\end{align}
or, in terms of the heat current $\nQ= \nT-\mu \nJ$,
\begin{align}
&\pm \omega \left<\nJ\nQ \right>_{\pm }+~(\pm \mu \omega -B) \left<\nJ \nJ \right>_{\pm }+\omega \,  n  \pm i \beta \left<\nJ \nS \right>_{\pm }= 0 \,, \\
&\pm \omega  \left<\nQ\nQ \right>_{\pm }+~(\pm \mu \omega -B) \left< \nJ\nQ \right>_{\pm }+\omega  (\epsilon - \mu \, n )   \mp  i \mu\beta \left<\nJ \nS \right>_{\pm }= 0 \,, \\
&\mp \omega  \left<\nS\nQ \right>_{\pm }+~(\mp \mu \omega -B) \left< \nS\nJ \right>_{\pm } \mp i \beta \left<\nS \nS \right>_{\pm }= 0 \,.
\end{align}
Finally, using the Kubo formulas for conductivities\footnote{The complexified conductivities are denoted by $X_\pm \equiv X_{xy} \pm i X_{xx}$, where $X = \sigma, \alpha, \bar{\alpha}, \bar{\kappa}$. }
\begin{equation} 
\sigma_\pm = \frac{1}{\omega}\left<\nJ \nJ \right>_{\pm }  \,,  \quad \alpha_\pm = \frac{1}{\omega T} \left<\nQ \nJ \right>_{\pm }  \,, \quad  \bar{\alpha}_\pm =\frac{1}{\omega T} \left<\nJ \nQ \right>_{\pm }  \,, \quad  \bar{\kappa}_\pm = \frac{1}{\omega T} \left<\nQ \nQ \right>_{\pm } \,,
\end{equation} 
we obtain the relations between the conductivities:
\begin{align} 
&\mathrm{Ward\ 1:}\quad \pm \omega^2 T \bar \alpha_{\pm }+~\omega(\pm \mu \omega -B) \sigma_{\pm }+\omega  \, n  \pm i \beta \left<\nJ \nS \right>_{\pm }= 0 \,,  \label{Ward001} \\
&\mathrm{Ward\ 2:}\quad \pm \omega^2 T \bar\kappa_{\pm }+~\omega(\pm \mu \omega -B)T \bar\alpha_{\pm }+\omega  (\epsilon' - \mu \, n  )  \mp  i \mu\beta \left<\nJ \nS \right>_{\pm }= 0  \,,  \label{Ward002} \\
&\mathrm{Ward\ 3:}\quad \mp \omega  \left<\nS\nQ \right>_{\pm }+~(\mp \mu \omega -B) \left< \nS\nJ \right>_{\pm } \mp i \beta \left<\nS \nS \right>_{\pm }= 0  \,, \label{Ward003}
\end{align}
where we redefined $\bar{\kappa}_{\pm}$ 
\begin{equation} \label{countt}
\bar \kappa_{\pm} \rightarrow \bar \kappa_{\pm} - \frac{\left<  \nT \nT  \right>_{\pm, \omega = 0} }{\omega  T} \,,
\end{equation}
to subtract a counter term and $\epsilon' \equiv \epsilon \pm  \left<  \nT \nT  \right>_{\pm, \omega = 0}$. In normal phase, if $\beta=0$ and $B \ne 0$, $\left<  \nT \nT  \right>_{\pm, \omega = 0} = \pm \epsilon/2$~\cite{Herzog:2009xv}.

If $B=0$, \eqref{tt1}-\eqref{tt2} is simplified as 
 \begin{align}
 &\omega\left< {\nJ}{\nT} \right> + \omega \, n  +i \beta \left< {\nJ}{\nS}\right>=0 \,, \\
  &\omega\left< {\nT}{\nT} \right> + \omega \, \epsilon+i \beta \left<{\nT}\nS\right>=0 \,, \\
  &\omega \left<\nS\nT \right> + i\beta \left< \nS \nS \right>=0 \,,
\end{align}
 where 
 \begin{equation}
 \begin{split}
 &\left< {\nJ}{\nT} \right>  \equiv \tilde G_R^{x,0x} ~,~\left< {\nJ}{\nS} \right>  \equiv \tilde G_R^{x,1} ~,~ \left< {\nT}{\nT} \right>  \equiv \tilde G_R^{0x,0x} \,, \\
 &\left< {\nT}{\nS} \right>  \equiv \tilde G_R^{0x,1}~,~ \left< {\nS}{\nT} \right>  \equiv \tilde G_R^{1,0x}~,~\left< {\nS}{\nS} \right>  \equiv  \tilde G_R^{1,1} \,,
 \end{split}
 \end{equation}
 since we don't need to consider $y$-coordinate.
Or, in terms of the heat current $\nQ= \nT-\mu \nJ$,
\begin{align} 
&\left< {\nJ}{\nQ} \right>  + \mu\left< {\nJ}{\nJ} \right> + n +i \frac{\beta}{\omega}\left< {\nJ}{\nS} \right>=0 \,,  \\
&\left< {\nQ}{\nQ} \right>  + \mu\left< {\nJ}{\nQ} \right>+ \mu\left< {\nQ}{\nJ} \right> + \mu^2 \left< {\nJ}{\nJ} \right> + \epsilon +i \frac{\beta}{\omega}\left( \left< {\nQ}{\nS} \right>+\mu \left< {\nJ}{\nS} \right>   \right)=0 \,, \\
&\left< {\nS}{\nQ} \right>  + \mu\left< {\nS}{\nJ} \right> + n +i \frac{\beta}{\omega}\left< {\nS}{\nS} \right>=0   \,.
\end{align}
  Using the Kubo formulas
\begin{equation}
\sigma = \frac{1}{i\omega}\left<\nJ \nJ \right> \,,  \quad \alpha = \frac{1}{i\omega T} \left<\nQ \nJ \right>  \,, \quad  \bar{\alpha}=\frac{1}{i\omega T} \left<\nJ \nQ \right> \,, \quad  \bar{\kappa} = \frac{1}{i\omega T} \left<\nQ \nQ \right>\,,
\end{equation}
we obtain the relations between the conductivities:
\begin{align}
&\mathrm{Ward\ 4:}\quad  \alpha  + \frac{\mu}{T}  \sigma  -i \frac{n}{\omega T}   +  \beta \frac{\left< \nJ\nS \right>}{\omega^2 T} =0 \,, \label{Ward1} \\
&\mathrm{Ward\ 5:}\quad  {\bar \kappa} + 2 \mu \alpha  + \frac{\mu^2\sigma}{T}  -i\frac{  \epsilon'  }{\omega T}  +\beta \frac{\left< \nQ\nS\right>}{\omega^2 T}     +\beta \frac{\mu \left< \nJ\nS\right>}{\omega^2 T}=0\label{Ward2} \,, \\
&\mathrm{Ward\ 6:}\quad  \left< \nS\nT\right>   + i \beta \frac{\left<  \nS\nS \right>}{\omega}=0 \,, \label{Ward3}
\end{align}
where $\bar{\kappa}$ is redefined as \eqref{countt} to subtract a counter term and $\epsilon' = \epsilon +  \left<  \nT \nT  \right>_{\omega = 0}$. In normal phase,  $\left<  \nT \nT  \right>_{\omega = 0} = -\epsilon$ and $\epsilon/2$ for $\beta=0$ and $\beta \ne 0$ respectively.  In superconducting phase for $\beta = 0$, $\left<  \nT \nT  \right>_{\omega = 0}$ = $-\epsilon$.\footnote{{If $W$ is invariant under $U(1)$ gauge transformations, $ \bar{A}_\mu  \rightarrow  \bar{A}_\mu + \partial_\mu \Lambda$, the Ward identity for one-point function yields current conservation $\partial_\mu \left<  J^\mu (x)  \right> = 0$. The Ward identities for two point functions are $-\omega G^{t,t} + k G^{t,x} = 0, -\omega G^{t,x} + k G^{x,x} = 0, -\omega G^{\Phi,t} + k G^{\Phi,x} = \langle {\mathcal O^\Phi}\rangle, -\omega G^{\bar{{\Phi}}, t} + k G^{\bar{{\Phi}},x} = - \langle \mathcal{O}^\Phi \rangle$,~\cite{Herzog:2011ec} and  $-\omega G^{1, t} + k G^{1,x} = 0$.}}

 \subsection{Numerical confirmation: holography}
 
 In the previous subsection we have derived the Ward identities  for two-point functions from field theory perspective.  Here we show that those Ward identities indeed hold in our holographic model studied in~\cite{Kim:2014bza,Kim:2015dna,Kim:2015wba}. More concretely, our goal is to compute $\sigma, \alpha, \bar{\kappa},\langle\nJ\nS \rangle, \langle\nQ\nS \rangle ,\langle\nS\nS \rangle$ numerically and plug them into the Ward identities \eqref{Ward001}-\eqref{Ward002}  and \eqref{Ward1}-\eqref{Ward3} to check if they add up to zero or not.
 
\begin{figure}[]
 \centering
     {\includegraphics[width=4.5cm]{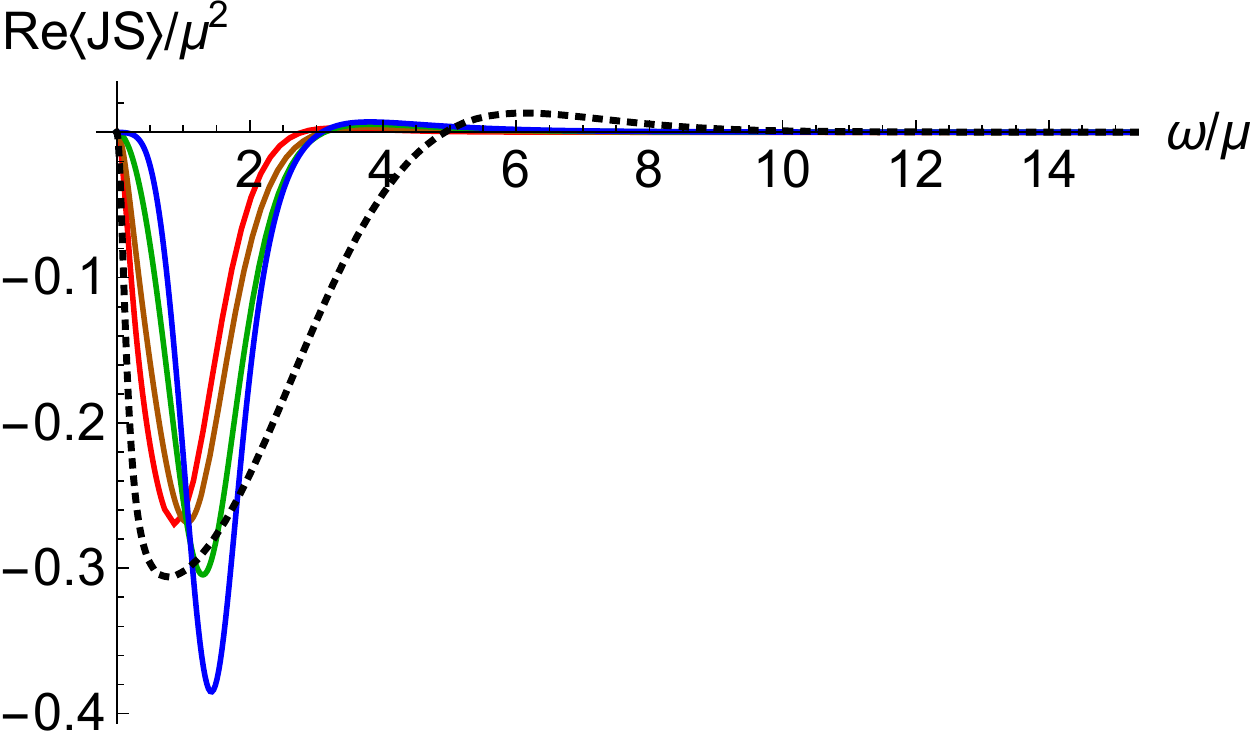} \label{}}\hspace{3mm}
   {\includegraphics[width=4.5cm]{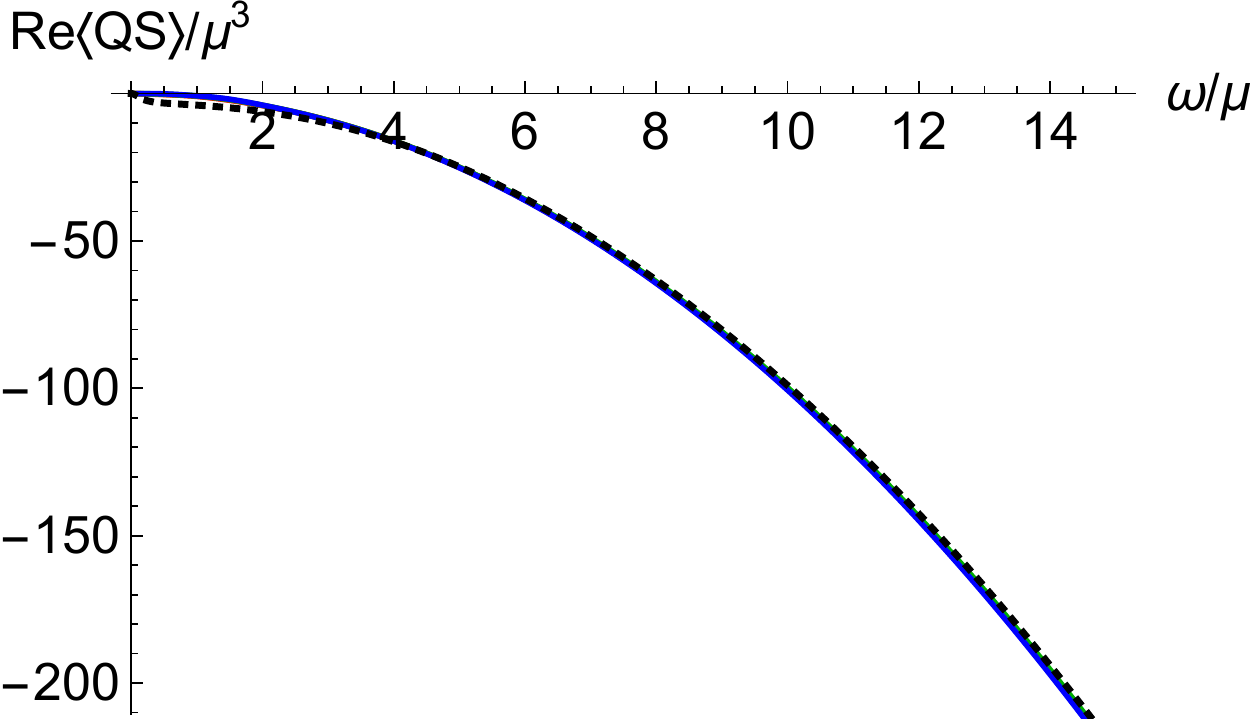} \label{}}\hspace{3mm}
   {\includegraphics[width=4.5cm]{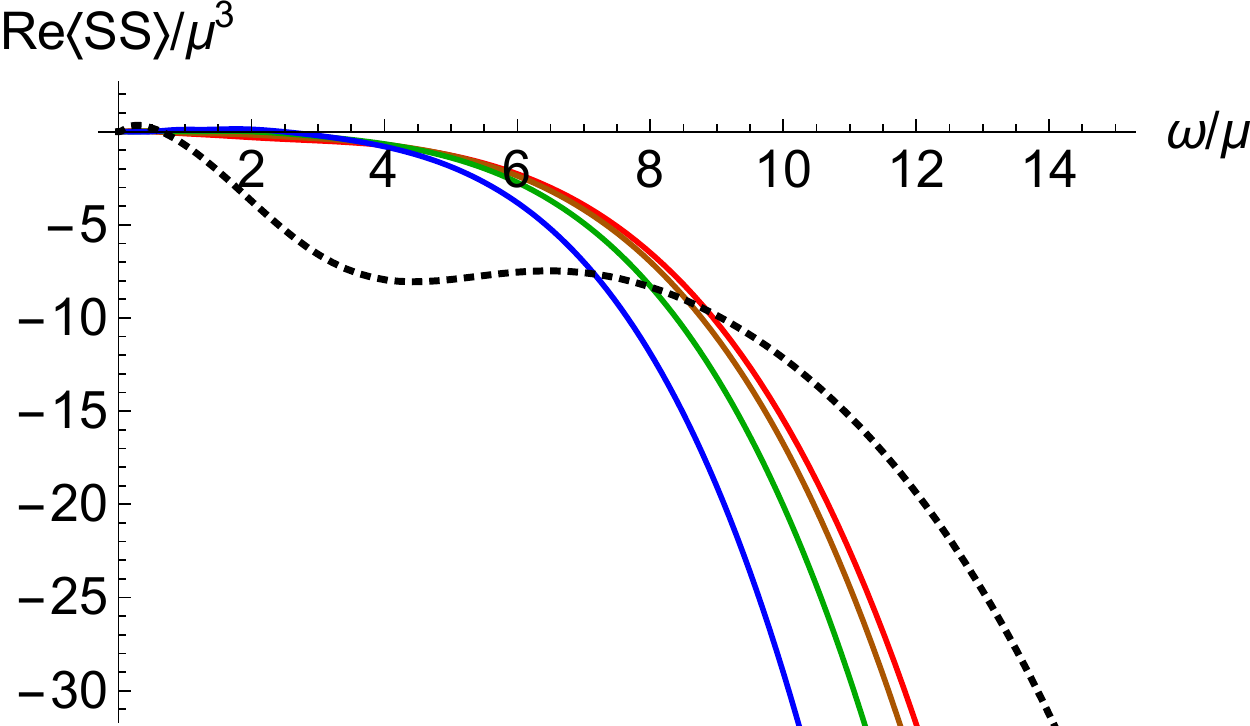} \label{}}
 {\includegraphics[width=4.5cm]{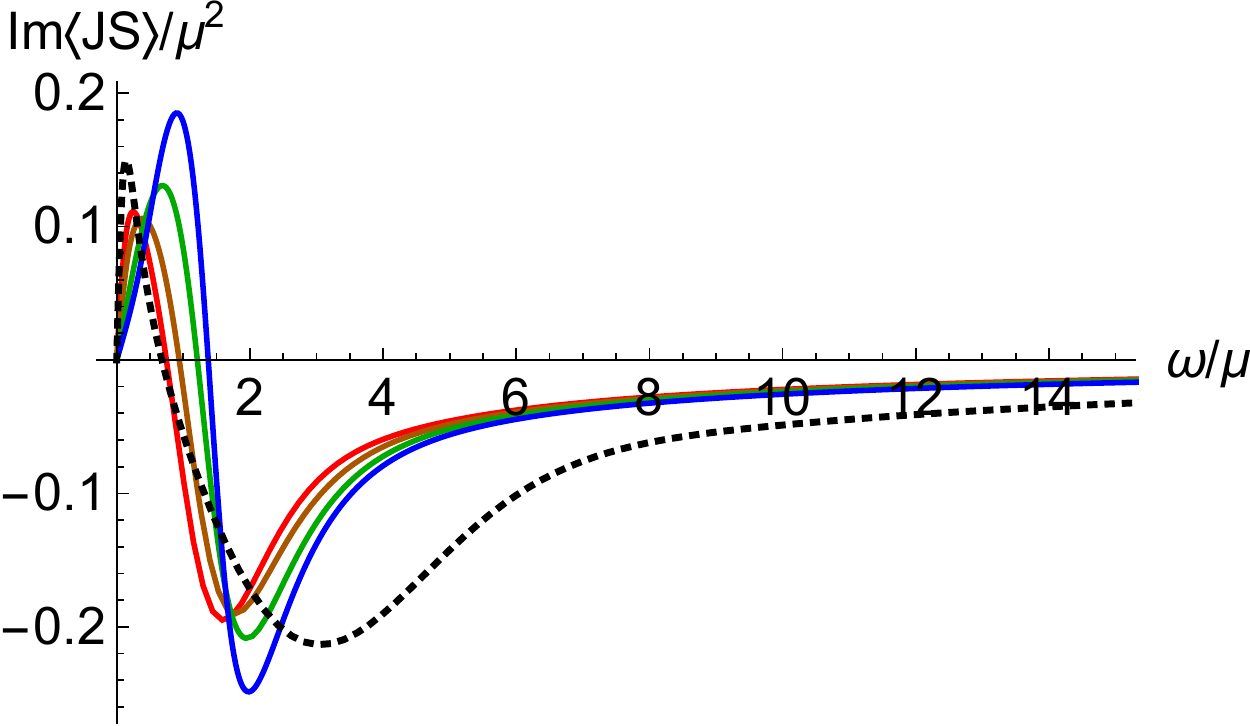} \label{}}\hspace{3mm}
   {\includegraphics[width=4.5cm]{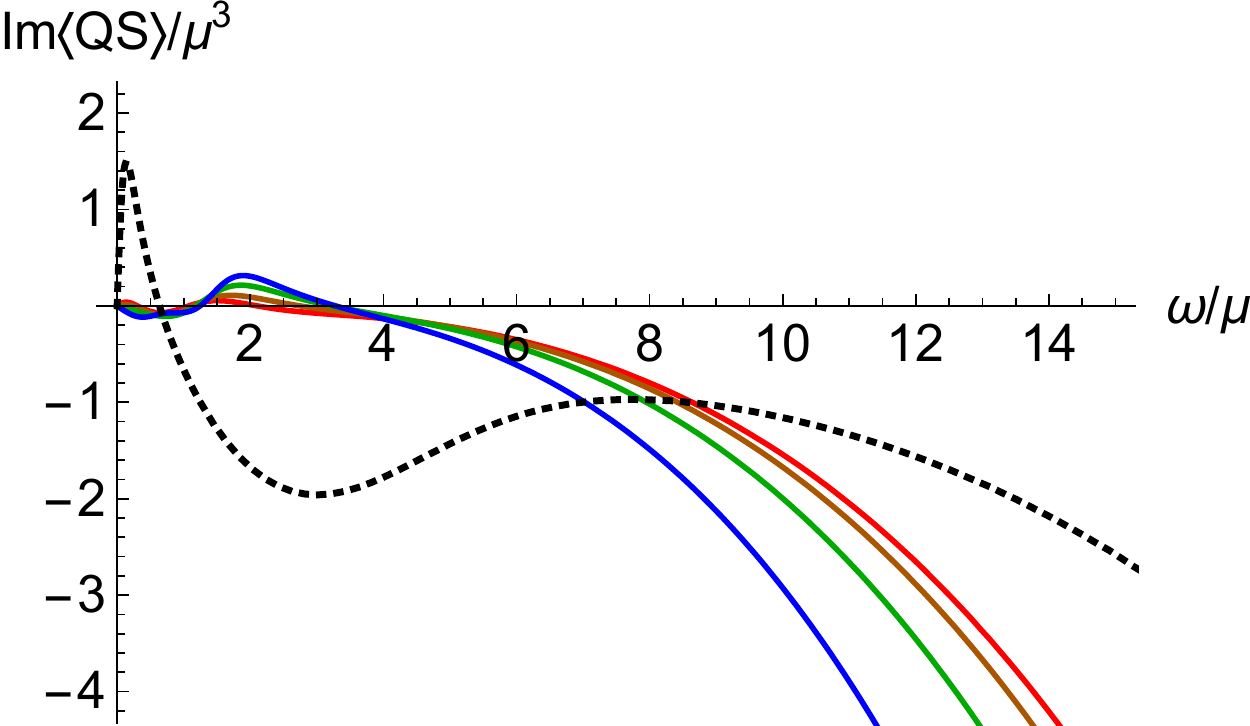} \label{}}\hspace{3mm}
   {\includegraphics[width=4.5cm]{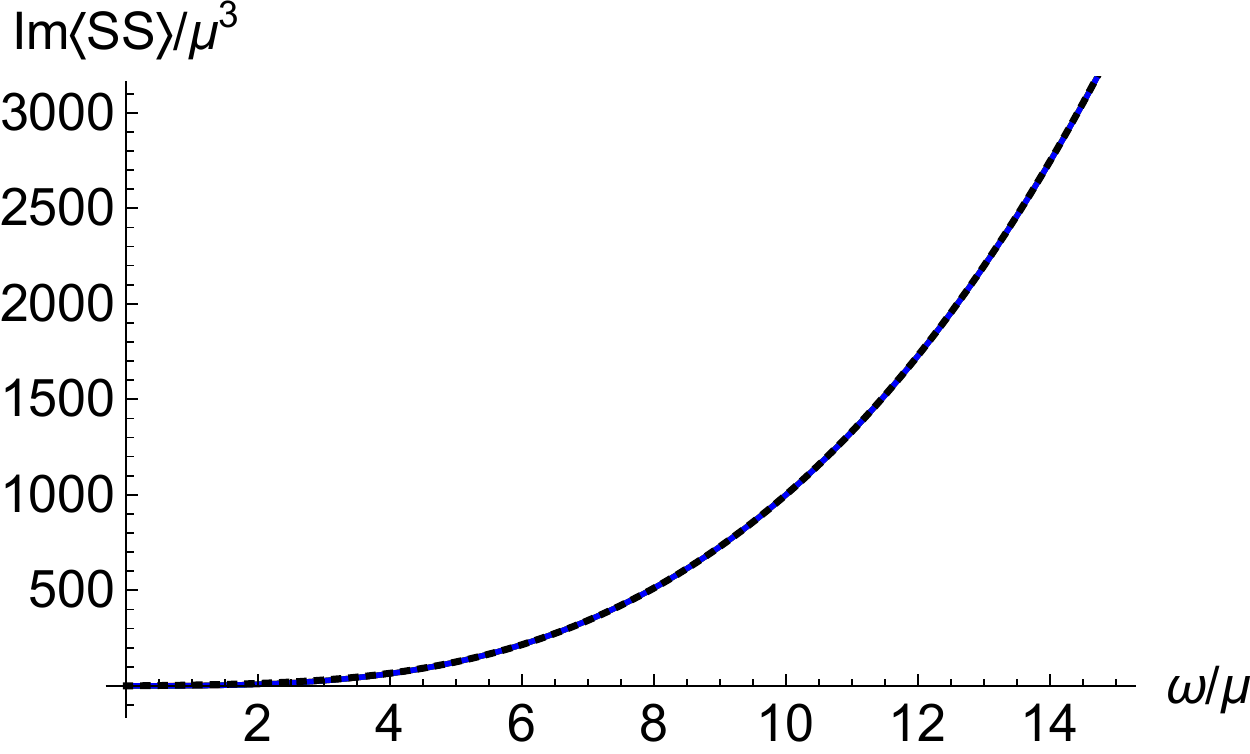} \label{}}
  \caption{ $\langle \nJ\nS \rangle$, $\langle \nQ\nS \rangle$, $\langle \nS\nS \rangle$ for $\mu/\beta=1$ at $T/T_c = 3.2, 1, 0.89, 0.66, 0.27$  (dotted, red, orange, green, blue).} 
            \label{2ptScalar1}
\end{figure}

\begin{figure}[] 
 \centering
     {\includegraphics[width=4.5cm]{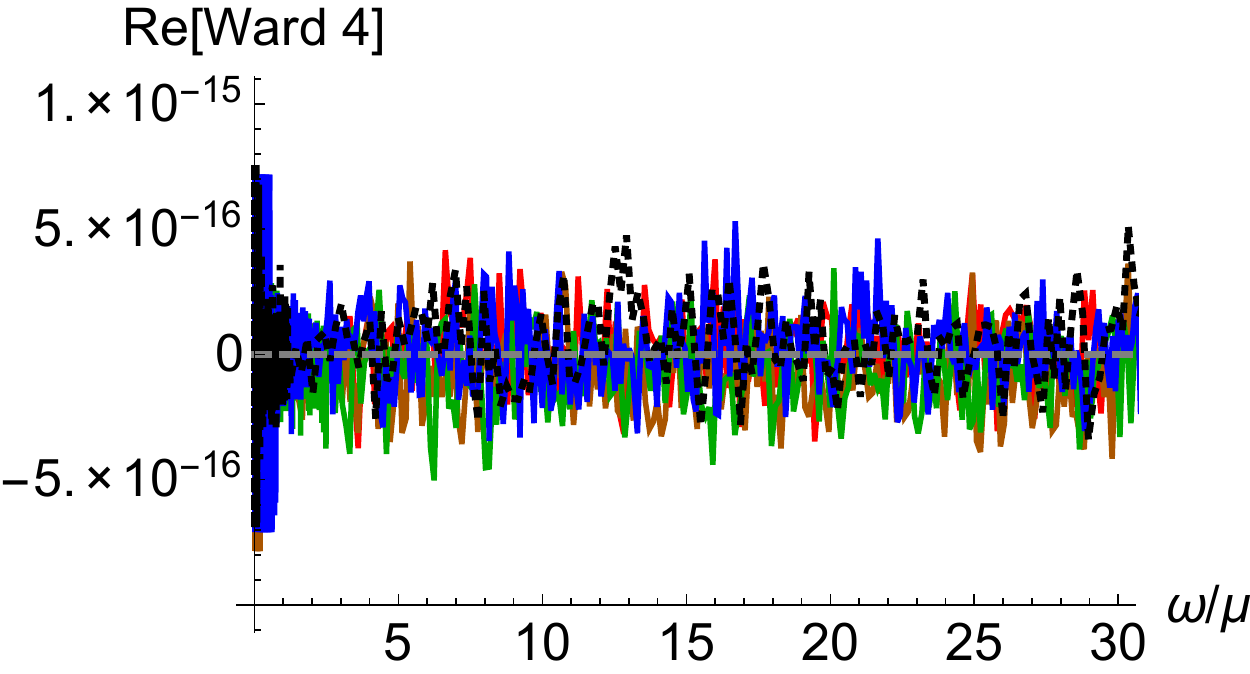} \label{}}\hspace{3mm}
   {\includegraphics[width=4.5cm]{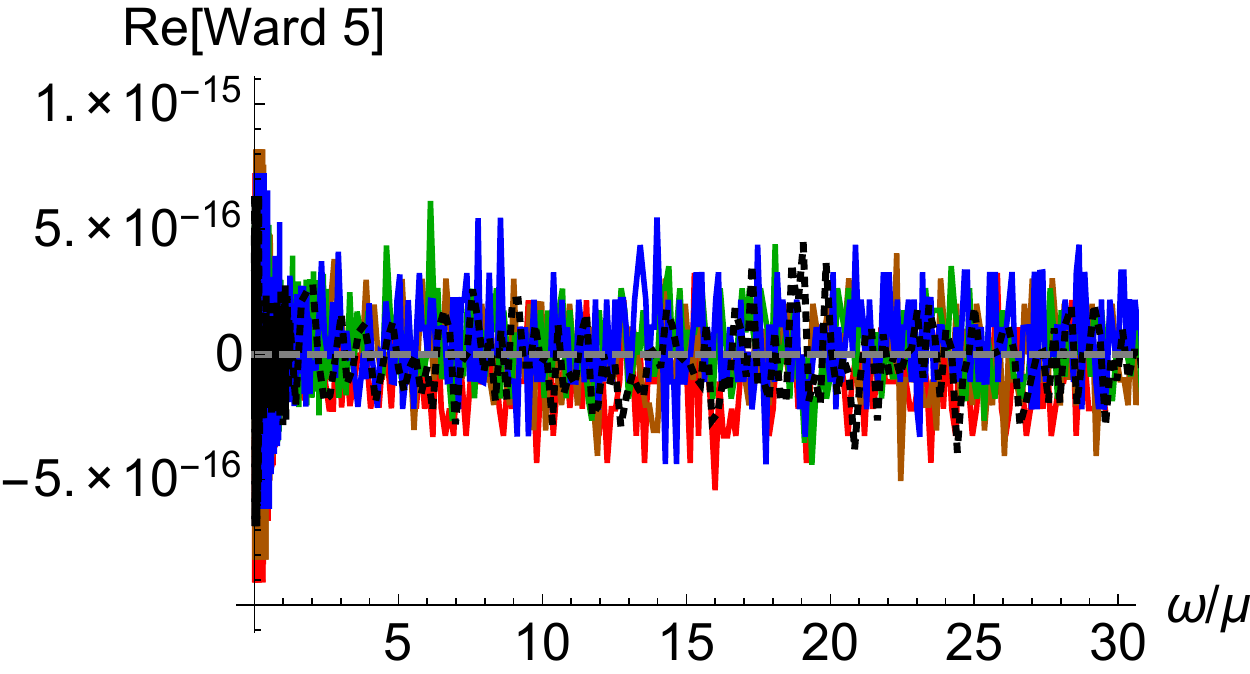} \label{}}\hspace{3mm}
   {\includegraphics[width=4.5cm]{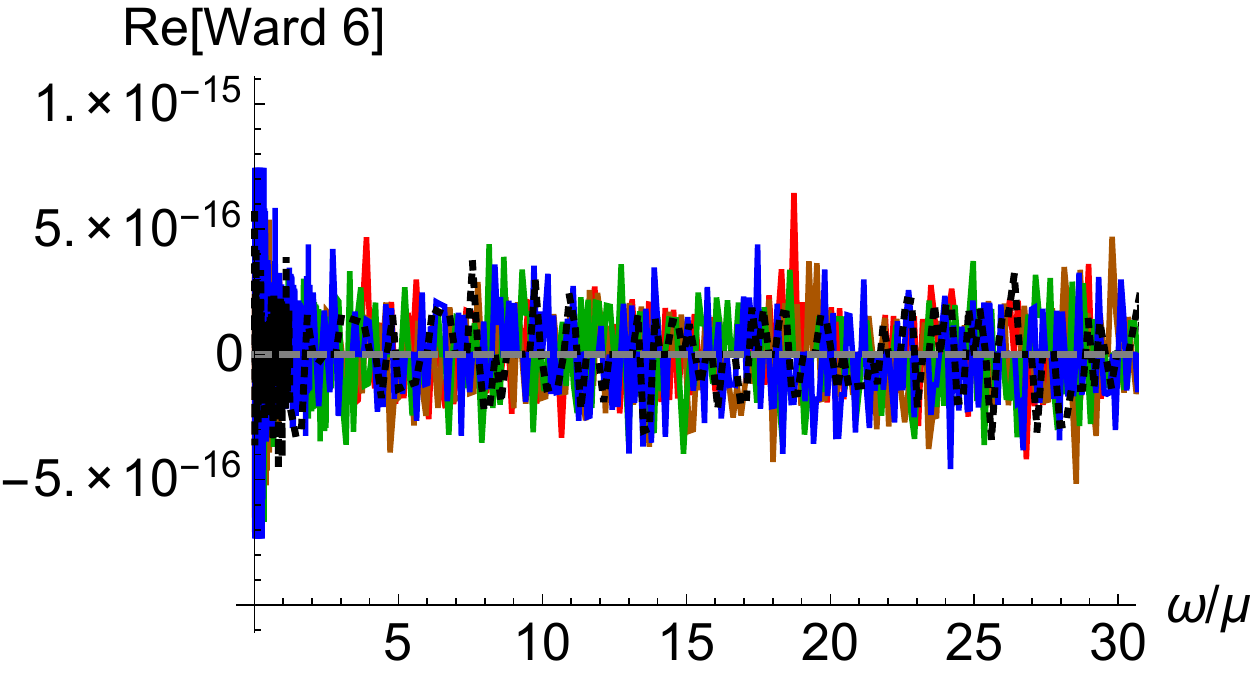} \label{}}
     \subfigure[ Ward 4: \eqref{Ward1}]
 {\includegraphics[width=4.5cm]{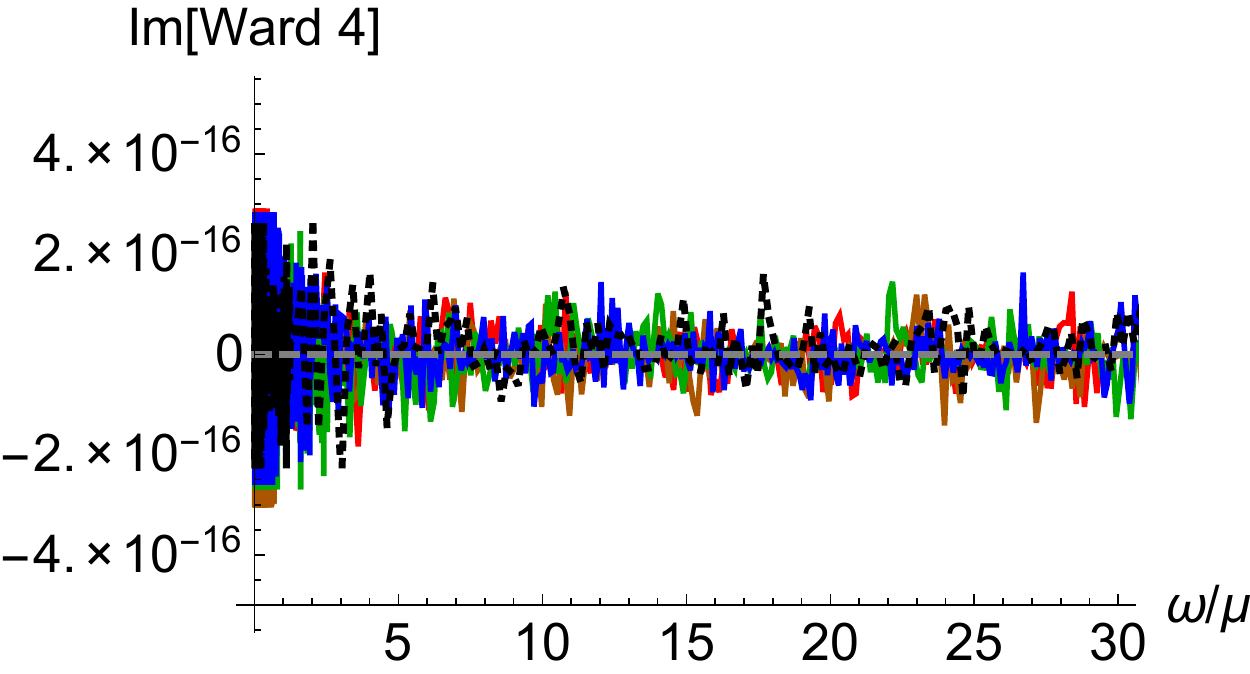} \label{}}\hspace{3mm}
    \subfigure[Ward 5: \eqref{Ward2}  ]
   {\includegraphics[width=4.5cm]{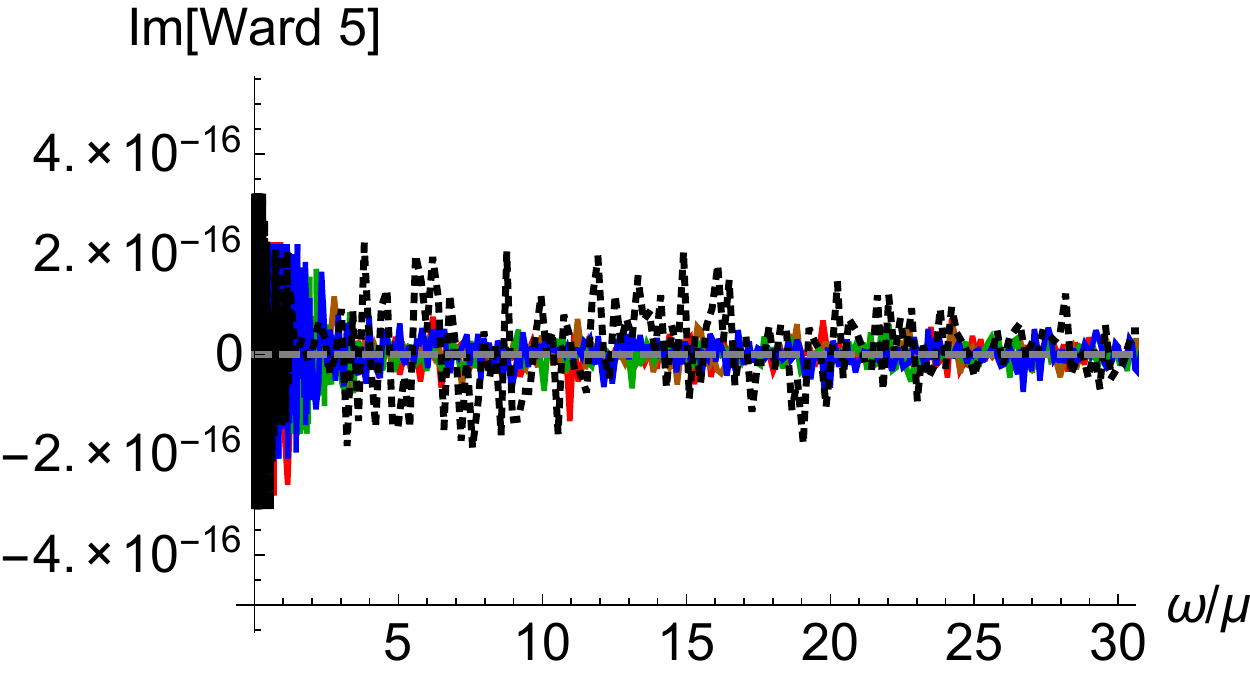} \label{}}\hspace{3mm}
     \subfigure[ Ward 6: \eqref{Ward3} ]
   {\includegraphics[width=4.5cm]{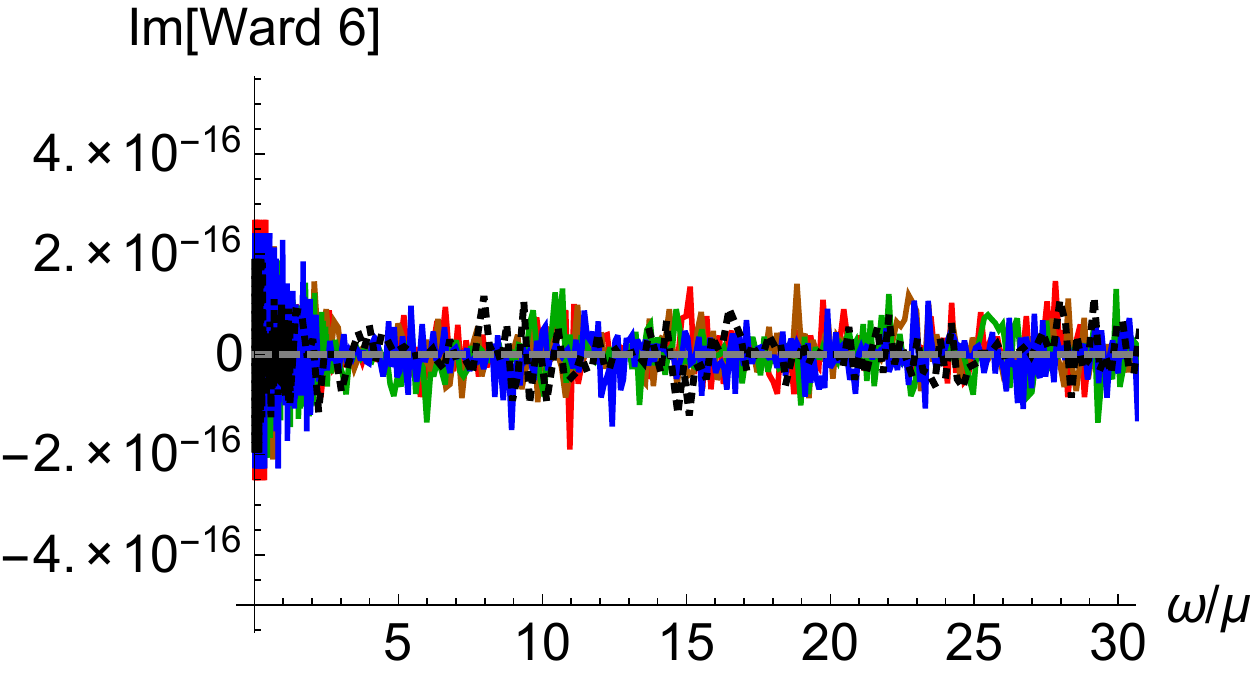} \label{}}
  \caption{Confirmation of Ward identities. The left hand side of \eqref{Ward1}-\eqref{Ward3} is plotted for the temperatures shown in Fig.\ref{2ptScalar1} all together. They are almost zero, less than $10^{-15}$. } 
            \label{WC1}
\end{figure}

For $B=0$, the conductivities $\sigma, \alpha, \bar{\kappa}$  were reported in \cite{Kim:2015wba} and reproduced in Figure \ref{oldresult}.  Here in Figure \ref{2ptScalar1} we display  the other two-point correlation functions related to the real scalar operator, $\left< \nJ\nS \right>$,  $\left< \nQ\nS \right>$, and  $\left< \nS\nS \right>$.  
Contrary to Figure \ref{oldresult} there is no divergence at $\omega = 0$, which is also shown in their small $\omega$ expressions \eqref{poleJSsp}-\eqref{poleSSsp}. By using the data in Figure \ref{oldresult} and \ref{2ptScalar1} we numerically compute the left hand side of three Ward identities \eqref{Ward1}-\eqref{Ward3}. The numerical sums for all considered temperatures are shown together in Figure \ref{WC1}. All of them vanish($ <10^{-15}$), confirming analytic formulas. We have also checked three other cases: 1) $B=0$ and $\beta/\mu=0.1$, 2) $B=0$ and $\mu =0$, 3) $B\ne0$.  It turned out that all numerical sums vanish too.  For completeness, we show the numerical data for these three cases in the appendix \ref{appA}.
%
 
\begin{sloppypar}
\end{sloppypar}
%
%

\section{Homes' law and Uemura's law} \label{sec5}

We first analyse small $\omega$ behaviours of the two-point correlation functions based on our numerical results and Ward Identities both in superconducting and normal phase. After identifying superfluid density and normal component density in the two fluid model of superconductor we check Homes' law and Uemura's law.  

\subsection{Conductivities at small $\omega$}

For $\beta=0$, the Ward identities  \eqref{Ward1}-\eqref{Ward3} become simplified
\begin{equation} \label{wardb0}
\begin{split}
&\text{Re}[\alpha] = - \frac{\mu}{T} \text{Re}[\sigma] \,, \qquad \qquad \text{Re}[\bar{\kappa}] = - \mu \text{Re}[\alpha] \,, \\
&\text{Im}[\alpha] + \frac{\mu}{T} \text{Im}[\sigma]  = \frac{ n }{\omega T}\,, \qquad  \text{Im}[\bar{\kappa}] + \mu \text{Im}[\alpha] =  \frac{ \epsilon' -\mu \,  n  }{\omega T}  \,. 
\end{split}
\end{equation}
This relation was reported in ~\cite{Hartnoll:2009sz} for normal phase and here we have shown it still holds for superconducting phase. 
By  these relations, once $\sigma$ is obtained, $\alpha$ and $\bar{\kappa}$ are completely determined.  In both normal  and superconducting phase,  ${\rm Im}[\sigma]$ turns out to have $1/\omega$ pole by numerical computation so  ${\rm Re}[\sigma]$ is infinite by the  Kramers-Kronig relation~\cite{Hartnoll:2009sz, Hartnoll:2008kx, Kim:2014bza}.  Therefore,  by \eqref{wardb0}, ${\rm Re}[\alpha]$ and ${\rm Re}[\bar{\kappa}]$ are also infinite and ${\rm Im}[\alpha]$ and ${\rm Im}[\bar{\kappa}]$ have $1/\omega$ poles.  In normal phase it is due to the absence of momentum relaxation and in superconducting phase there is another contribution due to condensate.

For  $\beta \neq 0$, the Ward identities  \eqref{Ward1}-\eqref{Ward3} may be rewritten as  
\begin{align}
&\text{Re}[\alpha] + \frac{\mu}{T} \text{Re}[\sigma]  = -\frac{\beta}{T} \frac{\text{Re}[\left< \nJ \nS \right>]}{\omega^{2}}  \label{ward1re} \,, \\
&\text{Im}[\alpha] + \frac{\mu}{T} \text{Im}[\sigma] =  \frac{ n }{\omega T} - \frac{\beta}{T} \frac{\text{Im}[\left< \nJ \nS \right>]}{\omega^{2}} \,,\label{ward1im} \\
& \text{Re}[\bar{\kappa}] + \mu \text{Re}[\alpha] =- \frac{\beta}{T} \frac{\text{Re}[\left< \nQ \nS \right>]}{\omega^{2}}  \,,\label{ward2re}\\
& \text{Im}[\bar{\kappa}] + \mu \text{Im}[\alpha]=   \frac{ \epsilon' }{\omega T} - \frac{\mu \, n }{\omega T} - \frac{\beta}{T} \frac{\text{Im}[\left< \nQ \nS \right>]}{\omega^{2}} \,, \label{ward2im} \\
& \left< \nQ \nS \right> + \mu \left< \nJ \nS \right> + \beta \frac{ \left<\nS \nS \right>}{\omega} = 0 \,, \label{ward3re}
\end{align}
where \eqref{ward2re} and \eqref{ward2im} are obtained by combining (\ref{Ward1}) and (\ref{Ward2}), and we used 
 $\nQ= \nT-\mu \nJ$. 
Contrary to the case of $\beta=0$, $\alpha$ and $\bar{\kappa}$ are not determined by $\sigma$ only, because there are other correlators $\left< \nJ \nS \right>$, $\left< \nQ \nS \right>$, and $\left< \nS \nS \right>$ involved in the Ward Identities. For exmaple, once we know $\sigma$, $\alpha$, and $\bar{\kappa}$, we can read off $\left< \nJ \nS \right>$, $\left< \nQ \nS \right>$, and $\left< \nS \nS \right>$ by the Ward identities. 

In  normal phase (see, for example, the dotted curve in Figure \ref{oldresult}), the real and imaginary part of $\sigma$, $\alpha$, and $\bar{\kappa}$ are all finite at $\omega = 0$ due to the momentum relaxation ($\beta \ne 0$).   At small $\omega$, it is inferred that Re$[\left< \nJ \nS \right>] \sim \omega^{2}$ from (\ref{ward1re}) and Im$[\left< \nJ \nS \right>] \sim \omega$ from (\ref{ward1im}).  Also Re$[\left< \nQ \nS \right>] \sim \omega^{2}$ from (\ref{ward2re}) and Im$[\left< \nQ \nS \right>] \sim \omega$ from (\ref{ward2im})\footnote{It is possible that the power of $\omega$ could be bigger than what are inferred.  We have fixed them from numerical data.}.
Finally, the small $\omega$ behaviour of $\left< \nS \nS \right>$ is determined by $\left< \nJ \nS \right>$ and $\left< \nQ \nS \right>$ via (\ref{ward3re}).  
In superconducting phase (see for example the solid curves in Figure \ref{oldresult}), unlike normal phase, ${\rm Im}[\sigma]$ and ${\rm Im}[\bar{\kappa}]$ have $1/\omega$ poles, which implies the existence of  delta functions at $\omega=0$ in the corresponding real parts. In summary, the small $\omega$ behaviours can be written as
%
\begin{align}
&\sigma  \sim K_s\frac{\pi}{2}\delta(0)+ \sigma_{DC}   + i \left( \frac{K_s}{\omega} +  \omega \sigma_{I} \right), \label{poleSsp}\\ 
&\alpha  \sim \alpha_{DC}   +  i \omega \alpha_{I} , \label{poleAsp}\\
&\bar{\kappa}  \sim -\frac{\mu^2 K_s}{T} \frac{\pi}{2}\delta(0) + \bar{\kappa}_{DC}  + i  \left( - \frac{\mu^2 K_s}{ T} \frac{1}{\omega}+ \omega  \bar{\kappa}_{I}  \right), \label{poleKsp}
\end{align}
\begin{align}
&\left< \nJ\nS\right> \sim \frac{1}{\beta}\left(-\mu K_{s} \frac{\pi}{2} \delta(0) - \mu \sigma_{DC} - T \alpha_{DC}  \right) \omega^{2}+  i \omega  \left( \frac{ n -\mu K_s}{\beta}  \right), \label{poleJSsp}\\
&\left< \nQ\nS\right> \sim \frac{1}{\beta} \left( \mu^{2} K_{s} \frac{\pi}{2} \delta(0) - T \kappa_{DC} - \mu T \alpha_{DC} \right) \omega^{2}+   i \omega  \left(\frac{\epsilon' - \mu (  n  - \mu K_s)}{\beta}  \right), \label{poleQSsp}\\
&\left< \nS\nS\right> \sim - \frac{\epsilon'}{\beta^{2}} \omega^{2}+   i \omega^{3}  \frac{T}{\beta^{2}} \left(- \kappa_{DC} - 2 \mu \alpha_{DC} - \frac{\mu^{2}}{T} \sigma_{DC} \right) \,,  \label{poleSSsp}
\end{align}
where $\sigma_{DC}, \alpha_{DC}, \bar{\kappa}_{DC}$ are real value of conductivities at $\omega=0$, while $\sigma_{I}, \alpha_{I}, \bar{\kappa}_{I}$ are imaginary values linear to $\omega$.  $K_s$ is introduced as a strength of the pole of ${\rm Im}[\sigma]$, 
\begin{equation}
K_s = \lim_{\omega \rightarrow 0} \omega \Im [\sigma] \,,
\end{equation}
which can be identified with the superfluid density~\cite{Horowitz:2013jaa}.  
In \eqref{poleSsp}-\eqref{poleSSsp}, $K_s$ is the only parameter characterising  the superconducting phase and if we set $K_s=0$ the expressions works for the normal phase.  
The two-point functions related to the scalar operator $\nS$ ($\left< \nJ \nS \right>$, $\left< \nQ \nS \right>$, $\left< \nS \nS \right>$) diverge when $\beta$ goes to zero at small $\omega$.
We have confirmed that \eqref{poleSsp}-\eqref{poleSSsp} agree to the numerical results in Figure \ref{2ptScalar1},\ref{fig:app1} and \ref{fig:app2}.

\begin{figure}[]
 \centering
      \subfigure[ $\frac{\mu}{\beta}=0.3,0.4, 10$ (red, orange, gray) ]
 {\includegraphics[width=5.5cm]{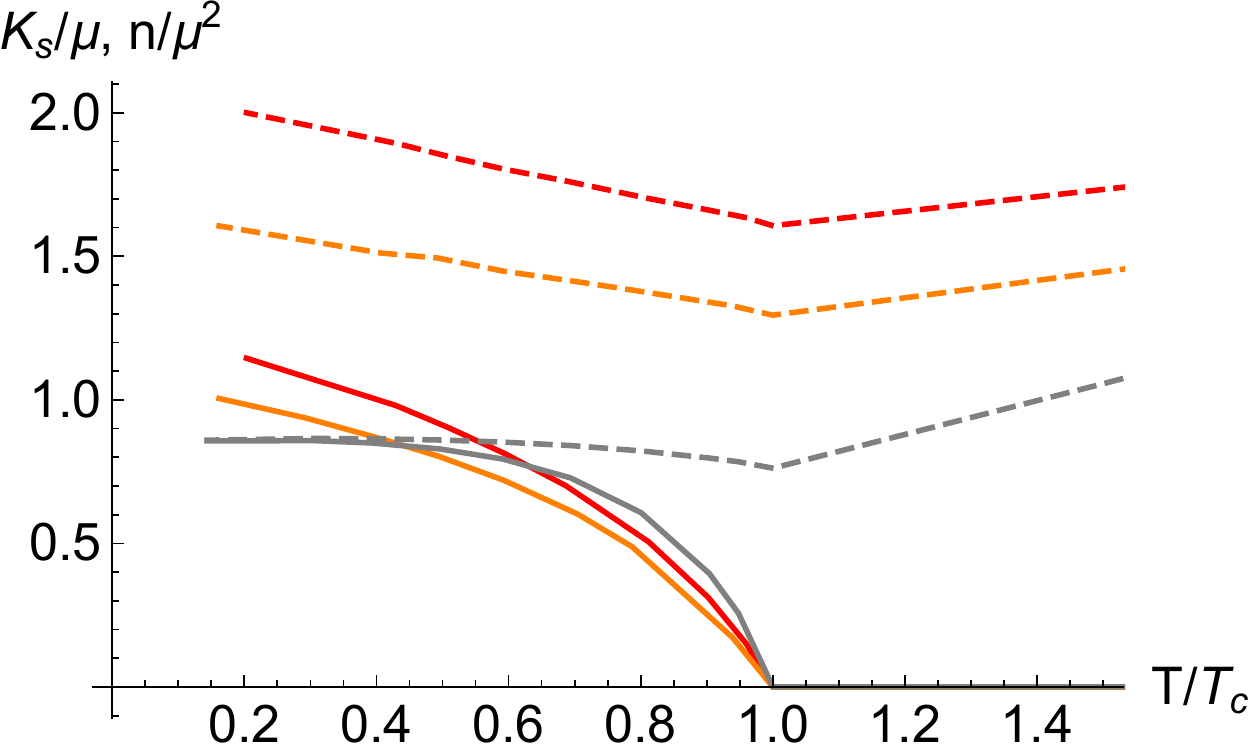} \label{}}\hspace{3mm}
    \subfigure[ $\frac{\mu}{\beta}=3,5,7,10$ (green, blue, purple, gray)]
   {\includegraphics[width=5.5cm]{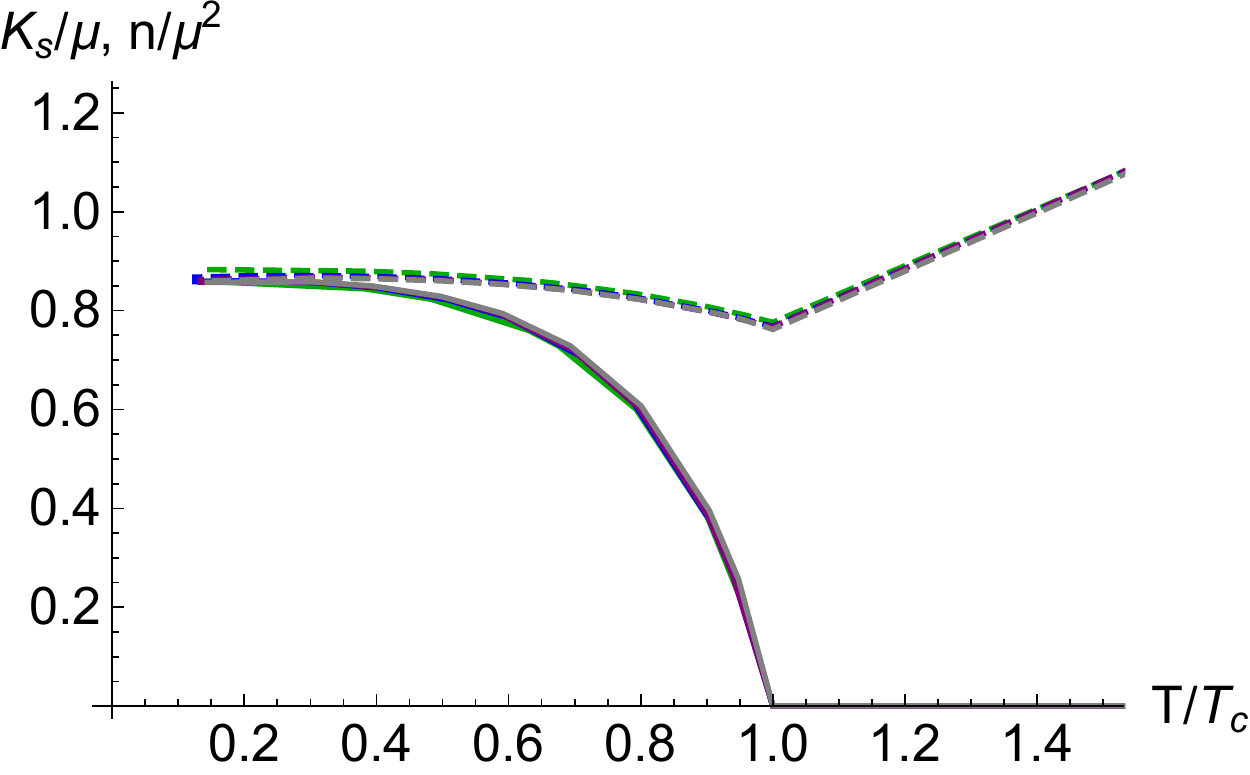} \label{}}\hspace{3mm}
  \caption{ $\frac{n}{\mu^2}$(dotted curves) and $\frac{K_s}{\mu}$(solid curves) for $q = 3$: The normal density $K_n/\mu$ at a given $\mu/\beta$ is the difference between the dotted and solid curve. The $K_n$ at low temperature vanishes when  $\frac{\mu}{\beta} \gtrsim 2$.  }
            \label{Ks}
\end{figure}

If we define  a normal fluid density ($K_n$) as 
\begin{equation}
K_{n} \equiv \frac{\beta}{\mu}\lim_{\omega \rightarrow 0} \frac{\mathrm{Im}[\left< \nJ \nS \right>]}{\omega} \,, 
\end{equation}
the Ward identity \eqref{ward1im} yields the charge conservation $n = \mu K_{s} + \mu K_{n}$. 
In Figure \ref{Ks}, we plot $n/ \mu^2$(dotted curves) and $K_{s}/\mu$(solid curves) versus $T/T_{c}$ at several $\mu/\beta\,$s. The difference between the dotted and solid curve at a given $\mu/\beta$ is $K_n/\mu$.
As temperature approaches to zero\footnote{Our numerics becomes unstable near zero temperature, so we present data up to the lowest possible temperature in our numerics.}, $K_n$ vanishes for $\mu/\beta \gtrsim 2$ (Figure \ref{Ks}(b)) while $K_n$ is nonzero for $\mu/\beta \lesssim 2$ (Figure \ref{Ks}(a)). 
Interestingly, it seems that this transition conicide with the coherent/incoherent metal transition studied in~\cite{Kim:2014bza}, where the metal state of this model was classified as coherent state with a well defined Drude peak in AC conductivity for $\mu/\beta \lesssim 2$ and incoherent  state without a Drude peak for $\mu/\beta \lesssim 2$ \footnote{Here, the metal state means both normal phase and the normal component of the two fluid model in superconductor phase.}.
In coherent state, the normal fluid density $K_n$ can be used as an input parameter to fit the Drude formula in the two fluid model of holographic superconductors~\cite{Horowitz:2013jaa}.
The non-zero $K_n$ at zero temperature for large momentum relaxation has been also observed in a holographic superconductor dual to a helical lattice~\cite{Erdmenger:2015qqa}.

\subsection{Homes' law and Uemura's law}
 Homes' law and Uemura's law are material independent universal scaling relations observed in high temperature superconductor as well as conventional superconductors~\cite{Homes:2005aa,Homes:2004wv,Erdmenger:2012ik, Erdmenger:2015qqa}. Uemura's law appearing in underdoped cuprates is 
\begin{equation}
\trho_{s}(\tT=0) = B\, \tT_{c} \,,
\end{equation} 
and Homes' law satisfied in a broader class of materials is 
\begin{equation} \label{Homeslaw1}
\trho_{s}(\tT=0) = C \sigma_{DC}(\tT_{c})\, \tT_{c} \,,
\end{equation}
where $B$ and $C$ are material independent universal constants. Here, the superfluid density ($\trho_{s}$), temperature($\tT$), and conductivity($\sigma_{DC}$) are all dimensionless~\cite{Erdmenger:2015qqa}. In this subsection we use momentum relaxation strength parameter($\beta$) as our scale so we choose $\trho_s = K_s/\beta$ and $\tT = T/\beta$. 
In our model,  there are two free parameters, $\mu/\beta$ and $q$. Thus universality of $B$ and $C$ means that $B$ and $C$ are independent of $\mu/\beta$ and $q$.  To check this it is convenient to fix $q$ first, and make plots of $B$ and  $C$ vs $\mu/\beta$ for Uemura's law and Homes' law respectively. 

\begin{figure}[]
 \centering
\subfigure[$\trho_{s}$] {\includegraphics[scale=0.35]{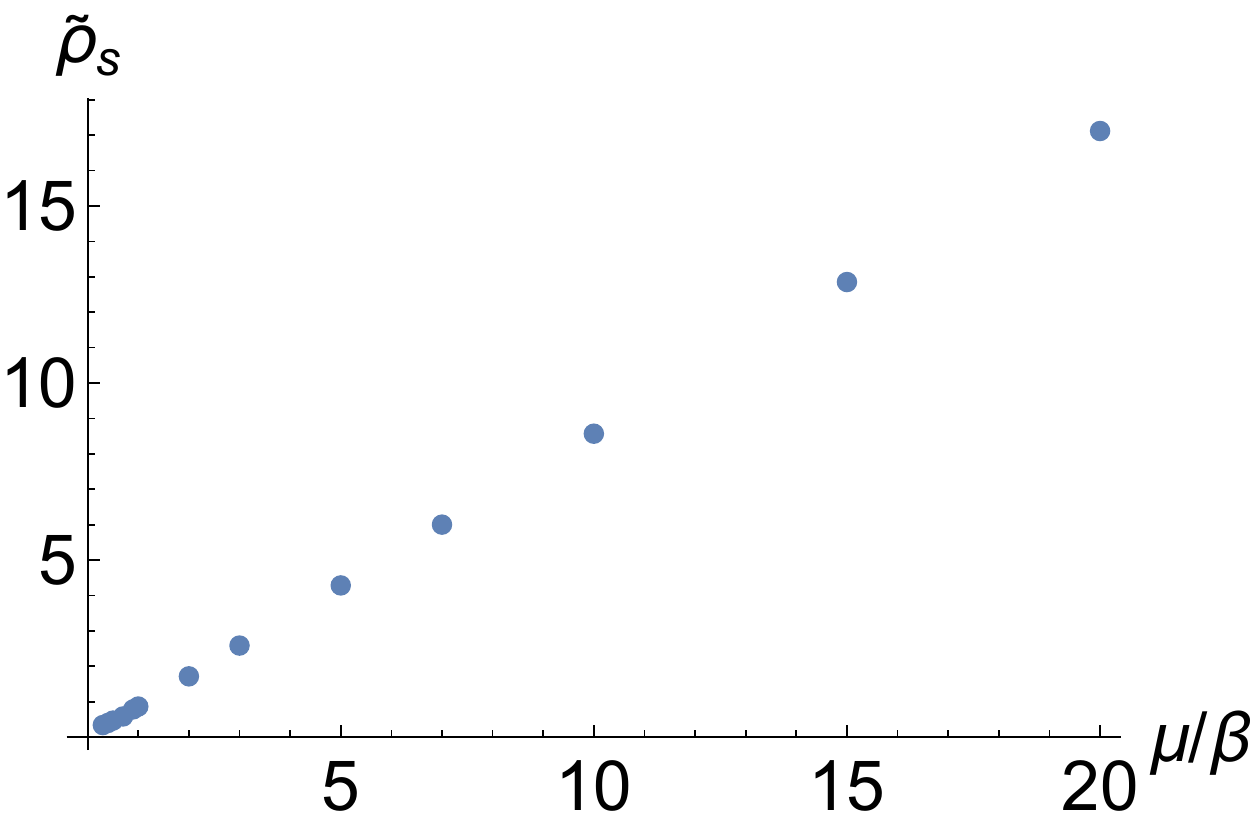} \label{}} 
\subfigure[$\tT_{c}$]{\includegraphics[scale=0.35]{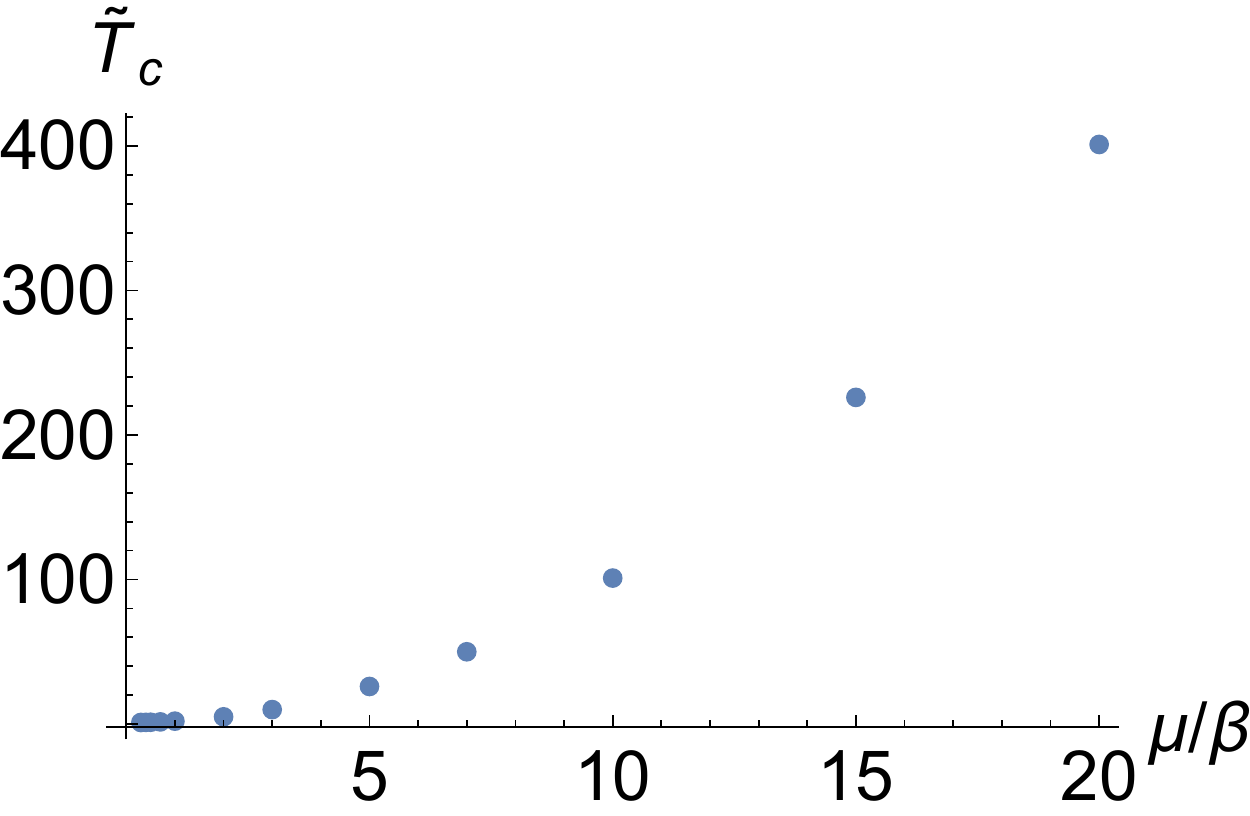} \label{}}
\subfigure[$\sigma_{DC}$]{\includegraphics[scale=0.35]{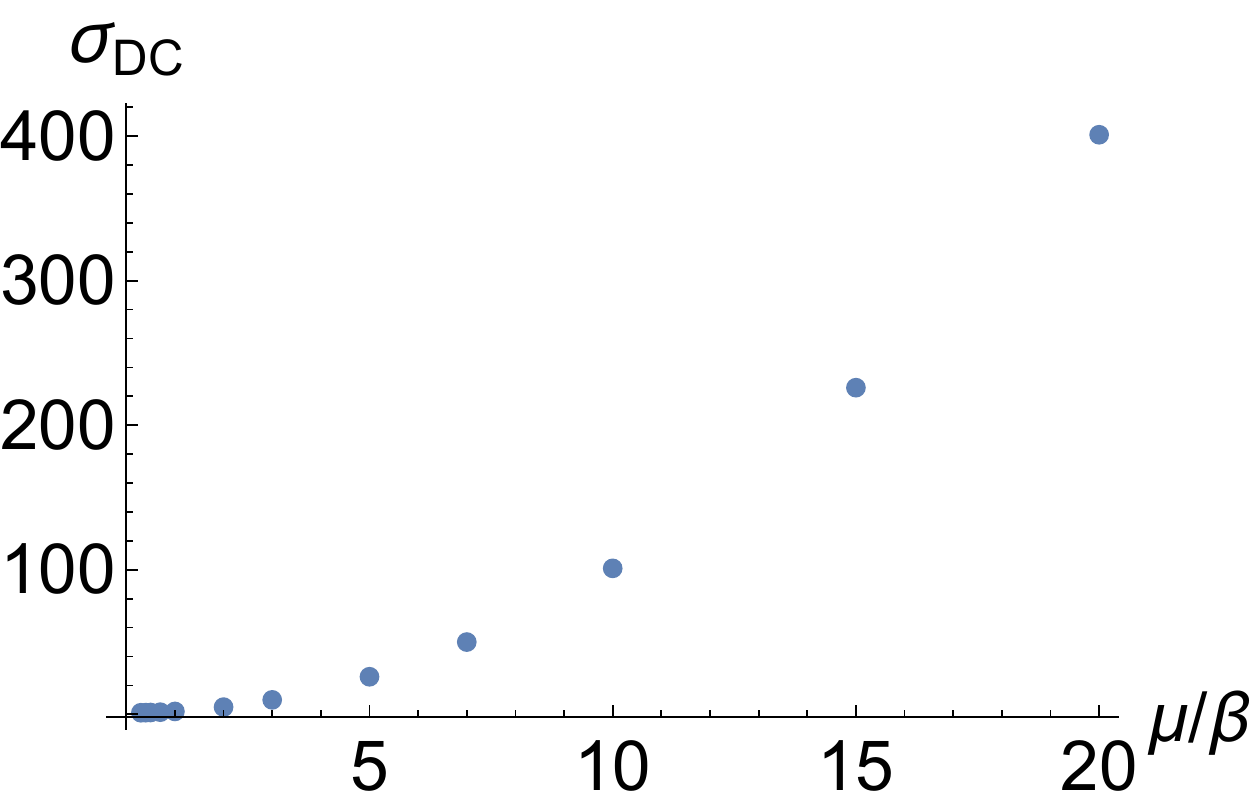} \label{}}
  \caption{$\trho_{s}, \tT_{s}$, and $\sigma_{DC} $ for $q=3$ } 
            \label{fig:each}
\end{figure}

To compute $B$ and $C$, 
\begin{equation}
B = \frac{\trho_s}{\tT_c} = \frac{K_s}{T_c} \,, \qquad  C =\frac{\trho_s}{\sigma_{DC} \tT_c} =\frac{K_s}{\sigma_{DC} T_c} \,,
\end{equation}
the superfluid density $\trho_s$($= \mu/\beta\cdot K_s/\mu$) can be read off from the solid curves in Figure \ref{Ks}, where the curves do not reach to $T=0$ because of instability of numerical analysis.  Therefore, we extrapolated the curves up to zero temperature to read $\trho_s$ at $T=0$. The conductivity $\sigma_{DC}$ can be read in Figure \ref{oldresult} or analytically $\sigma_{DC} = 1+\mu^2/\beta^2$ in our model~\cite{Andrade:2013gsa}. The transition temperature $\tT_c$ has  been computed numerically in \cite{Kim:2015dna}. Our numerical results of $\trho_s, \tT_c$ and $\sigma_DC$ for $q=3$ are shown in Figure \ref{fig:each}.

\begin{figure}[]
 \centering
\subfigure[$B(= \trho_{s}/\tT_{c})$ , $q=3$] {\includegraphics[scale=0.45]{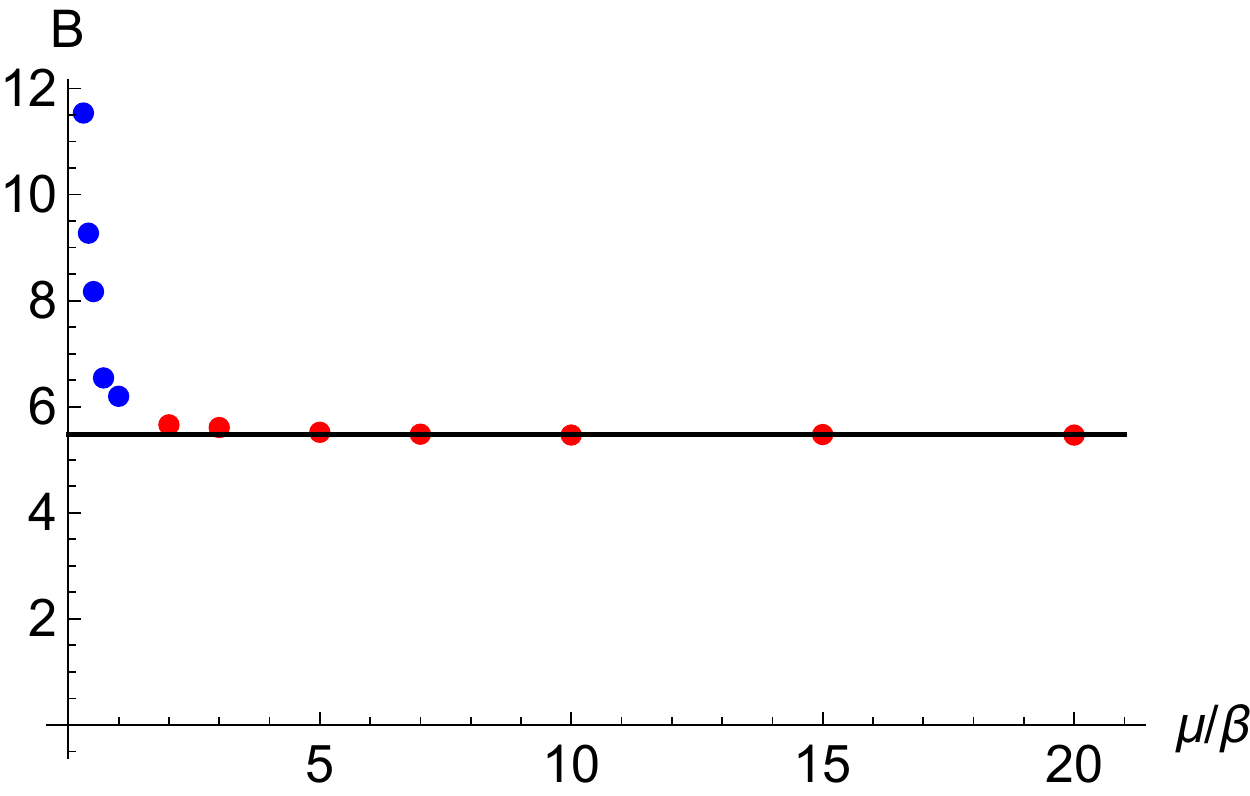} \label{}}  $\qquad $
\subfigure[$B(=\trho_{s}/\tT_{c})$ for $q=2,3,6$]{\includegraphics[scale=0.45]{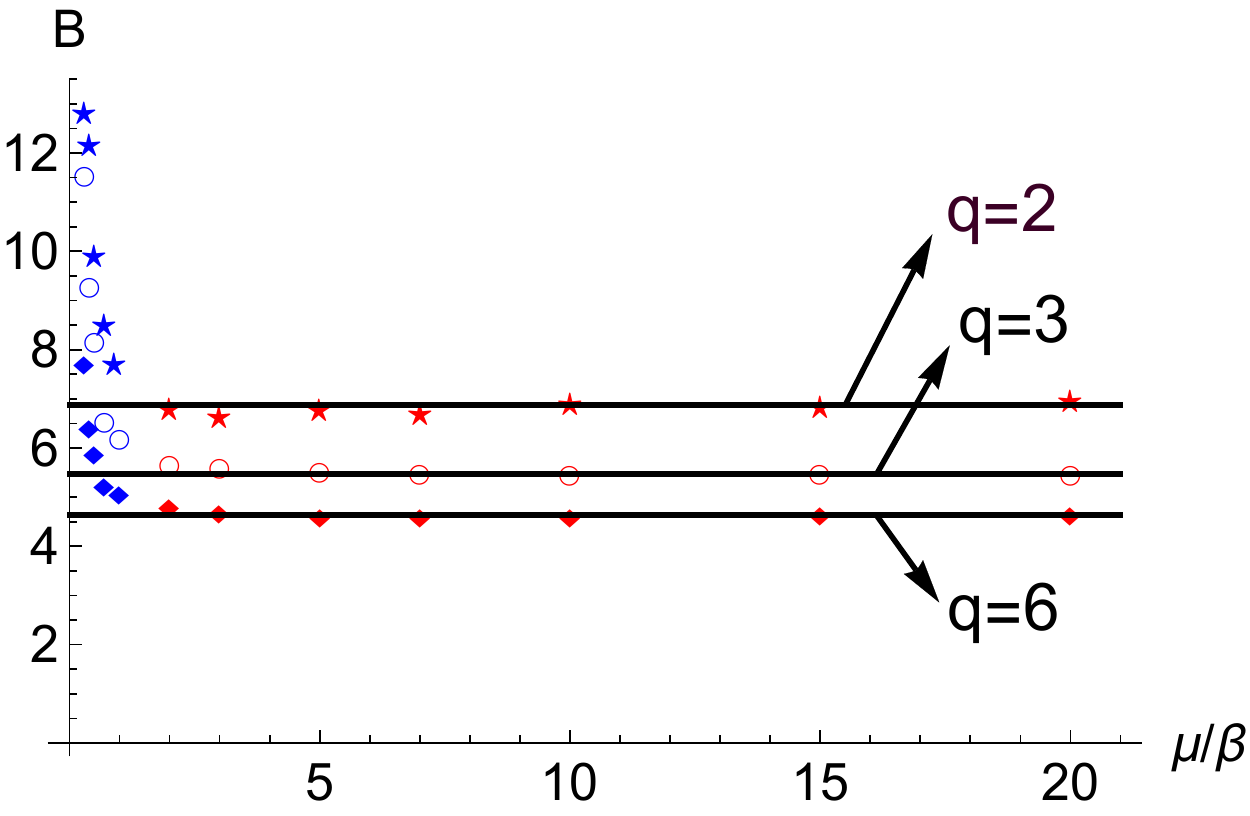} \label{}}
  \caption{Checking Uemura's law. Uemura's law holds in coherent regime (red dots: $\mu/\beta=2,3,5,7,10,15,20$) while it does not hold in incoherent regime (blue dots: $\mu/\beta=0.3,0.4,0.5,0.7,1$).   In (a) the black  line is drawn for $B \sim 5.47$, and in (b) the black lines are drawn for $B \sim 6.87, 5.47, 4.64$ for $q=2,3,6$ respectively. } 
            \label{results4}
\end{figure}

From Figure  \ref{fig:each} we may expect that there is  a linear relation between $\trho_s$ and $\tT_c$ at least for large $\mu/\beta$, which supports Uemura's law. To see if this is the case also for small $\mu/\beta$ we make a plot of $B$ vs $\mu/\beta$ in Figure \ref{results4}(a), where we find that Uemura's law holds only for $\mu/\beta \gtrsim 2$, of which data are red dots. 
 Interestingly, the parameter regime $\mu/\beta \gtrsim 2$ (red dots) belongs to coherent metal regime, where the optical conductivity of normal component shows a Drude peak behaviour. Furthermore, this regime corresponds to Figure \ref{Ks}(b), where charge density is the same as superfluid density at zero temperature. 
The blue dots belong to incoherent regime, where a optical conductivity loses a Drude behaviour. They correspond to Figure \ref{Ks}(a) and there is a gap between charge density and superfluid density at zero temperature. Also, for different values of $q$, we find that Uemura's law is satisfied for large $\mu/\beta$ but with a different constant $B$.
For example, for $q=2$, $B \approx 6.87$ and for $q=6$, $B \approx 4.64$ in the regime of $\mu/\beta \gtrsim 2$ (Figure \ref{results4}(c)).  
Since Uemura's law is observed in underdoped regimes, if $\beta$ can be interpreted as a doping parameter our result will be consistent with phenomena.

 \begin{figure}[]
 \centering
\subfigure[$C(=\trho_{s}/(\sigma_{DC} \tT_{c}))$, $q=3$]{\includegraphics[scale=0.45]{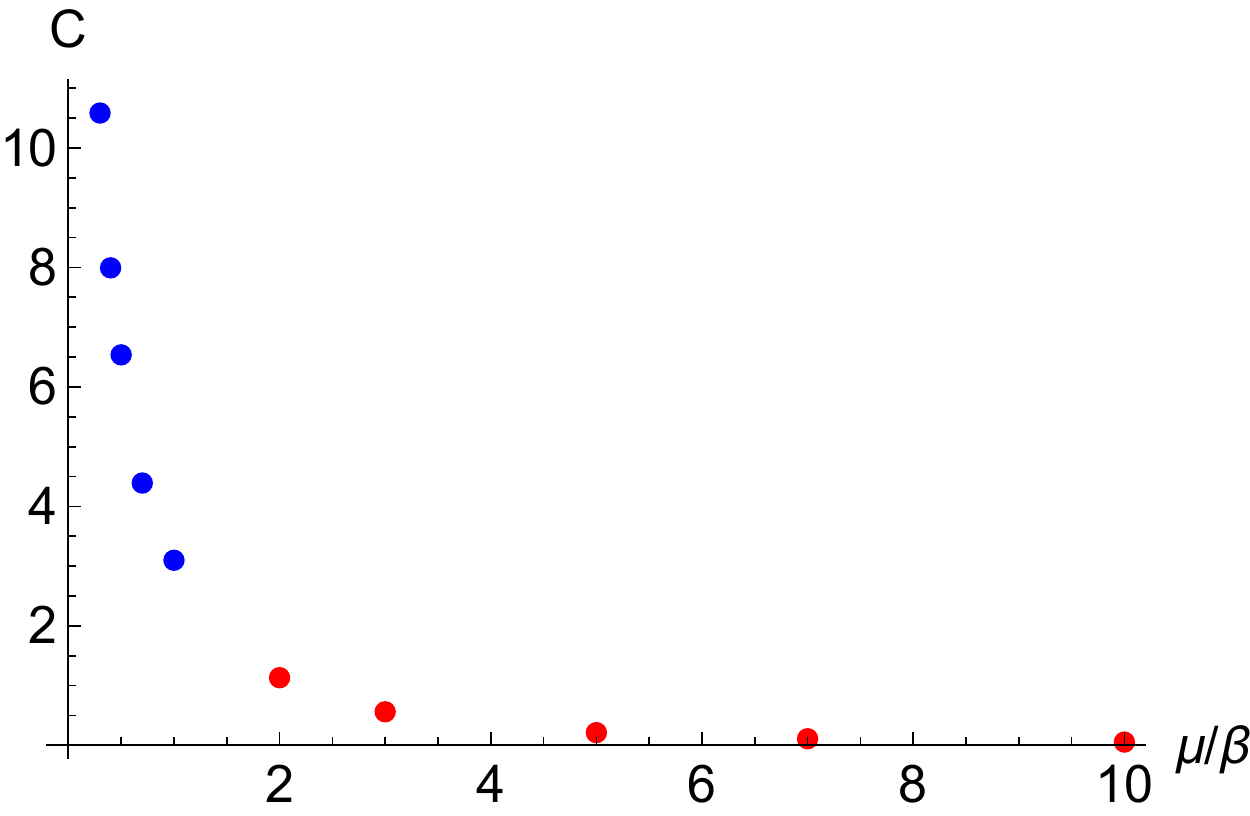} \label{}} $\qquad $
\subfigure[$\trho_{s}$ vs $\sigma_{DC} \tT_{c}$, $q=3$]{\includegraphics[scale=0.45]{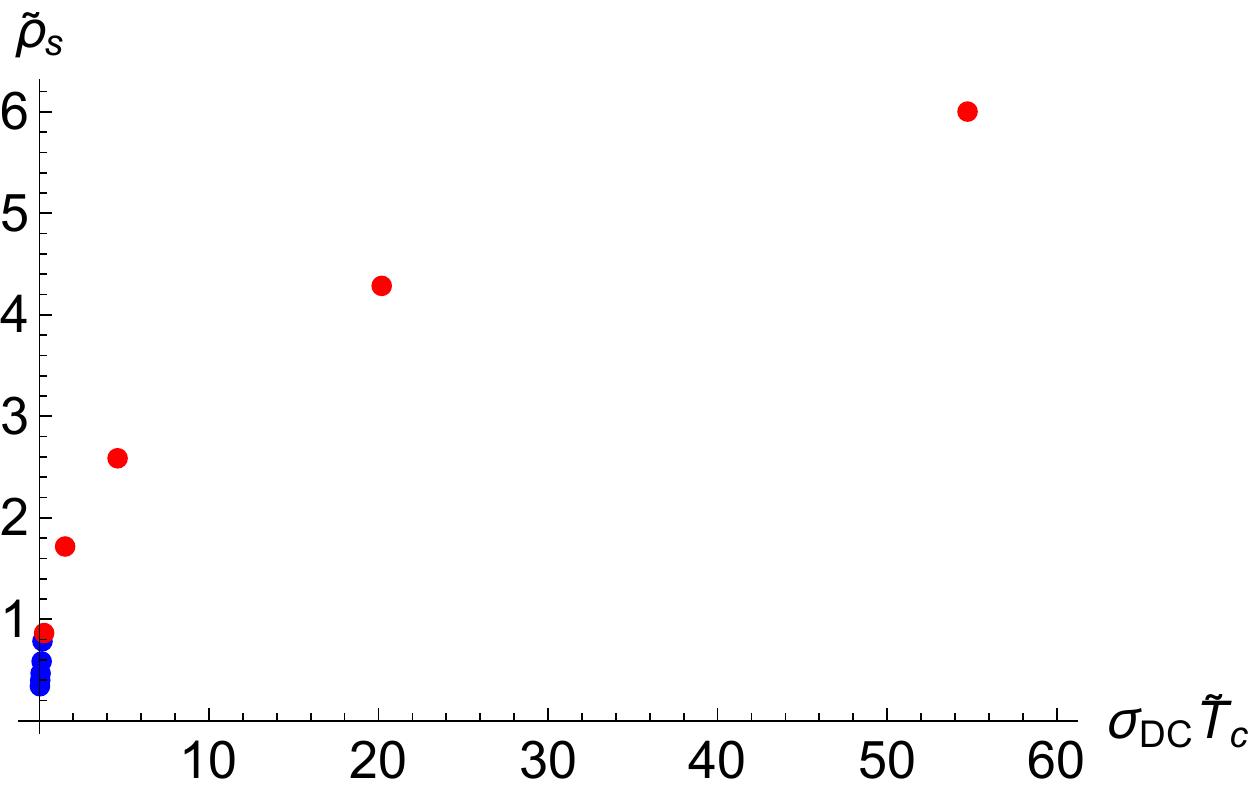} \label{}}
  \caption{Checking Homes' law. Homes' law does not hold.  The blue dots are for incoherent regime ($\mu/\beta=0.3,0.4,0.5,0.7,1$) and the red dots are for coherent regime ($\mu/\beta=2,3,5,7,10$).  In (a) the data for  $C$ do not align on a constant value and
   in (b) the data do not yield a linear relation. } 
            \label{results5}
\end{figure}

Based on our results on Uemura's law(Figure \ref{results4}(a)) and $\sigma_{DC}$(Figure \ref{fig:each}(c)), we may anticipate if Homes' law is satisfied. If $\sigma_{DC}$ is quickly decreasing function approaching to constant for $\mu/\beta \gtrsim 2$ we may have a chance to obtain  Homes' law. However, our $\sigma_{DC}$ does not show that behaviour. 
Therefore, as shown in Figure \ref{results5}, Home's law does not hold in both coherent regime (red dots) and incoherent regime (blue dots).  In Figure \ref{results5}(a), for large $\mu/\beta$, $C(=\trho_s/(\sigma_{DC} \tT_c))$ approaches to a constant value, but it is zero. It simply means that $\sigma_{DC}$ goes to infinite as momentum relaxation goes to zero. 
Figure \ref{results5}(b) is another representation, a plot of $\trho_{s}$ versus $\sigma_{DC} \tT_{c}$, where 
it is also clear that there is no linear relation between between $\trho_{s}$ and $\sigma_{DC} \tT_{c}$. 
For different values of $q$, we considered  $q=2$ and $q=6$ and obtained figures qualitatively similar to Figure  \ref{results5}, so Homes' law seems not satisfied for different values of $q$ either.

Homes' law may be understood based on Planckian dissipation, for which the time scale of dissipation is shortest possible~\cite{Zaanen:2004aa} . In summary, the left hand side of \eqref{Homeslaw1}, superfluid density is proportional to density of mobile electrons in superconducting state ($n_S$). The right hand side of \eqref{Homeslaw1}, $\sigma_{DC}$ is proportional to density of mobile electrons in normal state ($n_N$) times relaxation time ($\tau$), and the relaxation time is inversely proportional to the temperature (Planckian dissipation):
\begin{equation} \label{Zaa}
\rho_s \sim  n_S \,, \qquad \sigma_{DC} \sim n_N \tau(T_c)\,, \qquad \tau(T_c) \approx \frac{\hbar}{k_B T_c} \,,
\end{equation}
where $k_B$ is Boltzmann's constant and proportionality constants of the relations are material independent.  
Notice that 
thanks to the Planckian dissipation $T_c$ is cancelled out in Homes' law, leaving universal constant $\hbar/k_B$.
Finally if we use another empirical law, Tanner's law, $n_S = n_N/4$, Homes' law is obtained. 

In our model, it turns out a kind of Tanner's law holds in coherent regime ($\mu/\beta \ge 2$). 
In Figure \ref{Ks}(b), all curves coincide and it means $n_s/n_N$ does not depend on $\mu/\beta$, which is the qualitative content of Tanner's law. Therefore, if our system were Planckian dissipator in coherent regime, we would have seen Homes' law. 
The relaxation time $\tau$ for our model can be written as
\begin{equation}
\tau = \frac{f(T/\beta, \mu/\beta ,q )}{T} \,,
\end{equation} 
where $T$ in the denominator is extracted to mimic the form of Planckian dissipation~\cite{Erdmenger:2015qqa}. 
Since our system does not show Homes' law it is not a Planckian dissipator, which means $f$ is not universal near $T_c$.
Indeed we may induce that $f \sim \mu^2/\beta^2$ because $T_c \sim \mu/\beta$ from Figure \ref{fig:each}(b) and $\tau \sim \mu/\beta$ from the analysis in \cite{Kim:2014bza}.

Our results on Uemura's law and Homes' law are different from the previous work~\cite{Erdmenger:2015qqa}, where  a superconductor model in a helical lattice was studied. In the model, there are two parameters corresponding to the strength of momentum relaxation effect: the lattice strength $\lambda$ and the helix pitch $p$, and it was found that Homes' law held for restricted parameter regime (not in small momentum relaxation, but for rather large values of $\lambda$ and $p$) while Uemura's law did not hold.  In particular, Homes' law was observed in insulating phase near phase transition. However, in our model there is no insulating phase and it may be a reason why two models show different results.  There are other differences between two models. The model in helix lattice is anisotropic five dimensional model, while our model is isotropic four dimensional.

\section{Conclusion and discussions}\label{sec6}

In this paper, we analysed a holographic superconductor model incorporating momentum relaxation.
Building on previous works~\cite{Kim:2014bza,Kim:2015dna,Kim:2015sma,Kim:2015wba}, we focused on three issues, where momentum relaxation plays an important role.
(1) Ward identities: constraints between conductivities, (2) conductivities with a neutral scalar hair instability,  (3) Homes' law and Uemura's law.

In holographic methods, we often need to solve complicated differential equations which do not allow analytic solutions, so it is important to develop reliable and systematic numerical methods. Computing AC conductivities is such an example, for which we have developed a numerical method. However, to make sure our numerics are reliable and robust, it will be good to have a cross-check. The Ward identity serves as a nice cross-check of our numerical method since we can compare our numerical results with the independently derived analytic formula. 
When there is a neutral scalar instability we explicitly showed that the DC electric conductivity is finite, while it is infinite for a complex scalar instability. This shows that the neutral scalar instability has nothing to do with superconductivity as expected. 

Homes' law is very interesting and important not only because of its material independent universality but also a possible relation to quantum criticality and Planckian dissipation, which also underpins the universal bound of the viscosity to entropy in strongly correlated systems such as quark-gluon plasma.
We have checked Homes' law and Uemura's law in our model. It turns out that Homes' law does not hold and Umeura's law holds for small momentum relaxation related to coherent metal regime.  Our results are different from \cite{Erdmenger:2015qqa}, where a holographic superconductor in a helical lattice was considered and it was shown that Homes' law is satisfied for some restricted parameter regime in insulating phase, while Uemuras' law is not satisfied at all. The difference may be due to the existence of insulating phase and/or anisotropy in a model with a helical lattice. To clarify it, it will be helpful to study Homes' law in different holographic superconductor models such as anisotropic massless scalar model, Q-lattice model, or massive gravity model~\cite{WIP}.

Regarding Homes' law and Uemura's law, there may be an issue in the identification of the superfulid density. 
We have found that superfluid density and total charge density at zero temperature do not agree at large momentum relaxation, similarly to the case in a helical lattice~\cite{Erdmenger:2015qqa}. 
Because the Ferrell-Glover-Tinkham (FGT) sum rule still holds in our model~\cite{Kim:2015dna}, it is possible that part of the low frequency spectral weight are transferred to intermediate frequencies instead of the superfluid pole.
Therefore, as a cross check, it will be good to compute the superfluid density from the transverse response by the magnetic/London penetration depth, for which we need to solve for the transverse propagator at small non-zero momentum~\cite{Erdmenger:2015qqa}. There is also another closely related quantity to superfluid density. 
By integrating a Maxwell's equation over the holographic coordinate $r$,
\begin{equation}
\nabla_M F^{tM} =  i q \left(  \Phi^* D^t \Phi - \Phi D^t \Phi^*  \right)
\end{equation}
we may define the charge density of hair outside the horizon, $n_{\mathrm{hair}}$, as
\begin{equation}
n_{\mathrm{hair}} \equiv n-n_h \equiv \sqrt{-g} F^{t r} |_{r=\infty} -\sqrt{-g} F^{t r} |_{r=r_h} = iq \int^\infty_{r_h}  dr \sqrt{-g} \left(  \Phi^* D^t \Phi - \Phi D^t \Phi^*  \right)~~,
\end{equation}
where $n$ is the charge density of the dual field theory and $n_h$ is interpreted as the charge density inside the horizon.  In normal phase $n_\mathrm{hair} = 0$ while in superconducting phase $n_\mathrm{hair} \ne 0$. Therefore, 
$n_\mathrm{hair}$ plays a role of order parameter of superconducting phase transition.  For $q=0$, $n_\mathrm{hair}$ is zero so $K_s$ is zero, which is consistent with our results in section \ref{sec3}. However, it turns out that the numerical value of $K_s$ is different from $n_\mathrm{hair}$. We have checked Homes' law and Uemura's law by using $n_\mathrm{hair}$ as the superfulid density, but it did not support Homes' law and Uemura's law. 
It will be interesting to find a physical meaning of $n_\mathrm{hair}$ in the dual field theory and the precise relation to superfluid density.

\acknowledgments
We would like to thank Johanna Erdmenger, Steffen Klug, Rene Meyer,  Yunseok Seo, Sang-Jin Sin for valuable discussions and correspondence. 
The work of K.Y.Kim and K.K.Kim was supported by Basic Science Research Program through the National Research Foundation of Korea(NRF) funded by the Ministry of Science, ICT \& Future Planning(NRF-2014R1A1A1003220)  and the GIST Research Institute(GRI) in 2016.  K.K.Kim was also supported
by the National Research Foundation of Korea(NRF) grant with the grant number
NRF-2015R1D1A1A 01058220. {{M. Park is supported by TJ Park Science Fellowship of POSCO TJ Park Foundation.}}

\appendix 

\section{Two-point functions related to the real scalar operator} \label{appA}

\begin{figure}[!h]
 \centering
     {\includegraphics[width=4.5cm]{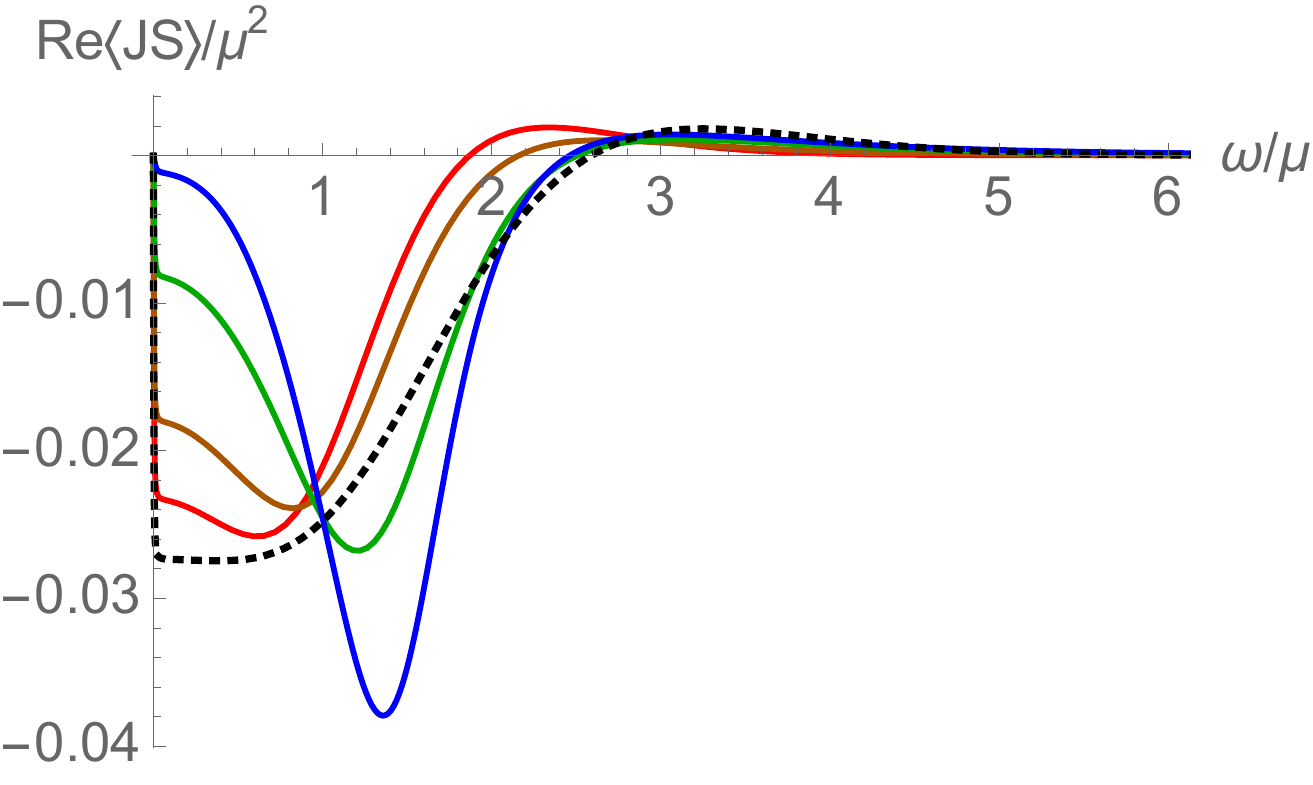} \label{}}\hspace{3mm}
   {\includegraphics[width=4.5cm]{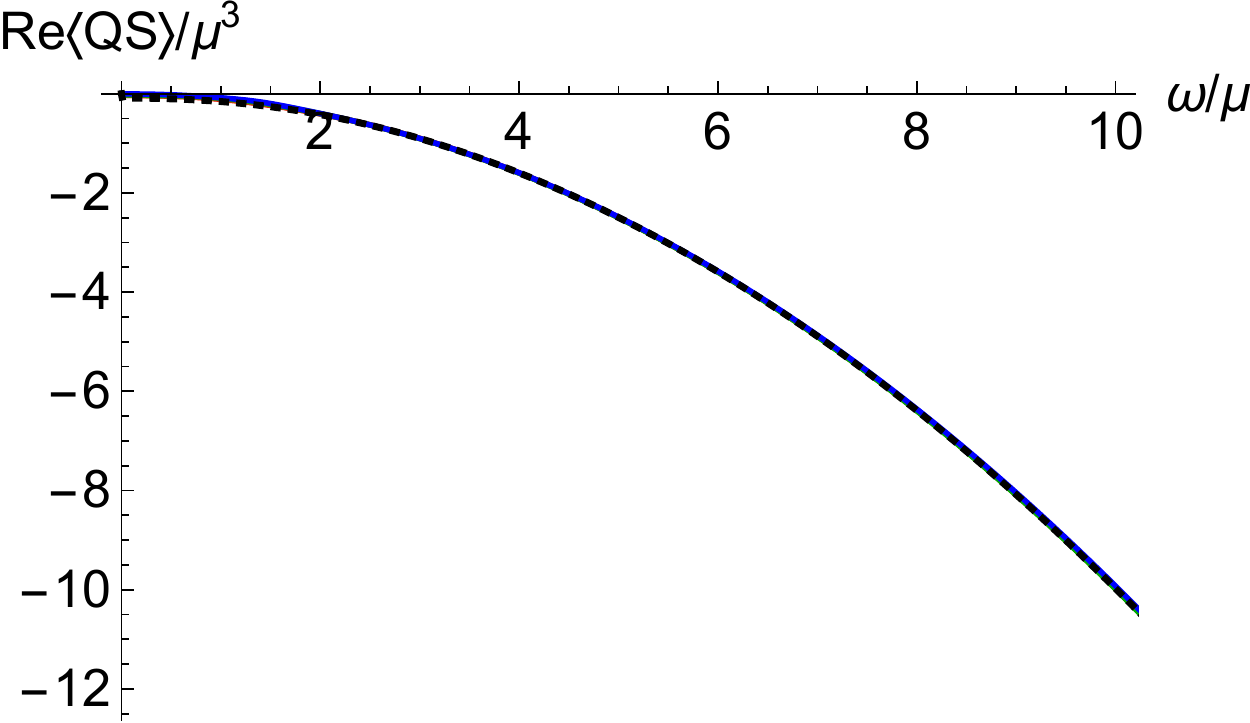} \label{}}\hspace{3mm}
   {\includegraphics[width=4.5cm]{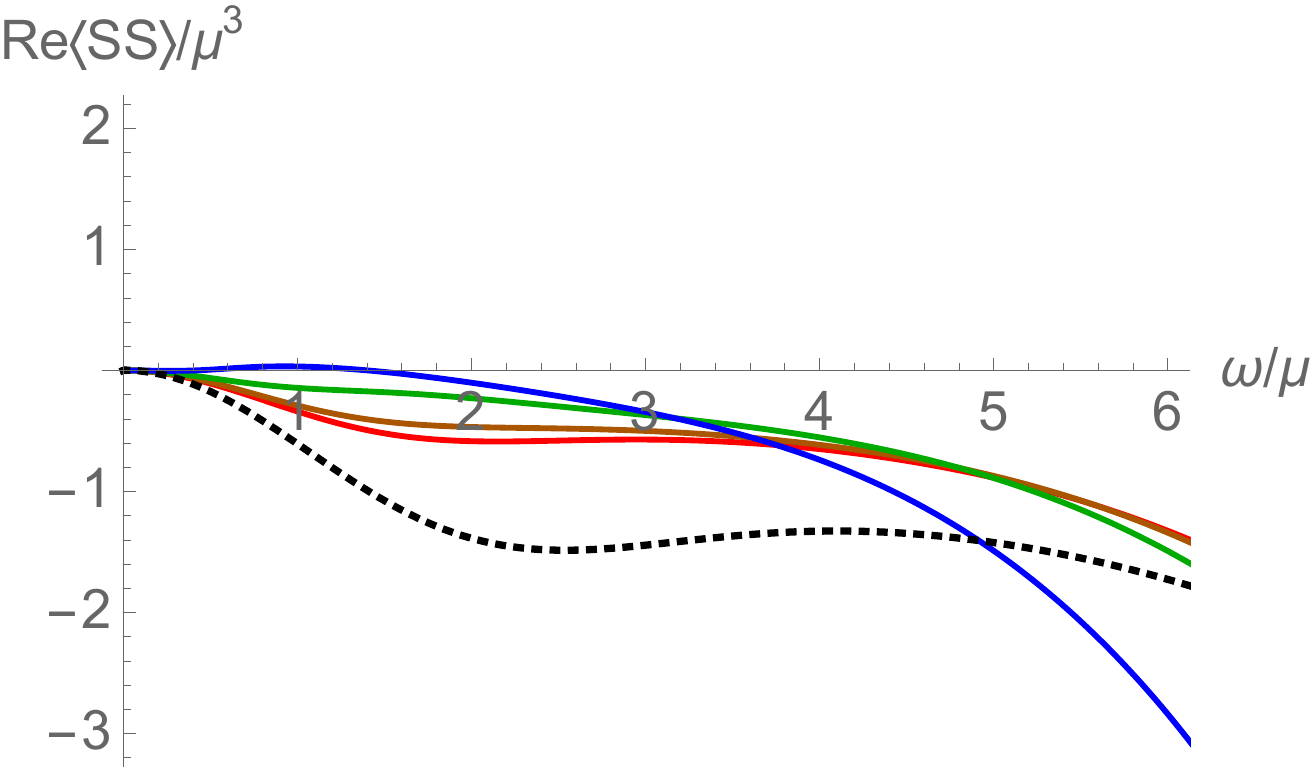} \label{}}
 {\includegraphics[width=4.5cm]{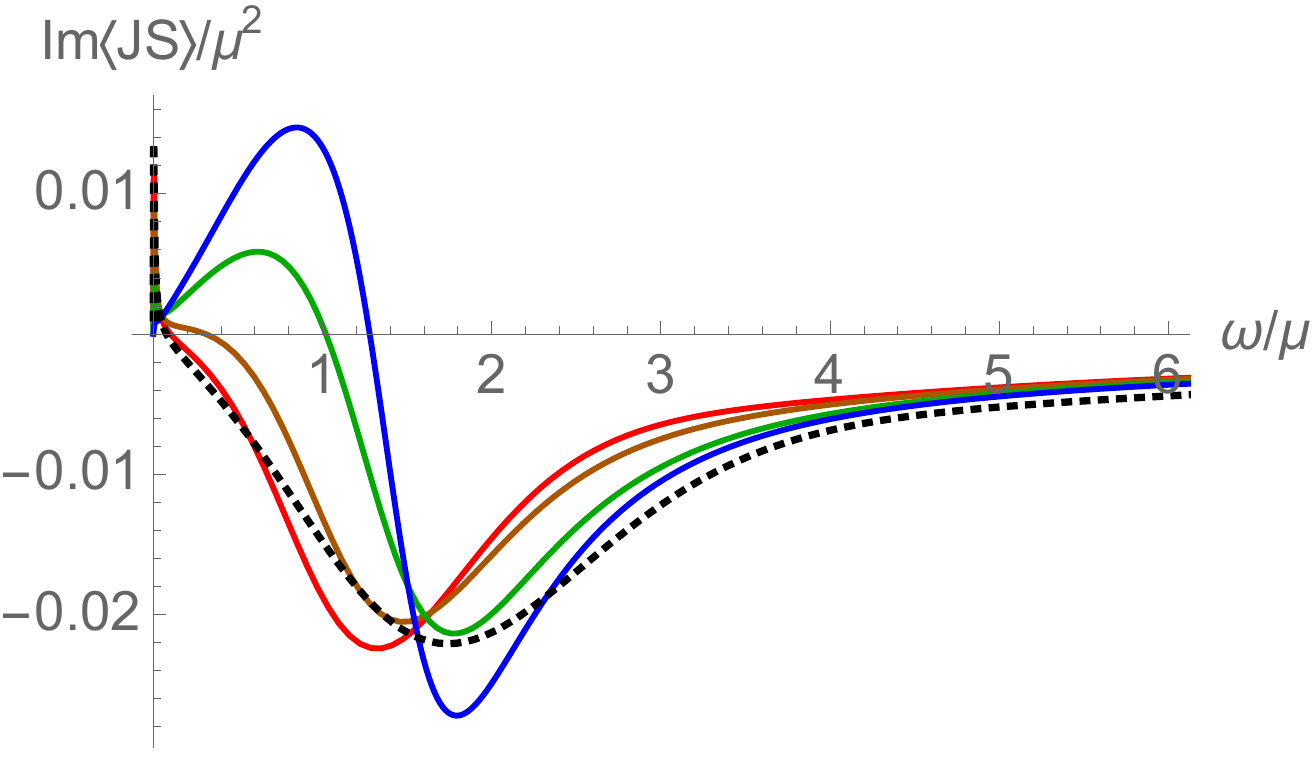} \label{}}\hspace{3mm}
   {\includegraphics[width=4.5cm]{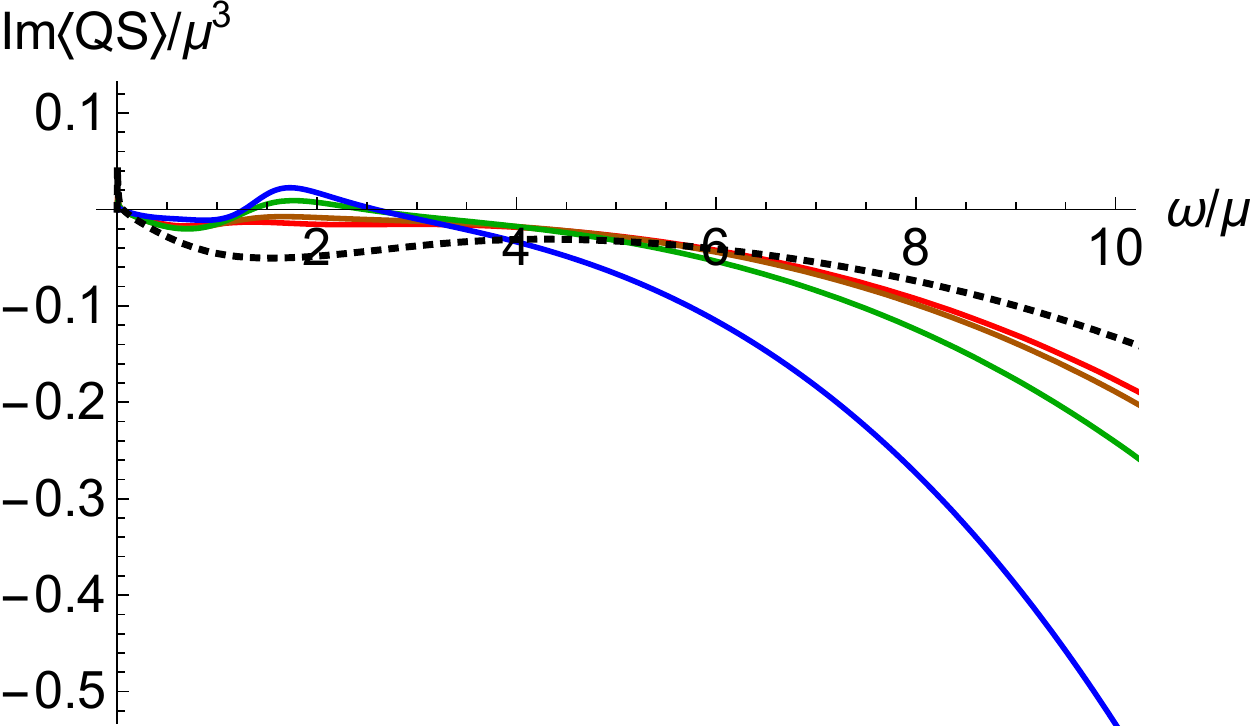} \label{}}\hspace{3mm}
   {\includegraphics[width=4.5cm]{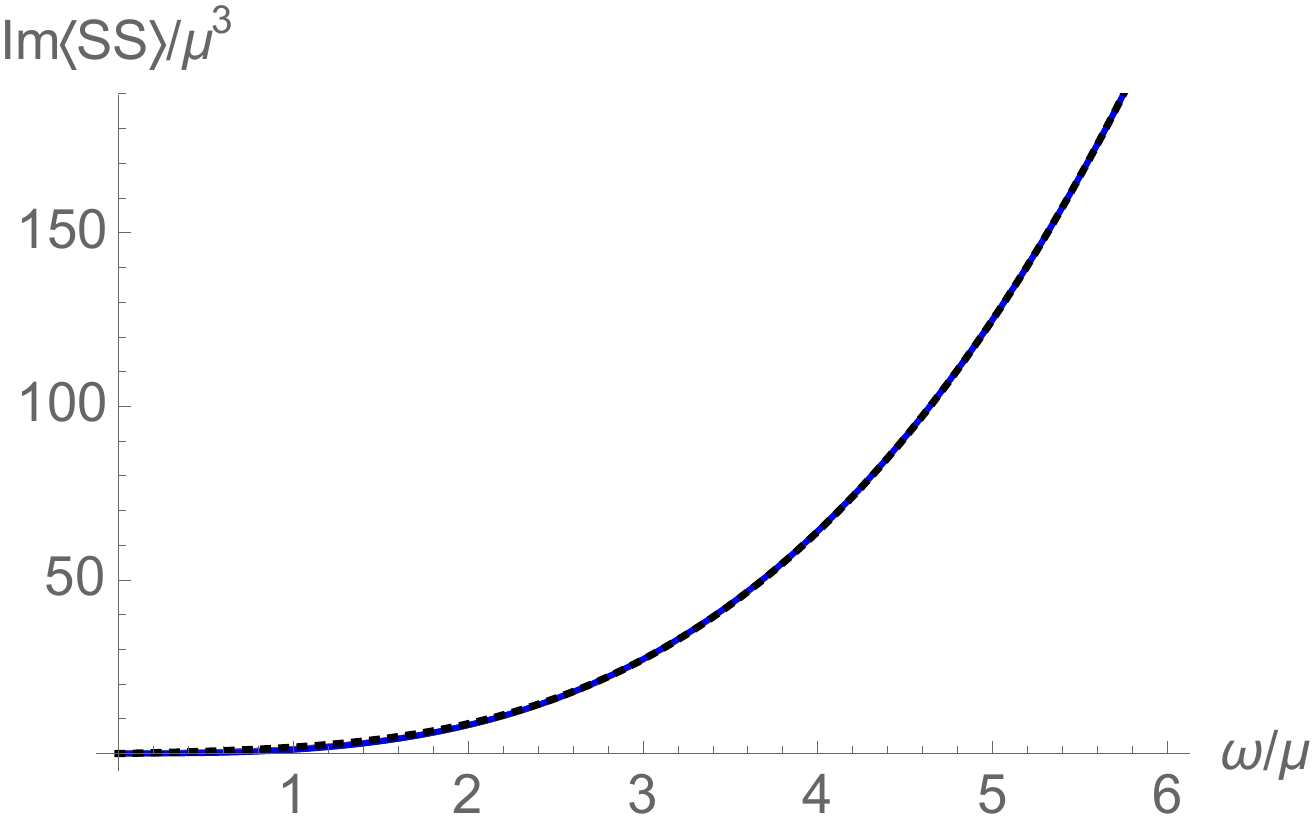} \label{}}
  \caption{{$\langle \nJ\nS \rangle$, $\langle \nQ\nS \rangle$, $\langle \nS\nS \rangle$ for $\beta/\mu=0.1$.  $T/T_c = 1.5, 1, 0.94, 0.76, 0.37$  (dotted, red, orange, green, blue)}} 
            \label{fig:app1}
\end{figure}

\begin{figure}[!h]
 \centering
     {\includegraphics[width=4.5cm]{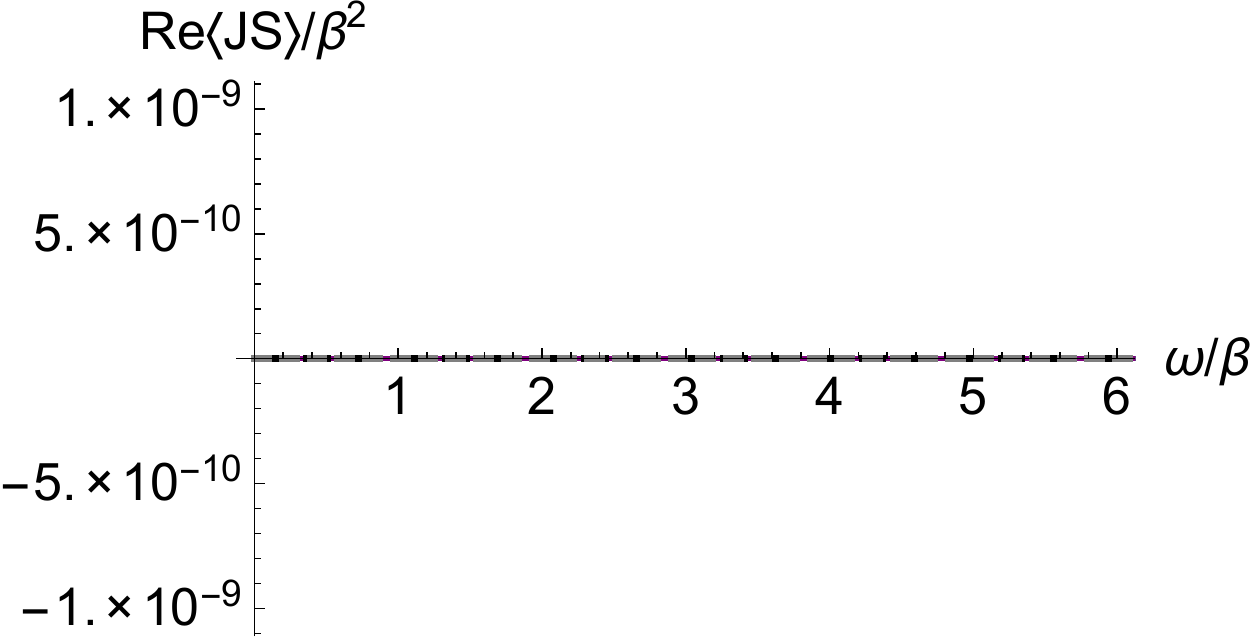} \label{}}\hspace{3mm}
   {\includegraphics[width=4.5cm]{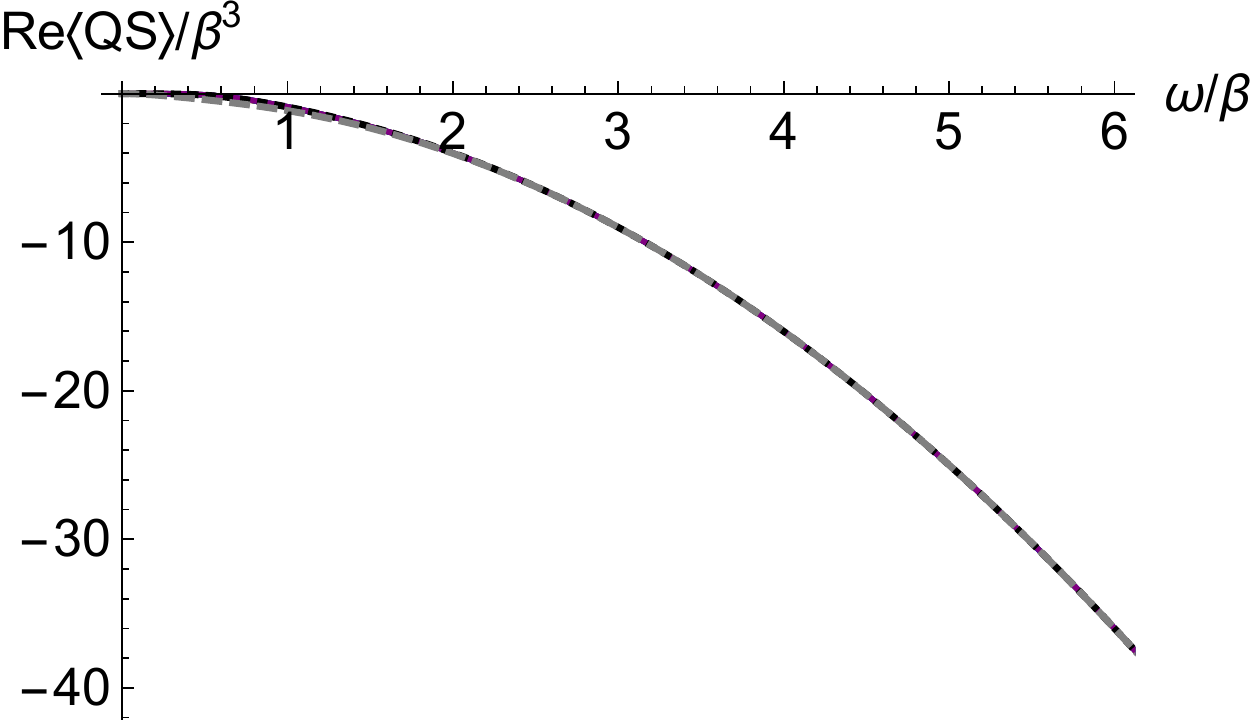} \label{}}\hspace{3mm}
   {\includegraphics[width=4.5cm]{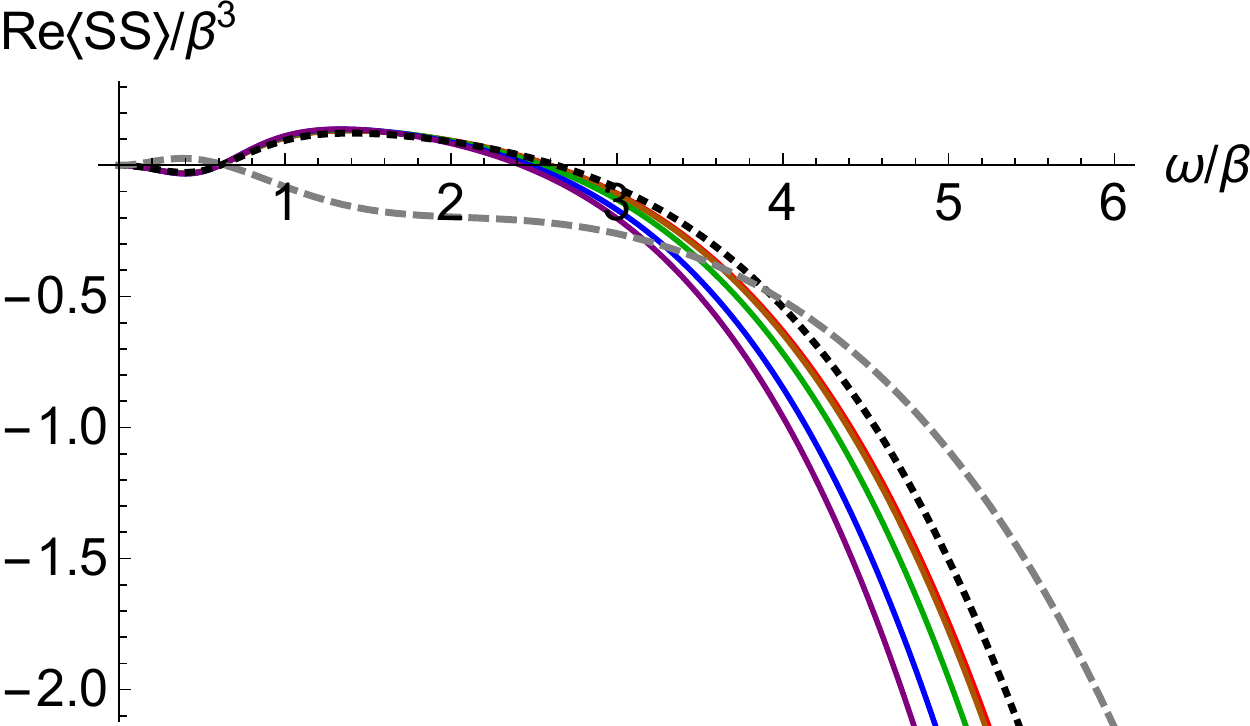} \label{}}
 {\includegraphics[width=4.5cm]{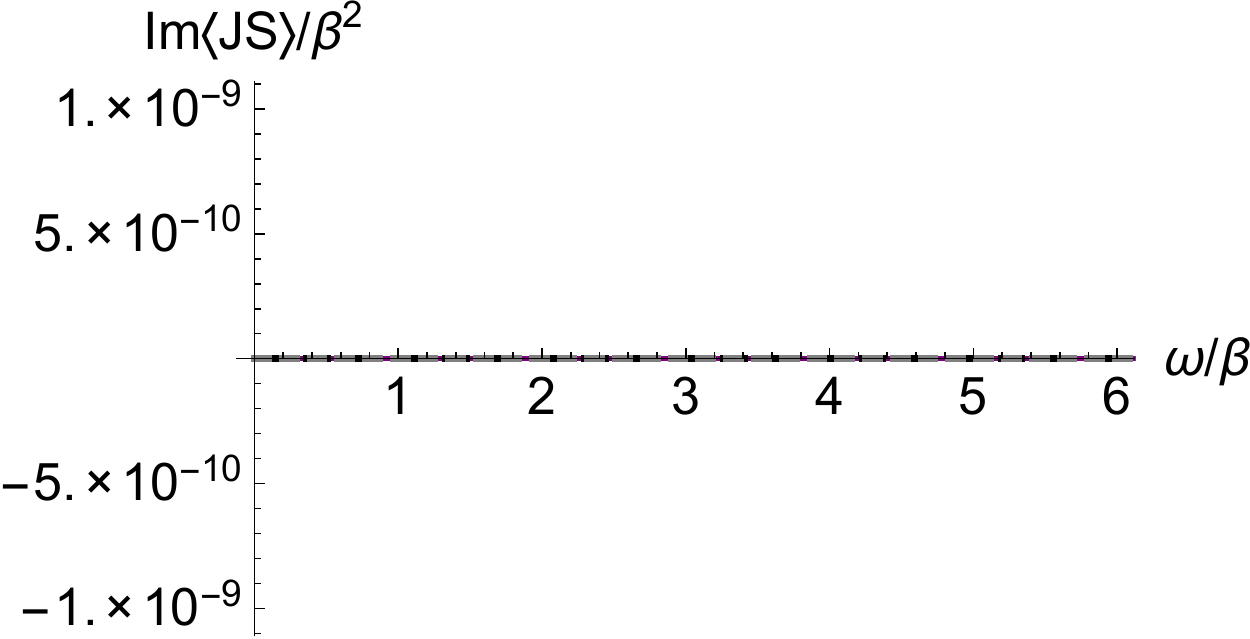} \label{}}\hspace{3mm}
   {\includegraphics[width=4.5cm]{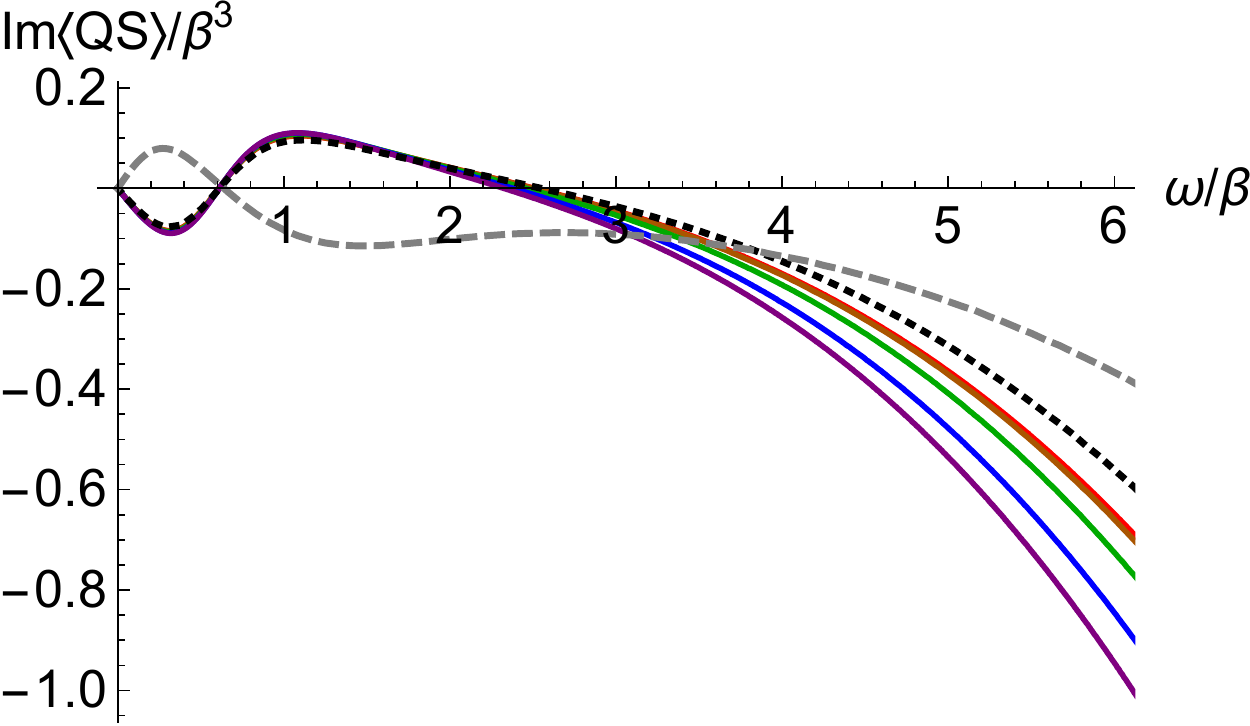} \label{}}\hspace{3mm}
   {\includegraphics[width=4.5cm]{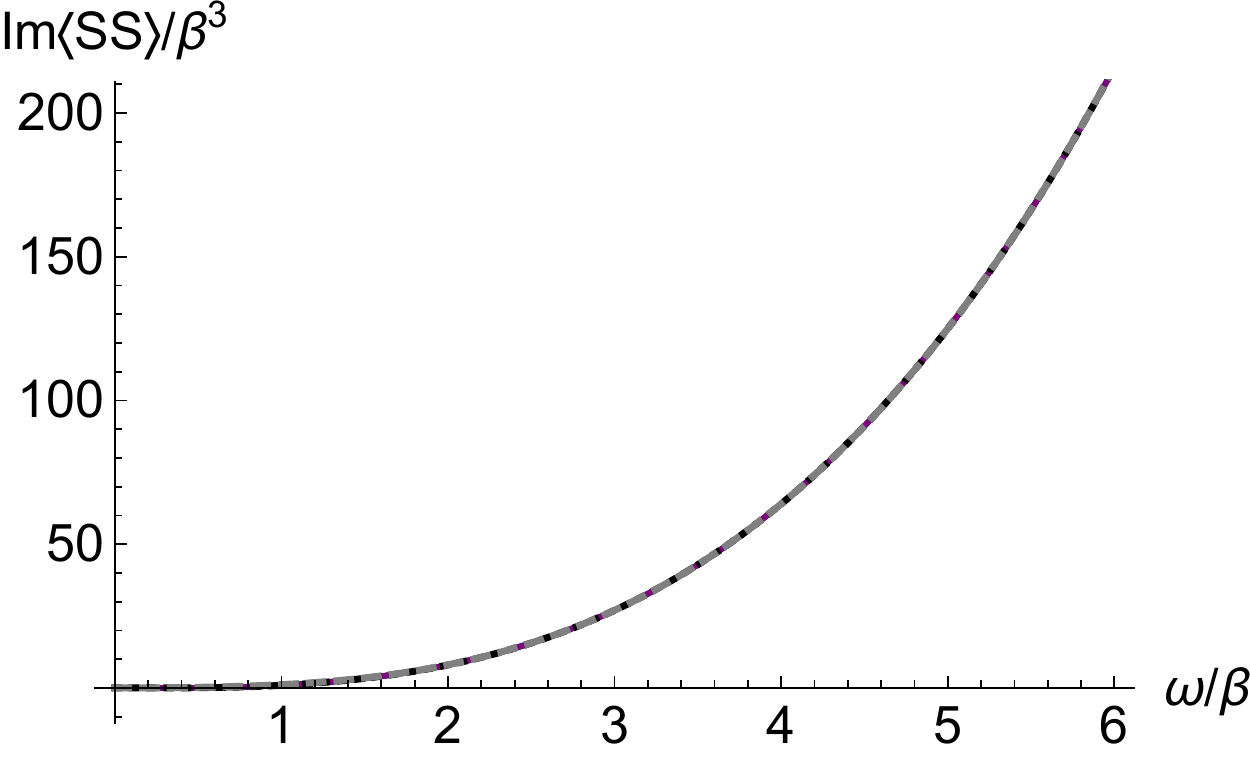} \label{}}
  \caption{$\langle \nJ\nS \rangle$, $\langle \nQ\nS \rangle$, $\langle \nS\nS \rangle$ for $\mu=0$.  $T/T_c = 13.2, 3.5, 1, 0.95, 0.7, 0.4, 0.25$  (dashed, dotted, red, orange, green, blue, purple) } 
            \label{fig:app2}
\end{figure}

\begin{figure}[!h]
 \centering
     {\includegraphics[width=4.5cm]{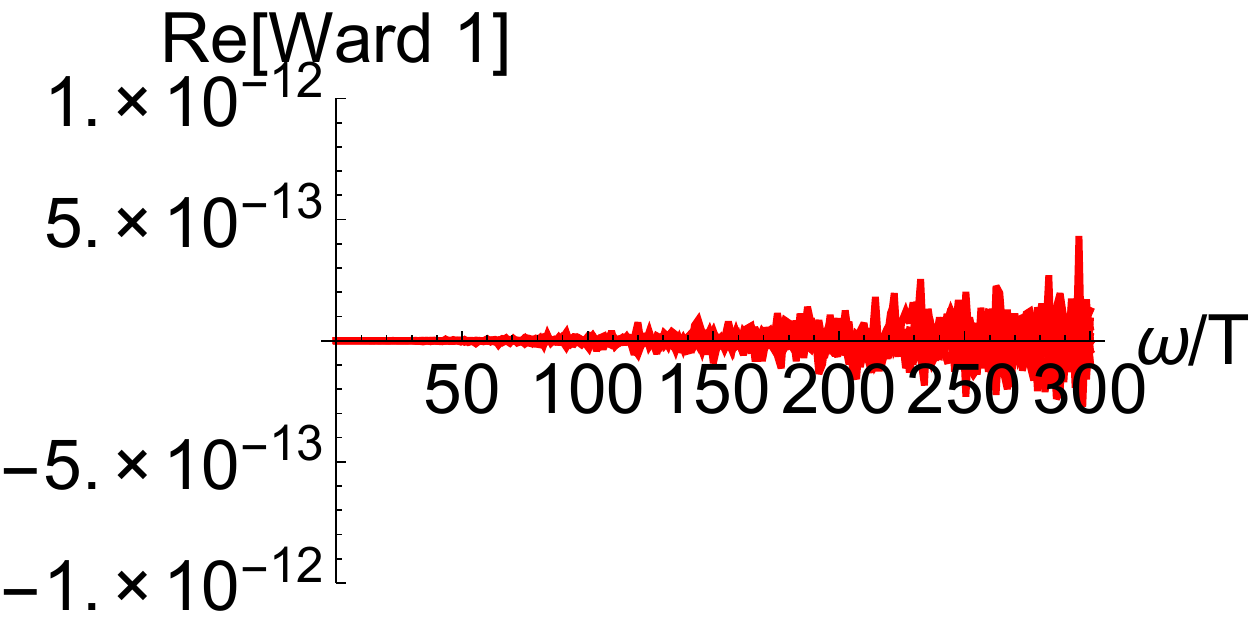} \label{}}\hspace{3mm}
   {\includegraphics[width=4.5cm]{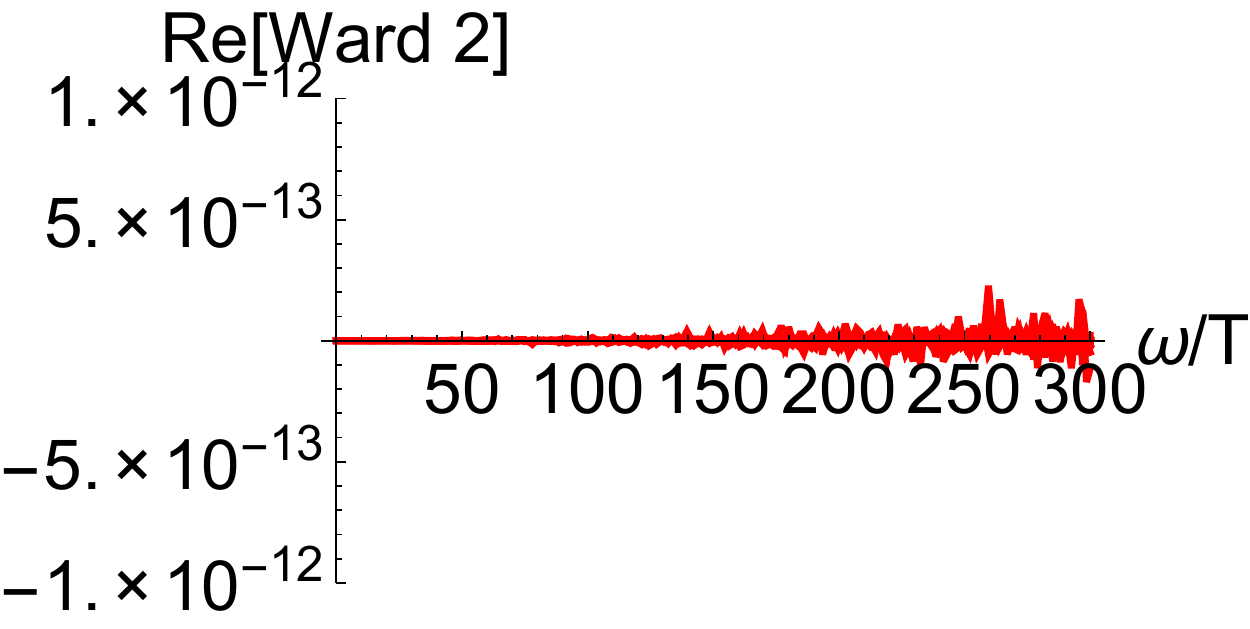} \label{}}\hspace{3mm}
   {\includegraphics[width=4.5cm]{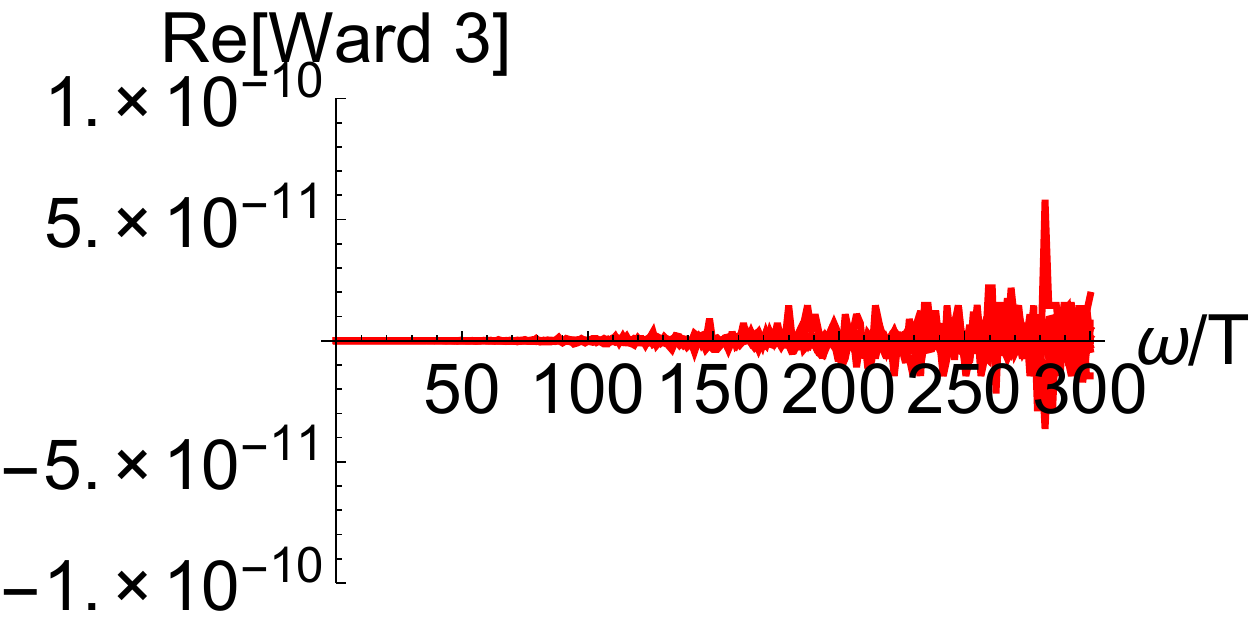} \label{}}
\subfigure[ Ward 1:  \eqref{Ward001}]
 {\includegraphics[width=4.5cm]{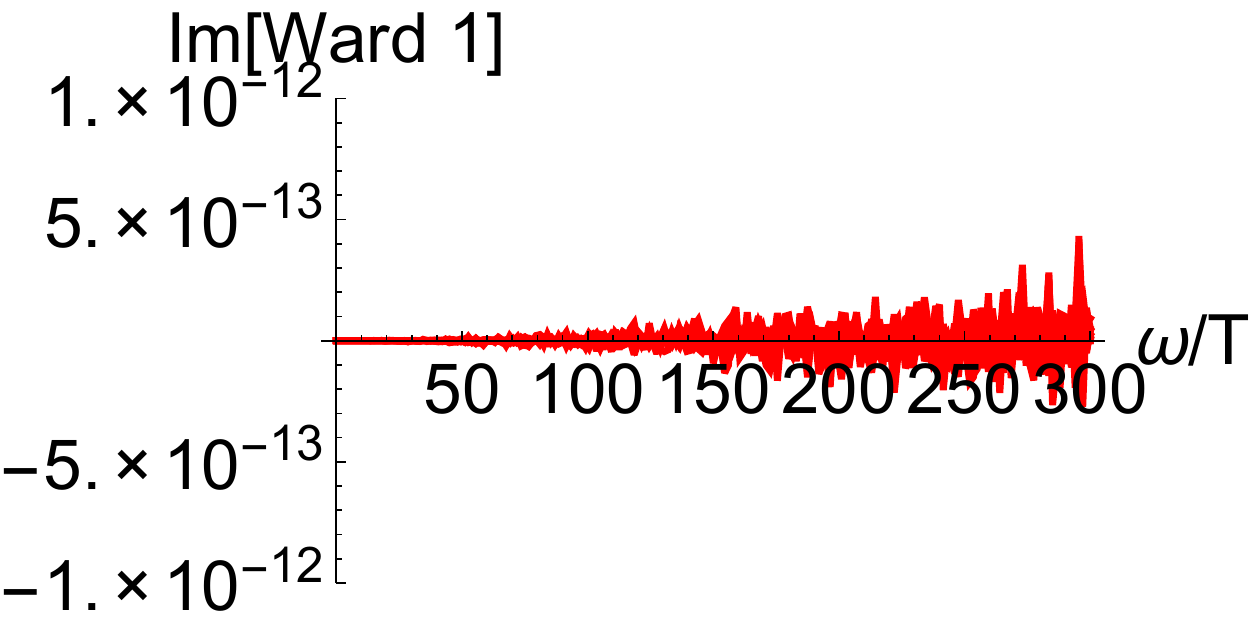} \label{}}\hspace{3mm}
 \subfigure[ Ward 2:  \eqref{Ward002} ]
   {\includegraphics[width=4.5cm]{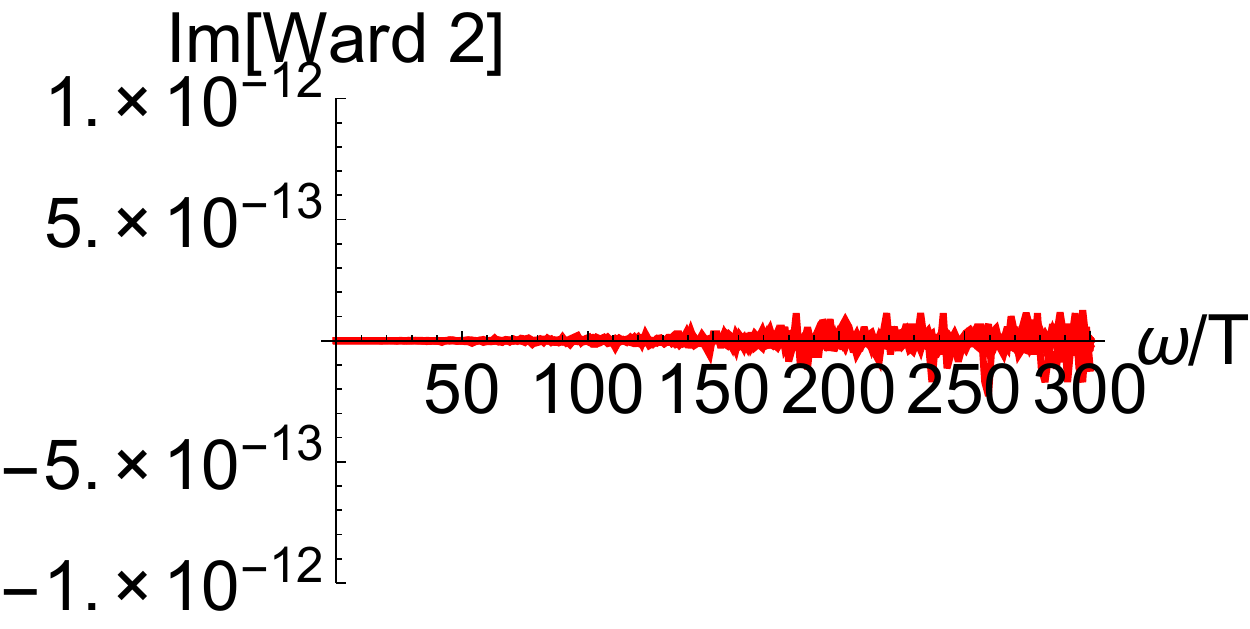} \label{}}\hspace{3mm}
 \subfigure[ Ward 3:  \eqref{Ward003} ]
   {\includegraphics[width=4.5cm]{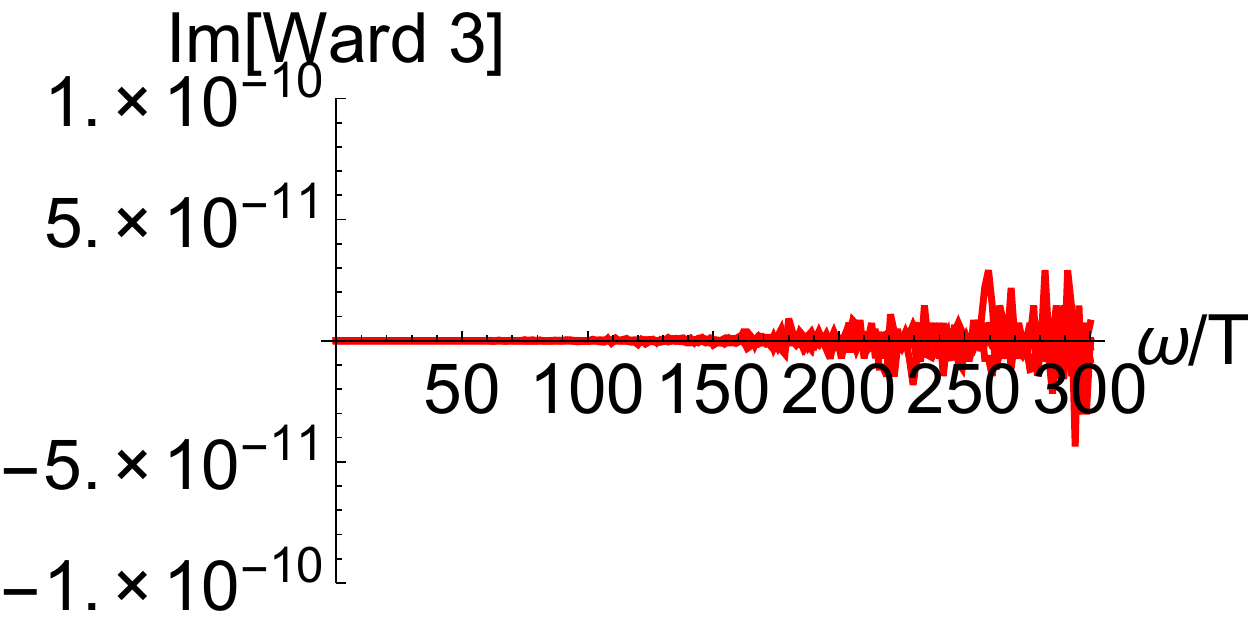} \label{}}
  \caption{Finite $B$ filed case: we plotted all components of the Ward identities \eqref{Ward001}-\eqref{Ward003} together for the case with $\mu/T=1$, $B/T=3$ and $\beta/\mu = 0,0.5,1,1.5$.   } 
            \label{fig:app3}
\end{figure}

As commented at the end of section \ref{sec4}, we have confirmed the Ward identities numerically for other cases too: 1) $B=0, \beta/\mu=0.1$, 2) $B=0, \mu =0$, 3) $B\ne0$. For completeness, we show here the numerical data of $\langle \nJ\nS \rangle$, $\langle \nQ\nS \rangle$, $\langle \nS\nS \rangle$ for (1) and (2) in Figure \ref{fig:app1} and \ref{fig:app2} respectively. For electric, thermoelectric and thermal conductivities we refer to \cite{Kim:2014bza, Kim:2015wba}. In Figure \ref{fig:app3} we show the numerical results of Ward identites for (3).
\newpage

\providecommand{\href}[2]{#2}\begingroup\raggedright\endgroup


\begin{thebibliography}{10}

\bibitem{CasalderreySolana:2011us}
J.~Casalderrey-Solana, H.~Liu, D.~Mateos, K.~Rajagopal and U.~A. Wiedemann,
  \emph{{Gauge/String Duality, Hot QCD and Heavy Ion Collisions}},
  \href{http://arxiv.org/abs/1101.0618}{{\tt 1101.0618}}.

\bibitem{Hartnoll:2009sz}
S.~A. Hartnoll, \emph{{Lectures on holographic methods for condensed matter
  physics}},
  \href{http://dx.doi.org/10.1088/0264-9381/26/22/224002}{\emph{Class.Quant.Grav.}
  {\bf 26} (2009) 224002}, [\href{http://arxiv.org/abs/0903.3246}{{\tt
  0903.3246}}].

\bibitem{Herzog:2009xv}
C.~P. Herzog, \emph{{Lectures on Holographic Superfluidity and
  Superconductivity}},
  \href{http://dx.doi.org/10.1088/1751-8113/42/34/343001}{\emph{J.Phys.A} {\bf
  A42} (2009) 343001}, [\href{http://arxiv.org/abs/0904.1975}{{\tt
  0904.1975}}].

\bibitem{Iqbal:2011ae}
N.~Iqbal, H.~Liu and M.~Mezei, \emph{{Lectures on holographic non-Fermi liquids
  and quantum phase transitions}},  \href{http://arxiv.org/abs/1110.3814}{{\tt
  1110.3814}}.

\bibitem{Hartnoll:2008vx}
S.~A. Hartnoll, C.~P. Herzog and G.~T. Horowitz, \emph{{Building a Holographic
  Superconductor}},
  \href{http://dx.doi.org/10.1103/PhysRevLett.101.031601}{\emph{Phys.Rev.Lett.}
  {\bf 101} (2008) 031601}, [\href{http://arxiv.org/abs/0803.3295}{{\tt
  0803.3295}}].

\bibitem{Hartnoll:2008kx}
S.~A. Hartnoll, C.~P. Herzog and G.~T. Horowitz, \emph{{Holographic
  Superconductors}},
  \href{http://dx.doi.org/10.1088/1126-6708/2008/12/015}{\emph{JHEP} {\bf 0812}
  (2008) 015}, [\href{http://arxiv.org/abs/0810.1563}{{\tt 0810.1563}}].

\bibitem{Horowitz:2010gk}
G.~T. Horowitz, \emph{{Introduction to Holographic Superconductors}},
  \href{http://arxiv.org/abs/1002.1722}{{\tt 1002.1722}}.

\bibitem{Cai:2015cya}
R.-G. Cai, L.~Li, L.-F. Li and R.-Q. Yang, \emph{{Introduction to Holographic
  Superconductor Models}},
  \href{http://dx.doi.org/10.1007/s11433-015-5676-5}{\emph{Sci. China Phys.
  Mech. Astron.} {\bf 58} (2015) 060401},
  [\href{http://arxiv.org/abs/1502.00437}{{\tt 1502.00437}}].

\bibitem{Horowitz:2012ky}
G.~T. Horowitz, J.~E. Santos and D.~Tong, \emph{{Optical Conductivity with
  Holographic Lattices}},
  \href{http://dx.doi.org/10.1007/JHEP07(2012)168}{\emph{JHEP} {\bf 1207}
  (2012) 168}, [\href{http://arxiv.org/abs/1204.0519}{{\tt 1204.0519}}].

\bibitem{Horowitz:2012gs}
G.~T. Horowitz, J.~E. Santos and D.~Tong, \emph{{Further Evidence for
  Lattice-Induced Scaling}},
  \href{http://dx.doi.org/10.1007/JHEP11(2012)102}{\emph{JHEP} {\bf 1211}
  (2012) 102}, [\href{http://arxiv.org/abs/1209.1098}{{\tt 1209.1098}}].

\bibitem{Ling:2013nxa}
Y.~Ling, C.~Niu, J.-P. Wu and Z.-Y. Xian, \emph{{Holographic Lattice in
  Einstein-Maxwell-Dilaton Gravity}},
  \href{http://dx.doi.org/10.1007/JHEP11(2013)006}{\emph{JHEP} {\bf 1311}
  (2013) 006}, [\href{http://arxiv.org/abs/1309.4580}{{\tt 1309.4580}}].

\bibitem{Chesler:2013qla}
P.~Chesler, A.~Lucas and S.~Sachdev, \emph{{Conformal field theories in a
  periodic potential: results from holography and field theory}},
  \href{http://dx.doi.org/10.1103/PhysRevD.89.026005}{\emph{Phys.Rev.} {\bf
  D89} (2014) 026005}, [\href{http://arxiv.org/abs/1308.0329}{{\tt
  1308.0329}}].

\bibitem{Donos:2014yya}
A.~Donos and J.~P. Gauntlett, \emph{{The thermoelectric properties of
  inhomogeneous holographic lattices}},
  \href{http://arxiv.org/abs/1409.6875}{{\tt 1409.6875}}.

\bibitem{Vegh:2013sk}
D.~Vegh, \emph{{Holography without translational symmetry}},
  \href{http://arxiv.org/abs/1301.0537}{{\tt 1301.0537}}.

\bibitem{Davison:2013jba}
R.~A. Davison, \emph{{Momentum relaxation in holographic massive gravity}},
  \href{http://dx.doi.org/10.1103/PhysRevD.88.086003}{\emph{Phys.Rev.} {\bf
  D88} (2013) 086003}, [\href{http://arxiv.org/abs/1306.5792}{{\tt
  1306.5792}}].

\bibitem{Blake:2013bqa}
M.~Blake and D.~Tong, \emph{{Universal Resistivity from Holographic Massive
  Gravity}},
  \href{http://dx.doi.org/10.1103/PhysRevD.88.106004}{\emph{Phys.Rev.} {\bf
  D88} (2013) 106004}, [\href{http://arxiv.org/abs/1308.4970}{{\tt
  1308.4970}}].

\bibitem{Blake:2013owa}
M.~Blake, D.~Tong and D.~Vegh, \emph{{Holographic Lattices Give the Graviton a
  Mass}},
  \href{http://dx.doi.org/10.1103/PhysRevLett.112.071602}{\emph{Phys.Rev.Lett.}
  {\bf 112} (2014) 071602}, [\href{http://arxiv.org/abs/1310.3832}{{\tt
  1310.3832}}].

\bibitem{Amoretti:2014zha}
A.~Amoretti, A.~Braggio, N.~Maggiore, N.~Magnoli and D.~Musso,
  \emph{{Thermo-electric transport in gauge/gravity models with momentum
  dissipation}},  \href{http://arxiv.org/abs/1406.4134}{{\tt 1406.4134}}.

\bibitem{Amoretti:2014mma}
A.~Amoretti, A.~Braggio, N.~Maggiore, N.~Magnoli and D.~Musso, \emph{{Analytic
  DC thermo-electric conductivities in holography with massive gravitons}},
  \href{http://arxiv.org/abs/1407.0306}{{\tt 1407.0306}}.

\bibitem{Amoretti:2015gna}
A.~Amoretti and D.~Musso, \emph{{Magneto-transport from momentum dissipating
  holography}}, \href{http://dx.doi.org/10.1007/JHEP09(2015)094}{\emph{JHEP}
  {\bf 09} (2015) 094}, [\href{http://arxiv.org/abs/1502.02631}{{\tt
  1502.02631}}].

\bibitem{Donos:2013eha}
A.~Donos and J.~P. Gauntlett, \emph{{Holographic Q-lattices}},
  \href{http://dx.doi.org/10.1007/JHEP04(2014)040}{\emph{JHEP} {\bf 1404}
  (2014) 040}, [\href{http://arxiv.org/abs/1311.3292}{{\tt 1311.3292}}].

\bibitem{Donos:2014uba}
A.~Donos and J.~P. Gauntlett, \emph{{Novel metals and insulators from
  holography}}, \href{http://dx.doi.org/10.1007/JHEP06(2014)007}{\emph{JHEP}
  {\bf 1406} (2014) 007}, [\href{http://arxiv.org/abs/1401.5077}{{\tt
  1401.5077}}].

\bibitem{Ling:2014bda}
Y.~Ling, P.~Liu, C.~Niu, J.-P. Wu and Z.-Y. Xian, \emph{{Holographic fermionic
  system with dipole coupling on Q-lattice}},
  \href{http://dx.doi.org/10.1007/JHEP12(2014)149}{\emph{JHEP} {\bf 12} (2014)
  149}, [\href{http://arxiv.org/abs/1410.7323}{{\tt 1410.7323}}].

\bibitem{Ling:2015exa}
Y.~Ling, P.~Liu and J.-P. Wu, \emph{{A novel insulator by holographic
  Q-lattices}}, \href{http://dx.doi.org/10.1007/JHEP02(2016)075}{\emph{JHEP}
  {\bf 02} (2016) 075}, [\href{http://arxiv.org/abs/1510.05456}{{\tt
  1510.05456}}].

\bibitem{Ling:2016ewj}
Y.~Ling, P.~Liu, C.~Niu and J.-P. Wu, \emph{{The pseudo-gap phase and the
  duality in holographic fermionic system with dipole coupling on Q-lattice}},
  \href{http://dx.doi.org/10.1088/1674-1137/40/4/043102}{\emph{Chin. Phys.}
  {\bf C40} (2016) 043102}, [\href{http://arxiv.org/abs/1602.06062}{{\tt
  1602.06062}}].

\bibitem{Andrade:2013gsa}
T.~Andrade and B.~Withers, \emph{{A simple holographic model of momentum
  relaxation}}, \href{http://dx.doi.org/10.1007/JHEP05(2014)101}{\emph{JHEP}
  {\bf 1405} (2014) 101}, [\href{http://arxiv.org/abs/1311.5157}{{\tt
  1311.5157}}].

\bibitem{Gouteraux:2014hca}
B.~Gout{\'e}raux, \emph{{Charge transport in holography with momentum
  dissipation}}, \href{http://dx.doi.org/10.1007/JHEP04(2014)181}{\emph{JHEP}
  {\bf 1404} (2014) 181}, [\href{http://arxiv.org/abs/1401.5436}{{\tt
  1401.5436}}].

\bibitem{Taylor:2014tka}
M.~Taylor and W.~Woodhead, \emph{{Inhomogeneity simplified}},
  \href{http://arxiv.org/abs/1406.4870}{{\tt 1406.4870}}.

\bibitem{Kim:2014bza}
K.-Y. Kim, K.~K. Kim, Y.~Seo and S.-J. Sin, \emph{{Coherent/incoherent metal
  transition in a holographic model}},
  \href{http://arxiv.org/abs/1409.8346}{{\tt 1409.8346}}.

\bibitem{Bardoux:2012aw}
Y.~Bardoux, M.~M. Caldarelli and C.~Charmousis, \emph{{Shaping black holes with
  free fields}}, \href{http://dx.doi.org/10.1007/JHEP05(2012)054}{\emph{JHEP}
  {\bf 1205} (2012) 054}, [\href{http://arxiv.org/abs/1202.4458}{{\tt
  1202.4458}}].

\bibitem{Iizuka:2012wt}
N.~Iizuka and K.~Maeda, \emph{{Study of Anisotropic Black Branes in
  Asymptotically anti-de Sitter}},
  \href{http://dx.doi.org/10.1007/JHEP07(2012)129}{\emph{JHEP} {\bf 1207}
  (2012) 129}, [\href{http://arxiv.org/abs/1204.3008}{{\tt 1204.3008}}].

\bibitem{Cheng:2014qia}
L.~Cheng, X.-H. Ge and S.-J. Sin, \emph{{Anisotropic plasma at finite $U(1)$
  chemical potential}},
  \href{http://dx.doi.org/10.1007/JHEP07(2014)083}{\emph{JHEP} {\bf 07} (2014)
  083}, [\href{http://arxiv.org/abs/1404.5027}{{\tt 1404.5027}}].

\bibitem{Fang:2015dia}
L.-Q. Fang, X.-M. Kuang, B.~Wang and J.-P. Wu, \emph{{Fermionic phase
  transition induced by the effective impurity in holography}},
  \href{http://dx.doi.org/10.1007/JHEP11(2015)134}{\emph{JHEP} {\bf 11} (2015)
  134}, [\href{http://arxiv.org/abs/1507.03121}{{\tt 1507.03121}}].

\bibitem{Seo:2015pug}
Y.~Seo, K.-Y. Kim, K.~K. Kim and S.-J. Sin, \emph{{Character of Matter in
  Holography: Spin-Orbit Interaction}},
  \href{http://arxiv.org/abs/1512.08916}{{\tt 1512.08916}}.

\bibitem{Andrade:2015iyf}
T.~Andrade and A.~Krikun, \emph{{Commensurability effects in holographic
  homogeneous lattices}},  \href{http://arxiv.org/abs/1512.02465}{{\tt
  1512.02465}}.

\bibitem{Andrade:2016tbr}
T.~Andrade, \emph{{A simple model of momentum relaxation in Lifshitz
  holography}},  \href{http://arxiv.org/abs/1602.00556}{{\tt 1602.00556}}.

\bibitem{Donos:2012js}
A.~Donos and S.~A. Hartnoll, \emph{{Interaction-driven localization in
  holography}}, \href{http://dx.doi.org/10.1038/nphys2701}{\emph{Nature Phys.}
  {\bf 9} (2013) 649--655}, [\href{http://arxiv.org/abs/1212.2998}{{\tt
  1212.2998}}].

\bibitem{Donos:2014oha}
A.~Donos, B.~Gout{\'e}raux and E.~Kiritsis, \emph{{Holographic Metals and
  Insulators with Helical Symmetry}},
  \href{http://arxiv.org/abs/1406.6351}{{\tt 1406.6351}}.

\bibitem{Donos:2014gya}
A.~Donos, J.~P. Gauntlett and C.~Pantelidou, \emph{{Conformal field theories in
  $d=4$ with a helical twist}},  \href{http://arxiv.org/abs/1412.3446}{{\tt
  1412.3446}}.

\bibitem{Horowitz:2013jaa}
G.~T. Horowitz and J.~E. Santos, \emph{{General Relativity and the Cuprates}},
  \href{http://arxiv.org/abs/1302.6586}{{\tt 1302.6586}}.

\bibitem{Zeng:2014uoa}
H.~B. Zeng and J.-P. Wu, \emph{{Holographic superconductors from the massive
  gravity}},
  \href{http://dx.doi.org/10.1103/PhysRevD.90.046001}{\emph{Phys.Rev.} {\bf
  D90} (2014) 046001}, [\href{http://arxiv.org/abs/1404.5321}{{\tt
  1404.5321}}].

\bibitem{Ling:2014laa}
Y.~Ling, P.~Liu, C.~Niu, J.-P. Wu and Z.-Y. Xian, \emph{{Holographic
  Superconductor on Q-lattice}},  \href{http://arxiv.org/abs/1410.6761}{{\tt
  1410.6761}}.

\bibitem{Andrade:2014xca}
T.~Andrade and S.~A. Gentle, \emph{{Relaxed superconductors}},
  \href{http://arxiv.org/abs/1412.6521}{{\tt 1412.6521}}.

\bibitem{Kim:2015dna}
K.-Y. Kim, K.~K. Kim and M.~Park, \emph{{A Simple Holographic Superconductor
  with Momentum Relaxation}},  \href{http://arxiv.org/abs/1501.00446}{{\tt
  1501.00446}}.

\bibitem{Erdmenger:2015qqa}
J.~Erdmenger, B.~Herwerth, S.~Klug, R.~Meyer and K.~Schalm, \emph{{S-Wave
  Superconductivity in Anisotropic Holographic Insulators}},
  \href{http://dx.doi.org/10.1007/JHEP05(2015)094}{\emph{JHEP} {\bf 05} (2015)
  094}, [\href{http://arxiv.org/abs/1501.07615}{{\tt 1501.07615}}].

\bibitem{Baggioli:2015zoa}
M.~Baggioli and M.~Goykhman, \emph{{Phases of holographic superconductors with
  broken translational symmetry}},
  \href{http://dx.doi.org/10.1007/JHEP07(2015)035}{\emph{JHEP} {\bf 07} (2015)
  035}, [\href{http://arxiv.org/abs/1504.05561}{{\tt 1504.05561}}].

\bibitem{Baggioli:2015dwa}
M.~Baggioli and M.~Goykhman, \emph{{Under The Dome: Doped holographic
  superconductors with broken translational symmetry}},
  \href{http://dx.doi.org/10.1007/JHEP01(2016)011}{\emph{JHEP} {\bf 01} (2016)
  011}, [\href{http://arxiv.org/abs/1510.06363}{{\tt 1510.06363}}].

\bibitem{Koga:2014hwa}
J.-i. Koga, K.~Maeda and K.~Tomoda, \emph{{Holographic superconductor model in
  a spatially anisotropic background}},
  \href{http://dx.doi.org/10.1103/PhysRevD.89.104024}{\emph{Phys.Rev.} {\bf
  D89} (2014) 104024}, [\href{http://arxiv.org/abs/1401.6501}{{\tt
  1401.6501}}].

\bibitem{Bai:2014poa}
X.~Bai, B.-H. Lee, M.~Park and K.~Sunly, \emph{{Dynamical Condensation in a
  Holographic Superconductor Model with Anisotropy}},
  \href{http://dx.doi.org/10.1007/JHEP09(2014)054}{\emph{JHEP} {\bf 1409}
  (2014) 054}, [\href{http://arxiv.org/abs/1405.1806}{{\tt 1405.1806}}].

\bibitem{Donos:2014cya}
A.~Donos and J.~P. Gauntlett, \emph{{Thermoelectric DC conductivities from
  black hole horizons}},  \href{http://arxiv.org/abs/1406.4742}{{\tt
  1406.4742}}.

\bibitem{Kim:2015sma}
K.-Y. Kim, K.~K. Kim, Y.~Seo and S.-J. Sin, \emph{{Gauge Invariance and
  Holographic Renormalization}},
  \href{http://dx.doi.org/10.1016/j.physletb.2015.07.058}{\emph{Phys. Lett.}
  {\bf B749} (2015) 108--114}, [\href{http://arxiv.org/abs/1502.02100}{{\tt
  1502.02100}}].

\bibitem{Kim:2015wba}
K.-Y. Kim, K.~K. Kim, Y.~Seo and S.-J. Sin, \emph{{Thermoelectric
  Conductivities at Finite Magnetic Field and the Nernst Effect}},
  \href{http://dx.doi.org/10.1007/JHEP07(2015)027}{\emph{JHEP} {\bf 07} (2015)
  027}, [\href{http://arxiv.org/abs/1502.05386}{{\tt 1502.05386}}].

\bibitem{Banks:2015aca}
E.~Banks and J.~P. Gauntlett, \emph{{A new phase for the anisotropic N=4 super
  Yang-Mills plasma}},
  \href{http://dx.doi.org/10.1007/JHEP09(2015)126}{\emph{JHEP} {\bf 09} (2015)
  126}, [\href{http://arxiv.org/abs/1506.07176}{{\tt 1506.07176}}].

\bibitem{Banks:2016fab}
E.~Banks, \emph{{Phase transitions of an anisotropic N=4 super Yang-Mills
  plasma via holography}},  \href{http://arxiv.org/abs/1604.03552}{{\tt
  1604.03552}}.

\bibitem{Hartnoll:2007ip}
S.~A. Hartnoll and C.~P. Herzog, \emph{{Ohm's Law at strong coupling: S duality
  and the cyclotron resonance}},
  \href{http://dx.doi.org/10.1103/PhysRevD.76.106012}{\emph{Phys.Rev.} {\bf
  D76} (2007) 106012}, [\href{http://arxiv.org/abs/0706.3228}{{\tt
  0706.3228}}].

\bibitem{Homes:2005aa}
C.~C. Homes, S.~V. Dordevic, T.~Valla and M.~Strongin, \emph{Scaling of the
  superfluid density in high-temperature superconductors}, {\emph{Phys. Rev. B
  72,} {\bf 134517} (2005) (8},
  [\href{http://arxiv.org/abs/cond-mat/0410719}{{\tt cond-mat/0410719}}].

\bibitem{Homes:2004wv}
C.~Homes, S.~Dordevic, M.~Strongin, D.~Bonn, R.~Liang et~al., \emph{{Universal
  scaling relation in high-temperature superconductors}},
  \href{http://dx.doi.org/10.1038/nature02673}{\emph{Nature} {\bf 430} (2004)
  539}, [\href{http://arxiv.org/abs/cond-mat/0404216}{{\tt cond-mat/0404216}}].

\bibitem{Erdmenger:2012ik}
J.~Erdmenger, P.~Kerner and S.~Muller, \emph{{Towards a Holographic Realization
  of Homes' Law}}, \href{http://dx.doi.org/10.1007/JHEP10(2012)021}{\emph{JHEP}
  {\bf 1210} (2012) 021}, [\href{http://arxiv.org/abs/1206.5305}{{\tt
  1206.5305}}].

\bibitem{Sachdev:2011cs}
S.~Sachdev and B.~Keimer, \emph{{Quantum Criticality}},
  \href{http://dx.doi.org/10.1063/1.3554314}{\emph{Phys. Today} {\bf 64N2}
  (2011) 29}, [\href{http://arxiv.org/abs/1102.4628}{{\tt 1102.4628}}].

\bibitem{Zaanen:2004aa}
J.~Zaanen, \emph{Superconductivity: Why the temperature is high},
  {\emph{Nature} {\bf 430} (07, 2004) 512--513}.

\bibitem{Bianchi:2001kw}
M.~Bianchi, D.~Z. Freedman and K.~Skenderis, \emph{{Holographic
  renormalization}},
  \href{http://dx.doi.org/10.1016/S0550-3213(02)00179-7}{\emph{Nucl.Phys.} {\bf
  B631} (2002) 159--194}, [\href{http://arxiv.org/abs/hep-th/0112119}{{\tt
  hep-th/0112119}}].

\bibitem{Hartnoll:2007ai}
S.~A. Hartnoll and P.~Kovtun, \emph{{Hall conductivity from dyonic black
  holes}}, \href{http://dx.doi.org/10.1103/PhysRevD.76.066001}{\emph{Phys.Rev.}
  {\bf D76} (2007) 066001}, [\href{http://arxiv.org/abs/0704.1160}{{\tt
  0704.1160}}].

\bibitem{Albash:2009iq}
T.~Albash and C.~V. Johnson, \emph{{Vortex and Droplet Engineering in
  Holographic Superconductors}},
  \href{http://dx.doi.org/10.1103/PhysRevD.80.126009}{\emph{Phys. Rev.} {\bf
  D80} (2009) 126009}, [\href{http://arxiv.org/abs/0906.1795}{{\tt
  0906.1795}}].

\bibitem{Maeda:2009vf}
K.~Maeda, M.~Natsuume and T.~Okamura, \emph{{Vortex lattice for a holographic
  superconductor}},
  \href{http://dx.doi.org/10.1103/PhysRevD.81.026002}{\emph{Phys. Rev.} {\bf
  D81} (2010) 026002}, [\href{http://arxiv.org/abs/0910.4475}{{\tt
  0910.4475}}].

\bibitem{Son:2002sd}
D.~T. Son and A.~O. Starinets, \emph{{Minkowski space correlators in AdS / CFT
  correspondence: Recipe and applications}}, {\emph{JHEP} {\bf 0209} (2002)
  042}, [\href{http://arxiv.org/abs/hep-th/0205051}{{\tt hep-th/0205051}}].

\bibitem{Lindgren:2015lia}
J.~Lindgren, I.~Papadimitriou, A.~Taliotis and J.~Vanhoof, \emph{{Holographic
  Hall conductivities from dyonic backgrounds}},
  \href{http://dx.doi.org/10.1007/JHEP07(2015)094}{\emph{JHEP} {\bf 07} (2015)
  094}, [\href{http://arxiv.org/abs/1505.04131}{{\tt 1505.04131}}].

\bibitem{Herzog:2011ec}
C.~P. Herzog, N.~Lisker, P.~Surowka and A.~Yarom, \emph{{Transport in
  holographic superfluids}},
  \href{http://dx.doi.org/10.1007/JHEP08(2011)052}{\emph{JHEP} {\bf 08} (2011)
  052}, [\href{http://arxiv.org/abs/1101.3330}{{\tt 1101.3330}}].

\bibitem{WIP}
K.-Y. Kim and C.~Niu, \emph{work in progress}, .

\end{thebibliography}
\end{document}